\documentclass{aa}
\usepackage{txfonts}
\usepackage{graphicx}
\usepackage{subfigure}
\usepackage{natbib}
\usepackage{longtable}
\usepackage{lscape}
\usepackage{rotating}
\usepackage{color}
\bibpunct{(}{)}{;}{a}{}{,}  

\begin{document}
\parindent 0pt

\newcommand{\noopsort}[1]{}
\newcommand{\singleletter}[1]{#1}

\title{A RAVE investigation on Galactic open clusters \\
       I. Radial velocities and metallicities}
\author{C. Conrad\inst{\ref{inst1}} \and R.-D. Scholz\inst{\ref{inst1}} \and  N.V. Kharchenko\inst{\ref{inst1},\ref{inst2},\ref{inst3}} \and A.E. Piskunov\inst{\ref{inst1},\ref{inst2},\ref{inst4}}
   \and E. Schilbach\inst{\ref{inst2}} \and S. R\"oser\inst{\ref{inst2}} \and C. Boeche\inst{\ref{inst2}} \and \\
        G. Kordopatis\inst{\ref{inst5}} \and A. Siebert\inst{\ref{inst6}} \and M. Williams\inst{\ref{inst1}} \and U. Munari\inst{\ref{inst7}} \and G. Matijevi{\v c}\inst{\ref{inst8}} 
   \and E.K. Grebel\inst{\ref{inst2}} \and  T. Zwitter \inst{\ref{inst9}, \ref{inst10}} \and R.S. de Jong\inst{\ref{inst1}} \and \\
        M. Steinmetz\inst{\ref{inst1}} \and G. Gilmore\inst{\ref{inst5}} \and G. Seabroke\inst{\ref{inst11}} \and K. Freeman\inst{\ref{inst12}} \and J.F. Navarro \inst{\ref{inst13}} 
   \and Q. Parker\inst{\ref{inst14},\ref{inst15},\ref{inst16}} \and  W. Reid\inst{\ref{inst14}, \ref{inst15}} \and \\
        F. Watson\inst{\ref{inst17}} \and B.K. Gibson\inst{\ref{inst18}} \and O. Bienaym\'{e}\inst{\ref{inst6}} \and R. Wyse\inst{\ref{inst19}} \and J. Bland-Hawthorn\inst{\ref{inst20}} 
   \and A. Siviero\inst{\ref{inst1}, \ref{inst21}}}

\institute{Leibniz-Institut f\"ur Astrophysik Potsdam (AIP), An der Sternwarte 16, 14482 Potsdam, Germany                                                      \label{inst1}
      \and Astronomisches Rechen-Institut, Zentrum f\"ur Astronomie der Universit\"at Heidelberg, M\"onchhofstra\ss e 12$-$14, 69120 Heidelberg, Germany       \label{inst2}
      \and Main Astronomical Observatory, 27 Academica Zabolotnogo Str., 03680 Kiev, Ukraine                                                                   \label{inst3}
      \and Institute of Astronomy, Russian Acad. Sci., 48 Pyatnitskaya Str., 109017 Moscow, Russia                                                             \label{inst4}
      \and Institute of Astronomy, Cambridge University, Madingley Road, Cambridge CB3 0HA, United Kingdom                                                     \label{inst5}
      \and Observatoire astronomique de Strasbourg, Universit\'e de Strasbourg, CNRS, UMR 7550, 11 rue de l'Universit\'e, 67000 Strasbourg, France             \label{inst6}
      \and INAF Astronomical Observatory of Padova, 36012 Asiago (VI), Italy                                                                                   \label{inst7}
      \and Dept. of Astronomy and Astrophysics, Villanova University, 800 E, Lancaster Ave, Villanova, PA 19085, USA                                           \label{inst8}
      \and Faculty of Mathematics and Physics, University of Ljubljana, Jadranska 19, Ljubljana, Slovenia                                                      \label{inst9}
      \and Center of excellence Space-SI, Askerceva 12, Ljubljana, Slovenia                                                                                    \label{inst10}
      \and Mullard Space Science Laboratory, University College London, Holmbury St Mary, Dorking, RH5 6NT, United Kingdom                                     \label{inst11}
      \and Research School of Astronomy and Astrophysics, Australian National University, Cotter Rd., Weston, ACT 2611, Australia                              \label{inst12}
      \and Department of Physics and Astronomy, University of Victoria, Victoria, BC, Canada V8P5C2                                                            \label{inst13}
      \and Department of Physics and Astronomy, Macquarie University, Sydney, NSW 2109, Australia                                                              \label{inst14}
      \and Research Centre for Astronomy, Astrophysics and Astrophotonics, Macquarie University, Sydney, NSW 2109, Australia                                   \label{inst15}
      \and Australian Astronomical Observatory, PO Box 296, Epping, NSW 1710, Australia                                                                        \label{inst16}
      \and Australian Astronomical Observatory, 105 Delhi Road, PO Box 915, North Ryde, NSW 1670, Australia                                                    \label{inst17}
      \and Jeremiah Horrocks Institute, University of Central Lancashire, Preston, PR1 2HE, United Kingdom                                                     \label{inst18}
      \and Johns Hopkins University, 3400 N Charles Street, Baltimore, MD 21218, USA                                                                           \label{inst19}
      \and Sydney Institute for Astronomy, School of Physics A28, University of Sydney, NSW 2006, Australia                                                    \label{inst20}
      \and Department of Physics and Astronomy, Padova University, Vicolo dell’Osservatorio 2, I-35122 Padova, Italy                                           \label{inst21}
}

\titlerunning{A RAVE investigation on Galactic open clusters I. Radial velocities and metallicities}
\authorrunning{C. Conrad et al.}

\abstract 
{Galactic open clusters (OCs) mainly belong to the young stellar population in the Milky Way disk, but are there groups and complexes of OCs that possibly define an additional level in hierarchical
star formation? Current compilations are too incomplete to address this question, especially regarding radial velocities (RVs) and metallicities ($[M/H]$).}
{Here we provide and discuss newly obtained RV and $[M/H]$ data, which will enable us to reinvestigate potential groupings of open clusters and associations.}
{We extracted additional RVs and $[M/H]$ from the RAdial Velocity Experiment (RAVE) via a cross-match with the Catalogue of Stars in Open Cluster Areas (CSOCA). For the identified OCs in RAVE we
derived $\overline{RV}$ and $\overline{[M/H]}$ from a cleaned working sample and compared the results with previous findings.}
{Although our RAVE sample does not show the same accuracy as the entire survey, we were able to derive reliable $\overline{RV}$ for 110 Galactic open clusters. For 37 OCs we publish $\overline{RV}$
for the first time. Moreover, we determined $\overline{[M/H]}$ for 81 open clusters, extending the number of OCs with $\overline{[M/H]}$ by 69.}
{}


\maketitle

\section{Introduction}

Open clusters (OCs) are birthplaces of stars \citep{Lada2003, Lada2006} and serve as convenient tracers of the young stellar population (age $\lesssim$ 2 Gyr) in the Galactic disk. Because OCs can
harbour up to a few thousand stars, certain parameters, such as age, distance, and velocities, can be derived more accurately for OCs than for isolated stars. In general, OC members are selected from
kinematics, that is, sharing a common motion (mainly proper motion is used), and photometry, that is, following the same isochrone in the colour-magnitude diagram. Cluster samples, reliably cleaned
from fore- and background stars, are ideal targets for systematic investigations of stellar systems and the Milky Way as a whole regarding structure, dynamics, formation, and evolution.\\
Throughout the past decades several comprehensive studies, observational and literature compilations, were carried out to identify and characterise Galactic OCs. One important study was conducted by
\citet{Lynga1987}, providing a catalogue of 1151 OCs partly equipped with distances, ages, and even more sparsely with metallicities. It is often referred to as the Lund catalogue. Another set of
catalogues was provided by \citet{Ruprecht1981}, containing solely central coordinates and identifiers for 137 globular clusters, 1112 open clusters, and 89 associations.\\
The Two Micron All Sky Survey \citep[2MASS;][]{MASS2003} provided a new source for cluster searches. \citet{Bica2003a, Bica2003b} identified 276 infrared clusters and stellar groups as well as 167
embedded clusters related to nebulae. In addition to the identifiers and coordinates, they list angular sizes measured by eye. \citet{Dutra2003} extended these catalogues to the southern hemisphere by
123 clusters, providing the same type of information. Another extensive infrared OC catalogue in 2MASS was generated by \citet{Froebrich2007} near the Galactic disk $(|b|<20^{\circ})$. They provide
coordinates, radii, and stellar densities for 1788 open and globular clusters, including 1021 new objects.\\
In the optical HIPPARCOS\footnote{HIPPARCOS - HIgh Precision PARallax COllecting Satellite} \citep{HIP1997} and TYCHO-2\footnote{The TYCHO catalogues are part of HIPPARCOS} \citep{TYCHO2000}
provided another opportunity for OC searches. \citet{Platais1998} published positions, distances, diameters, ages, and proper motions for 102 clusters and associations in HIPPARCOS, including 82 known
objects and 20 new discoveries. \citet{Alessi2003} detected 11 new OCs in the TYCHO-2 data and list positions, diameters, distances, ages, proper motions, and velocity dispersions.\\
Currently, most known OCs are summarised in two main online compilations. One is the collection of optically visible open clusters and candidates by \citet{Dias2002} (hereafter referred to as
DAML\footnote{DAML - http://www.astro.iag.usp.br/\~{}wilton/;\\ Version 3.3  provided on Jan/10/2013}). It contains all available parameters, such as positions, radii, distances, ages, and proper
motions for 2174 open clusters, including a few associations. Radial velocities (RVs) are given for 542 listings (25\%), and metallicities ($[M/H]$) or iron abundances ($[Fe/H]$) for 201 clusters
(9\%). The second is the WEBDA data base\footnote{WEBDA - http://www.univie.ac.at/webda} created by \citet{WEBDA1988} and maintained by \citet{WEBDA2012}, collecting information on 970 Galactic OCs
and 248 OCs in the Small Magellanic Cloud. For the Galactic OCs they list positions, diameters, distances, ages, proper motions, RVs, and colour excess, if available. The vast majority of WEBDA
entries (910) is included in the DAML.\\
These compilations are essential for comprehensive studies, being the most complete collections of open clusters and associations. However, the information therein is highly inhomogeneous, due to
different data sources and algorithms used for the membership selection and parameter determination. Furthermore, the provided parameters were not transferred to a uniform reference system, which
could induce additional systematic biases, which in turn could lead to false conclusions on the overall characteristics of the OC system.\\
\citet{Kharchenko2005a, Kharchenko2005b} presented the Catalogue of Open Cluster Data (COCD), comprising in total 650 Galactic open clusters and associations (OCs)\footnote{Since there are only seven
compact associations among the 650 entries in the COCD, we refer to all objects as OCs.}. The OCs were extracted from the DAML or were newly discovered by applying a uniform membership selection and
are provided with a mostly homogeneous set of parameters. \citet{Kharchenko2007} extended the RV information in COCD, based on the second edition of the Catalogue of Radial Velocities with Astrometric
Data \citep[CRVAD-2;][]{Kharchenko2007} and literature values. The results were published in the Catalogue of Radial Velocities of Open Clusters and Associations \citep[CRVOCA;][]{Kharchenko2007}.
Currently, this is the only global RV study for OCs.\\
Here we present an update and extension of RV and $[M/H]$ information on OCs in the southern hemisphere, using the RAdial Velocity Experiment \citep[RAVE;][]{RAVE1}. In a second publication (Conrad et
al. in prep.) we will use these additional and mostly homogeneous data, along with previous results, to reinvestigate the proposed OC groups and complexes \citep{Piskunov2006}. This may give us a
hint on how they formed.\\
This publication is structured as follows: in Sect. \ref{data} we briefly describe all catalogues used throughout the paper. In Sect. \ref{stars} we give a detailed description of our quality
requirements to ensure a good working sample and discuss the stellar parameters obtained for RAVE stars in OC areas. In Sect. \ref{cluster} we present the cluster mean values, and in Sect. \ref{disc}
we conclude with a discussion on our results and an outlook on our ongoing project.

\section{Catalogues} \label{data}

\subsection{Catalogue of Open Cluster Data} \label{COCD}

The All-Sky Compiled Catalogue of 2.5 million stars \citep[ASCC-2.5;][]{Kharchenko2001} contains relatively bright stars (V$_{Johnson}$ down to 12.5 mag) listed with proper motions. It was the source
catalogue for compiling the Catalogue of Open Cluster Data \citep[COCD;][]{Kharchenko2005a, Kharchenko2005b}. For the first part of the COCD \citet{Kharchenko2005a} identified ASCC-2.5 stars in areas
around 520 OCs taken from DAML. An independent search for OCs in ASCC-2.5 by \citet{Kharchenko2005b} extended the COCD by 109 previously unknown and 21 additional DAML clusters. The complete COCD
provides centre positions, core radii, tidal radii, distances, ages, and mean proper motions (PMs) for in total 650 OCs. Mean radial velocities ($\overline{RV}$s) are provided for about 50\% of the
listed objects.\\
In addition, \citet{Kharchenko2004b, Kharchenko2005b} published corresponding stellar catalogues for both parts of COCD, called the Catalogue of Stars in Open Cluster Areas (CSOCA). It provides
equatorial coordinates, proper motions, $B$ and $V$ magnitudes, angular distances to the OC centre, as well as RVs, trigonometric parallaxes, and spectral types, if available. For the membership
selection \citet{Kharchenko2004b, Kharchenko2005b} applied uniform procedures considering radial stellar density distributions, kinematics, and photometry, which typically converged after a few
iterations and provided three membership probabilities.\\
The spatial membership probability ($P_{pos}$) was set to unity for objects within the OC radius and zero otherwise. The kinematic membership probability ($P_{kin}$) can take values of 0$-$100\% and
is higher for stars sharing the common motion of the corresponding OC. The photometric membership probability ($P_{phot}$) also covers the range 0$-$100\% continuously and is higher for stars that are
closer to the corresponding OC-isochrone in the colour-magnitude diagram. Stars with $P_{phot}$ and $P_{kin} \ge 61$\% are called 1$\sigma$-members. Those with $P_{phot}$ and $P_{kin} \ge 14\%$ are
referred to as 2$\sigma$-members and targets with $P_{phot}$ and $P_{kin} \ge 1\%$ are considered as 3$\sigma$-members.\\
Moreover, CSOCA lists variability and binarity flags mainly from TYCHO-1 and -2 \citep{TYCHO1997, TYCHO2000}, HIPPARCOS \citep{HIP1997}, CMC\footnote{CMC - The Carlsberg Meridian Catalogs}
\citep{CMC1993}, GCVS\footnote{GCVS - The General Catalog of Variable Stars} \citep{GCVS1997}, NSV\footnote{NSV - The New Suspected Variables catalog} \citep{NSV1998}, and PPM \citep{PPM1991,
PPM1993}. The GCVS/NSV flags only indicate whether a star is variable or not, but do not specify the variability type. The CMC variability flag also does not provide specify the variable type, but
gives information on insufficient or missing magnitudes. The PPM binarity flag again only indicates binary candidates, but does not provide additional information on the system. More detailed
information on variability and binarity is provided by the TYCHO and HIPPARCOS flags. We found that about 10.4\% of the CSOCA stars are provided with flags indicating variability and about 4.1\% with
flags indicating binarity. Among the flagged stars we found 3336 (1.7\% of the CSOCA) that were indicated to be duplicity-induced variables.

\subsection{Previous RV data} \label{CRVAD}

The RV data in CSOCA were obtained from the Catalogue of Radial Velocities with Astrometric Data \citep[CRVAD;][]{Kharchenko2004a}, based primarily on the General Catalogue of mean Radial Velocities
\citep{Barbier2000}. \citet{Kharchenko2007} updated the CRVAD to a second version (CRVAD-2) using additional stellar RVs from the Geneva-Copenhagen survey \citep{Nordstrom2004}, the Pulkovo
Compilation of Radial Velocities \citep{Gontcharov2006} as well as CORAVEL and HIPPARCOS/TYCHO-2 kinematics on K and M giants \citep{Famaey2005}.\\
\citet{Kharchenko2007} stated that only 71\% of the CRVAD-2 entries are provided with RV uncertainties. Another 21.5\% have RV quality indices from \citet{Dufolt1995}, either indicating specific
standard errors or insufficient data. Only nine stars in CRVAD-2 show flags indicating insufficient data, which is negligible compared with the 7.5\% of CRVAD-2 entries with no available
uncertainties. We updated the RVs in CSOCA with CRVAD-2 information and found that 5\% of the 3$\sigma$-members, 6\% of the 2$\sigma$-members, and 9\% of the 1$\sigma$-members are equipped with RVs.\\
\citet{Kharchenko2007} updated the RV information in the COCD and presented their results in the Catalogue of Radial Velocities of Open Clusters and Associations (CRVOCA). It contains literature and
self-computed $\overline{RV}$ for 516 open clusters and associations, containing 395 COCD objects. The calculated $\overline{RV}$ are based on potential cluster members with $P_{kin}$ and $P_{phot}
\ge$ 1\%. For 32 clusters they found no such potential member and took one star with $P_{kin}>$1\% and its RV value as representative for the corresponding clusters. The literature values were
obtained from DAML for clusters and from \citet{Melnik1995}\footnote{http://lnfm1.sai.msu.ru/\~{ }anna/page3.html} for associations \citep[for a detailed list of references see][]{Kharchenko2007}.\\
Only 177 CRVOCA objects have both computed and literature values and agree well \citep[see Fig. 2 in][]{Kharchenko2007}. Of the 395 COCD clusters in CRVOCA, 363 have calculated $\overline{RV}$. The
remaining 32 OCs are provided with only literature values. Currently, the CRVOCA provides the most homogeneous RV reference sample for Galactic open clusters.

\subsection{Previous abundance data} \label{DAML}

The COCD itself does not provide any metallicity information for OCs. \citet{Dias2002}, on the other hand, provided metallicities or iron abundances for 96 COCD objects. Only 20 COCD entries have
abundance values derived from more than five individual measurements. The abundance uncertainties in DAML can reach 0.2 dex.\\
\citet{Dias2002} did not separate between mean metallicity ($[M/H]$) and iron abundance ($[Fe/H]$), but gave information on the photometric or spectroscopic technique used to derive the values and
literature references. When the abundance is directly derived spectroscopically from iron lines, we consider it representative for $[Fe/H]$, otherwise we expect it to be representative for $[M/H]$.
When no information on the technique or literature reference was given in DAML, we assumed the value to refer to $[M/H]$. Although the DAML metallicities are inhomogeneous, they provide a sufficient
reference sample with acceptable uncertainties.

\subsection{RAdial Velocity Experiment}

The RAdial Velocity Experiment \citep[RAVE; ][]{RAVE1} is a spectroscopic stellar survey in the southern hemisphere, observing primarily at high Galactic latitudes. The data were obtained with the
six-degree field (6dF) instrument at the Anglo-Australian Observatory, providing mid-resolution (R$=$7500) spectra in the spectral range of the CaII-triplet (8410$-$8795\AA). In addition to photometry
from TYCHO-2 \citep{TYCHO2000}, the DEep Near-Infrared southern sky Survey \citep[DENIS;][]{DENIS1997} and 2MASS \citep{MASS2003}, RAVE provides RVs, $[M/H]$, surface gravities ($\log g$), and
effective temperatures ($T_{eff}$) along with spectral quality parameters and flags.\\
Throughout the data releases the calibrations, especially regarding spectral parameters, were changed slightly. For details see the RAVE data release papers \citep{RAVE1, RAVE2, RAVE3,
Kordopatis2013}. For our project we used results from the most recently improved pipeline of RAVE DR4, containing in total 482430 entries for 425561 stars. DR4 combines pipeline results from DR3
with new stellar parameters from \citet{Kordopatis2013}. In addition, spectral classification flags by \citet{Matijevic2012} are included.\\
In RAVE studies on spectroscopic binaries were carried out by \citet{Matijevic2010,Matijevic2011}. Based on multiple measurements for about 8.7\% of DR3 stars, \citet{Matijevic2011} identified 1333
stars (6.6\% of RAVE DR3) with significantly varying RV data, which indicated them to be single-lined spectroscopic binaries (SB1). These authors also stated that for larger numbers of repetitions
(five or six measurements) the binary fraction for SB1 increases to about 10-15\%, which they referred to as the lower limit for the binary fraction in RAVE.\\
\citet{Matijevic2010}, on the other hand, investigated the cross-correlation function of observed to template spectra \citep{Munari2005} in DR2. They identified 123 double-lined spectroscopic binaries
(SB2), indicated either by more than one peak or an asymmetric central peak. From simulations, \citet{Matijevic2010} concluded that RAVE should be able to detect more than 2000 SB2 binaries. In their
recent work, \citet{Matijevic2012} not only updated the SB2 list, but also provided quality flags on RAVE spectra. These indicate problematic spectral features that might affect the reliability of the
stellar parameters.

\begin{figure*}[!ht]
\begin{center}
\includegraphics[width=15.7cm]{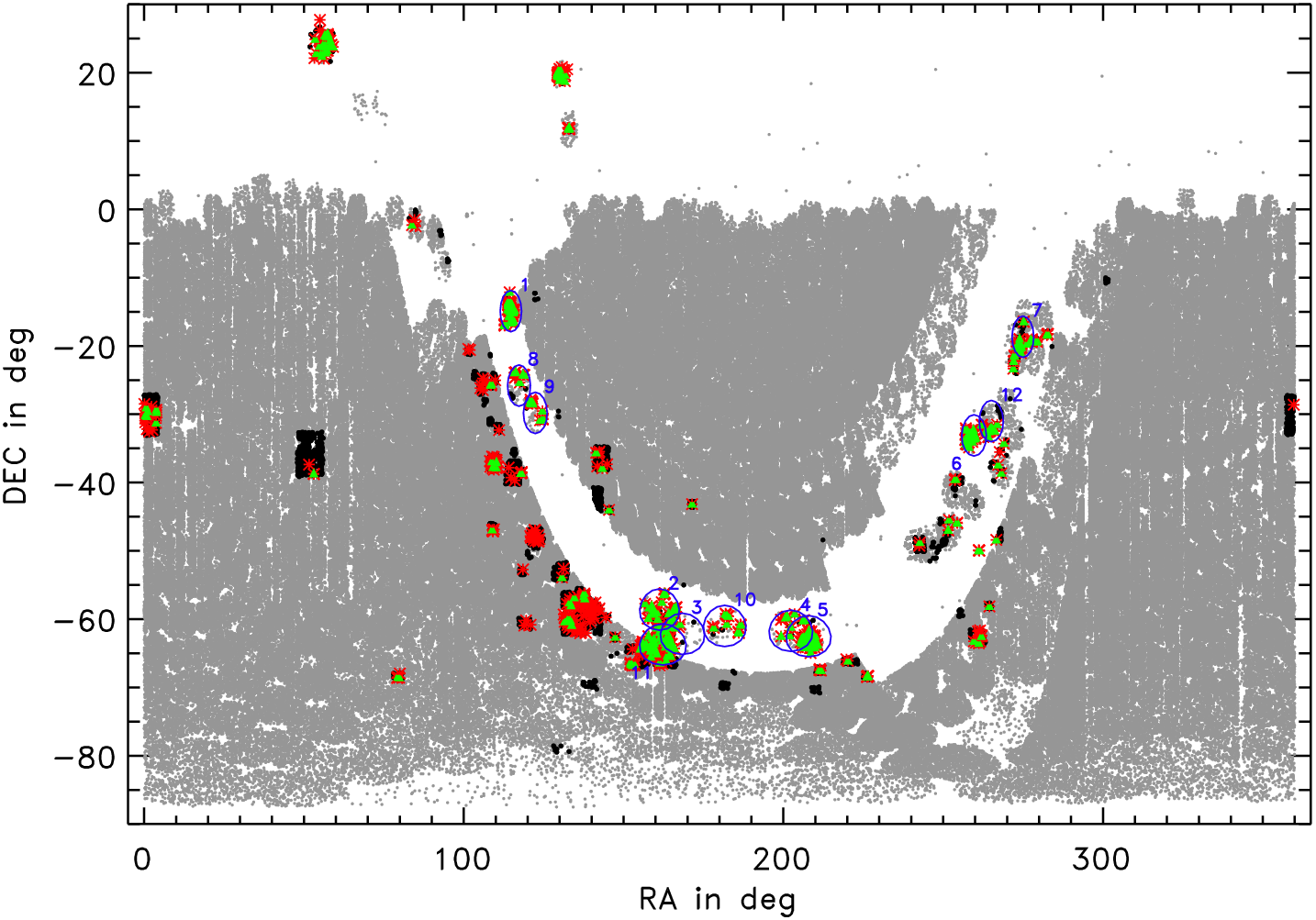}
\caption{Spatial distribution of stars in OC areas covered by RAVE. Black dots represent our high-quality RV sample. The entire RAVE DR4 is underlayed in grey. The good and best RV members are
overplotted as red asterisks and green triangles, respectively. The 12 dedicated OC fields are highlighted by blue circles.}
\label{spatial}
\end{center}
\end{figure*}

\subsection{Dedicated OC observations in RAVE}

In 2004, members of our research group proposed 12 observing fields to RAVE located in the Galactic plane (see Fig. \ref{spatial}). Each field contains at least 100 stars, and fields with more than
150 targets were suggested to be observed repeatedly with different fibre configurations to avoid allocation problems due to crowding. In total our dedicated OC fields in RAVE cover about 1500 stars
in areas around 85 known open clusters (OC areas\footnote{OC areas contain all stars in regions around known OCs \citep{Kharchenko2005a, Kharchenko2005b}, while our OCs contain only actual members.}),
including about 400 stars with known RVs from CRVAD-2 to ensure reliable $\overline{RV}$ determination for the observed OC. The observation sample was compiled from stars fainter than 9 mag in the SSS
$I$-band with no bright object within a radius of 10\arcsec and no star brighter than $I=$16 mag within a radius of 8\arcsec. The flux contamination of stars fainter than $I=$16 mag within a radius of
8\arcsec of the bright main target can be considered negligible. Hence, these objects were included in the observing sample. Up to the present, the overall number of OC areas covered by RAVE has
increased by almost a factor of three with respect to the 85 proposed areas, due to additional observations in regions around known OCs.

\section{Stellar parameters for stars in OC regions observed by RAVE} \label{stars}

\subsection{Sample selection and data quality} \label{quality}

To set up our working sample, we first updated the RV information in CSOCA with values from CRVAD-2 and then cross-matched the RV-updated CSOCA with RAVE DR4 based on a coordinate comparison with a
search radius of 3\arcsec. The spatial distribution of all COCD objects identified in RAVE is displayed in Fig. \ref{spatial}, with the 12 dedicated OC fields highlighted. The majority of our OCs are
located in or near the Galactic plane ($|b| \le$ 20 deg), usually avoided by RAVE.\\
In addition to the 85 OC areas from the dedicated cluster observations, we found 159 more regions covered by RAVE. In total, we identified 6402 measurements of 4865 stars in 244 OC areas, all
equipped with RV information in RAVE. We refer to this as our RV sample. Since $[M/H]$ determination requires spectra of higher quality, our metallicity sample comprises 6209 measurements of 4785
stars in 244 OC areas. \\
These two samples solely result from the cross-match between CSOCA and RAVE and still contain data of insufficient quality. To ensure good data quality in our working sample, we applied several
constraints in RAVE quality parameters and spectral classification flags. As a final step we included OC membership probabilities in our list of requirements to clean the working sample from
non-members.

\subsubsection*{Quality cut in signal-to-noise}

\begin{figure}[!ht]
\begin{center}
\includegraphics[width=8.5cm]{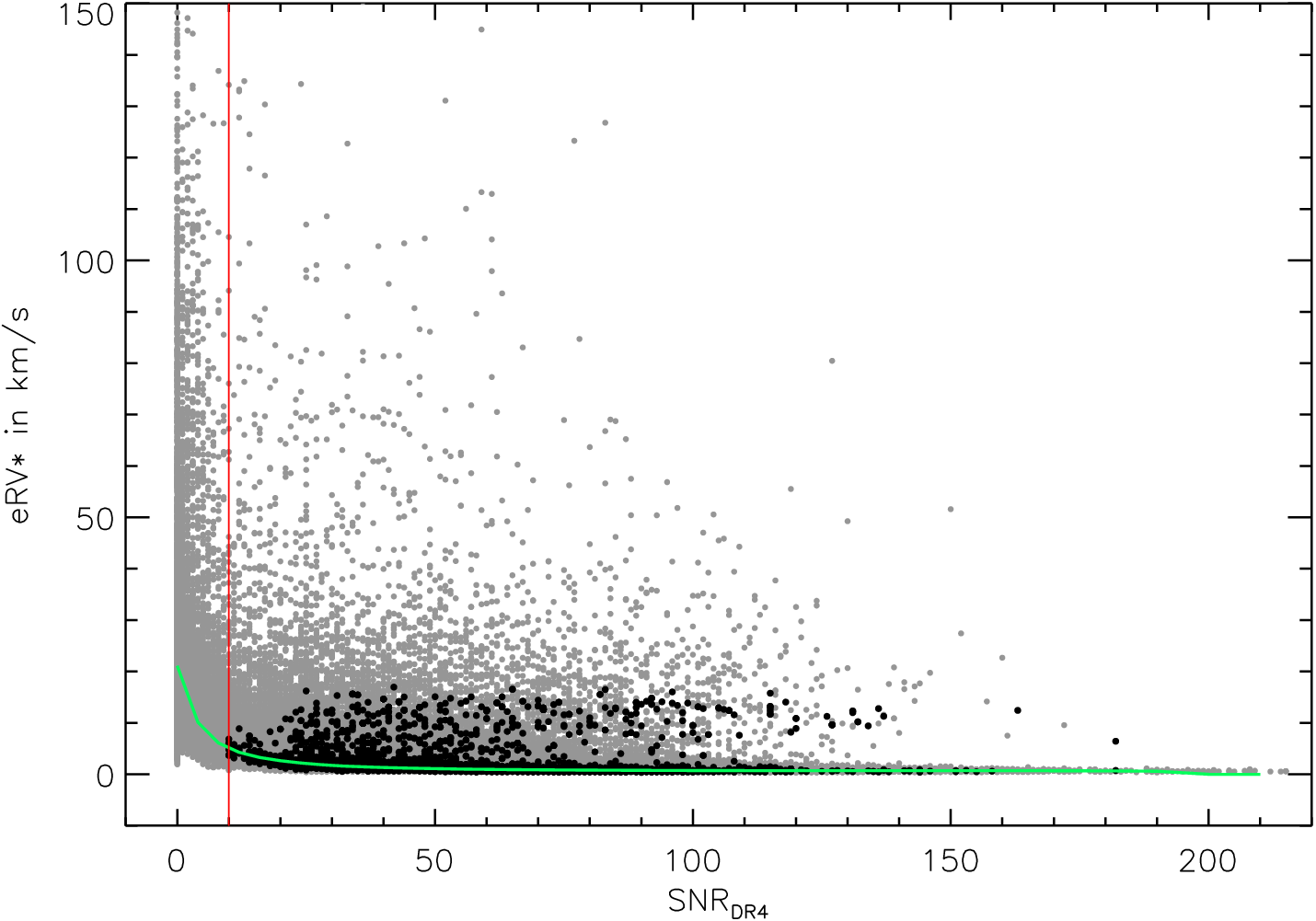}
\caption{$eRV^*$ vs. SNR distribution in RAVE DR4 (grey dots). Black dots show our high-quality RV sample. The green and red solid lines give the $\epsilon RV$ trend and cut at an SNR $\ge$ 10,
respectively.}
\label{SN-erv}
\end{center}
\end{figure}

One obvious parameter to define quality constraints is the spectral signal-to-noise ratio. Throughout this paper we use the listed SNR value in RAVE DR4 and show the distribution of RV uncertainties
($eRV^*$) with respect to the SNR in Fig. \ref{SN-erv}.\\
For the entire RAVE DR4 the distribution is very random. To better identify the overall trend we computed the median in $eRV^*$ ($\epsilon RV$) in bins along the SNR. For an SNR $<$ 100 we chose a
bin size of 4 and for an SNR $\ge$ 100 we changed it to 10, to include a sufficient number of data points. Typically, the overall trend is very flat and well below 5 km/s. Only for an SNR $\le$ 10 a
significant increase in $\epsilon RV$ is present. Thus, we defined our first cut at an SNR $\ge$ 10.

\subsubsection*{Quality cut in the spectral correlation coefficient}

\begin{figure}[!ht]
\begin{center}
\includegraphics[width=8.5cm]{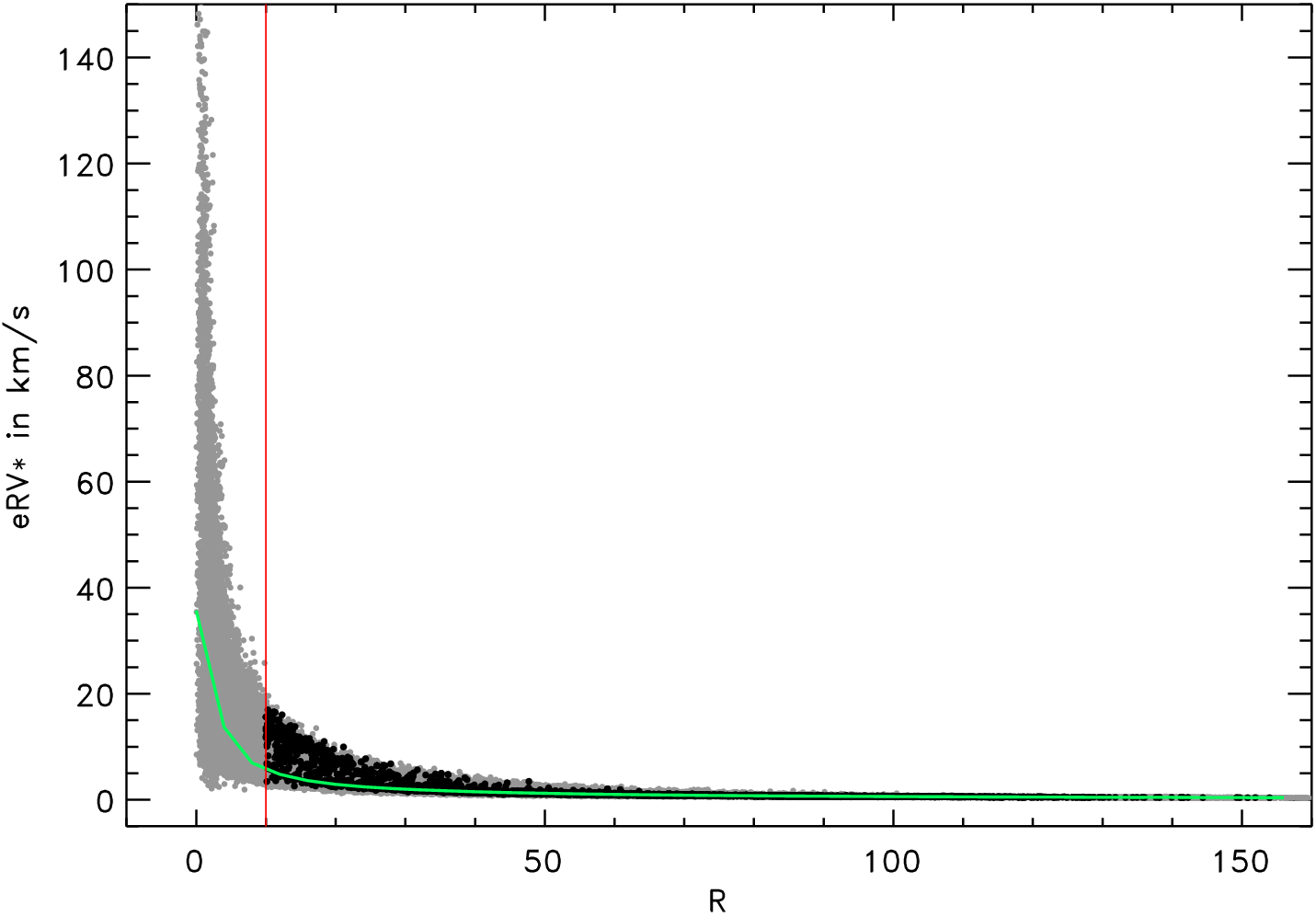}
\caption{$eRV^*$ vs. $R$ distribution in RAVE DR4 (grey dots) and our high-quality RV sample (black dots). The green and red solid lines represent the $\epsilon RV$ trend and our cut at $R \ge$
10, respectively.}
\label{tdc-erv}
\end{center}
\end{figure}

However, even at high SNR ($\ge$ 50) a considerable fraction of RAVE entries show $eRV^*$ of up to 40 km/s, making additional quality requirements necessary. Therefore, we checked the correlation
coefficient ($R$), which characterises the goodness-of-match between the observed and the template spectrum. The better the match, the higher is $R$, and the more reliable are the derived stellar
parameters.\\
The $eRV^*$ vs. $R$ distribution (Fig. \ref{tdc-erv}) is much tighter and appears to be more suited to ensure well-measured RV data than the SNR. Again we computed the overall trend in DR4 as
$\epsilon RV$ in bins of 4 along $R$. At $R<$ 10 the overall trend shows a significant increase, indicating poorly determined stellar parameters. Our second cut at $R \ge$ 10 cleans our working sample
from these unreliable targets and ensures $eRV^* \le 20$ km/s.

\subsubsection*{Quality cut in the RV correction parameter}

\begin{figure}[!ht]
\begin{center}
\includegraphics[width=8.5cm]{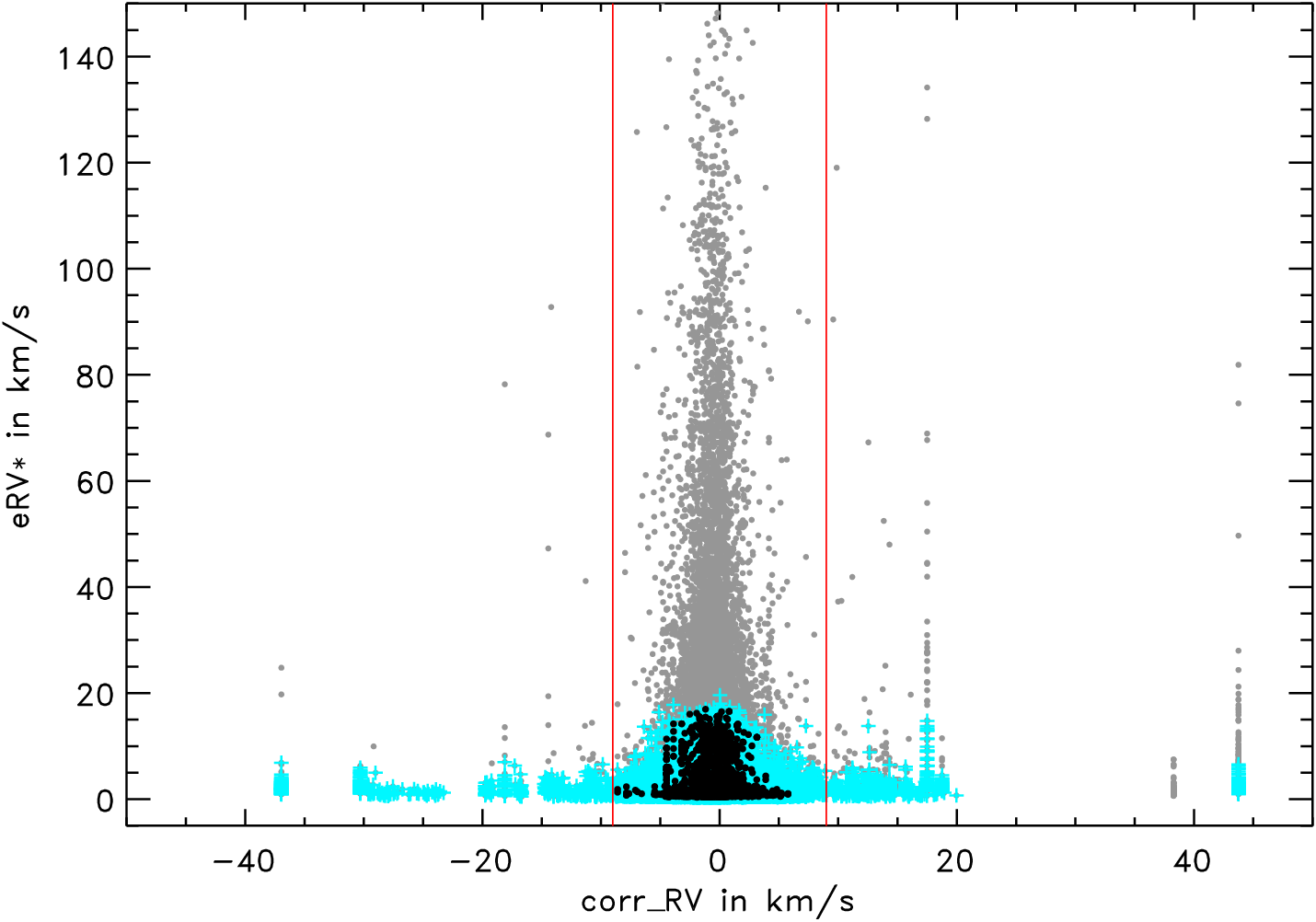}
\caption{$eRV^*$ vs. $corr\_RV$ distribution in RAVE DR4 (grey dots). Cyan crosses illustrate the subsample that matches an SNR $\ge$ 10 and $R \ge$ 10. Black dots show our high-quality RV sample and
the red solid lines illustrate our cuts at $|corr\_RV| \le$ 9 km/s.}
\label{corr-rv}
\end{center}
\end{figure}

Moreover, RAVE provides RV corrections ($corr\_RV$) based on systematic effects \citep[for details see][]{RAVE1, RAVE2, RAVE3}. The effect of $corr\_RV$ on the data quality, especially regarding
radial velocities, is shown as the $eRV^*$ vs. $corr\_RV$ distribution in Fig. \ref{corr-rv}.\\
Apparently, $corr\_RV$ can increase to 50 km/s and the distribution becomes more clumpy for higher $corr\_RV$ values. This is seen even for stars that match the first two criteria (SNR $\ge$ 10 and
$R \ge$ 10). Thus, our third cut we defined as $|corr\_RV| \le$ 9 km/s, where the distribution is very smooth.

\begin{table*}[!ht]
\begin{center}
\caption{Numbers for our different RV samples in RAVE and OC areas.}
\label{tab_rv}
\begin{tabular}{l| r r | r r r r }
\hline
             & \multicolumn{2}{|c|}{RAVE DR4} & \multicolumn{4}{|c}{OC sample} \\
\hline
Number of    & entire & high-quality &     RV & high-quality & good RV & best RV \\
             &   RAVE &      in RAVE & sample &    RV sample & members & members \\
\hline\hline
Measurements & 483849 &       405944 &   6402 &         4768 &     764 &     520 \\
Stars        & 426945 &       366922 &   4865 &         4064 &     664 &     443 \\
Clusters     &    --- &          --- &    244 &          217 &     120 &     105 \\
\hline
\end{tabular}
\end{center}
\end{table*}

\subsubsection*{Spectral flags and OC membership}

The study on the morphology of RAVE spectra by \citet{Matijevic2012} provides quality flags for the majority of RAVE spectra. The flags indicate SB2 binaries, too cool or too hot stars, problematic
spectral features, and reliable spectra. If an object is flagged reliable, we considered it for our working sample. If the RAVE target is not classified at all, we only applied the quality constraints
defined earlier (SNR $\ge$ 10, $R \ge$ 10 and $|corr\_RV| \le$ 9 km/s). These four constraints define our high quality RV sample in OC areas covered by RAVE.\\
Since we aim to investigate open clusters, we have to take into account the membership probabilities as well. Primarily we used $1\sigma$-members, and combined with the previous requirements, we
refer to these as our best RV members. In certain cases we also included $2\sigma$-members, which we call our good RV members.\\
In Tab. \ref{tab_rv} we summarise the samples considered in this work. Only about 1\% of the RAVE DR4 stars are located in OC areas from COCD and only 37.5\% of the COCD clusters are covered by RAVE.
After applying all quality requirements, we can only use about 12\% of the RAVE stars in OC areas to calculate $\overline{RV}$. The resulting OC sample is still larger than the sample covered by the
dedicated RAVE cluster fields.

\subsubsection*{Additional quality checks}

\begin{figure}[!ht]
\begin{center}
\includegraphics[width=8.5cm]{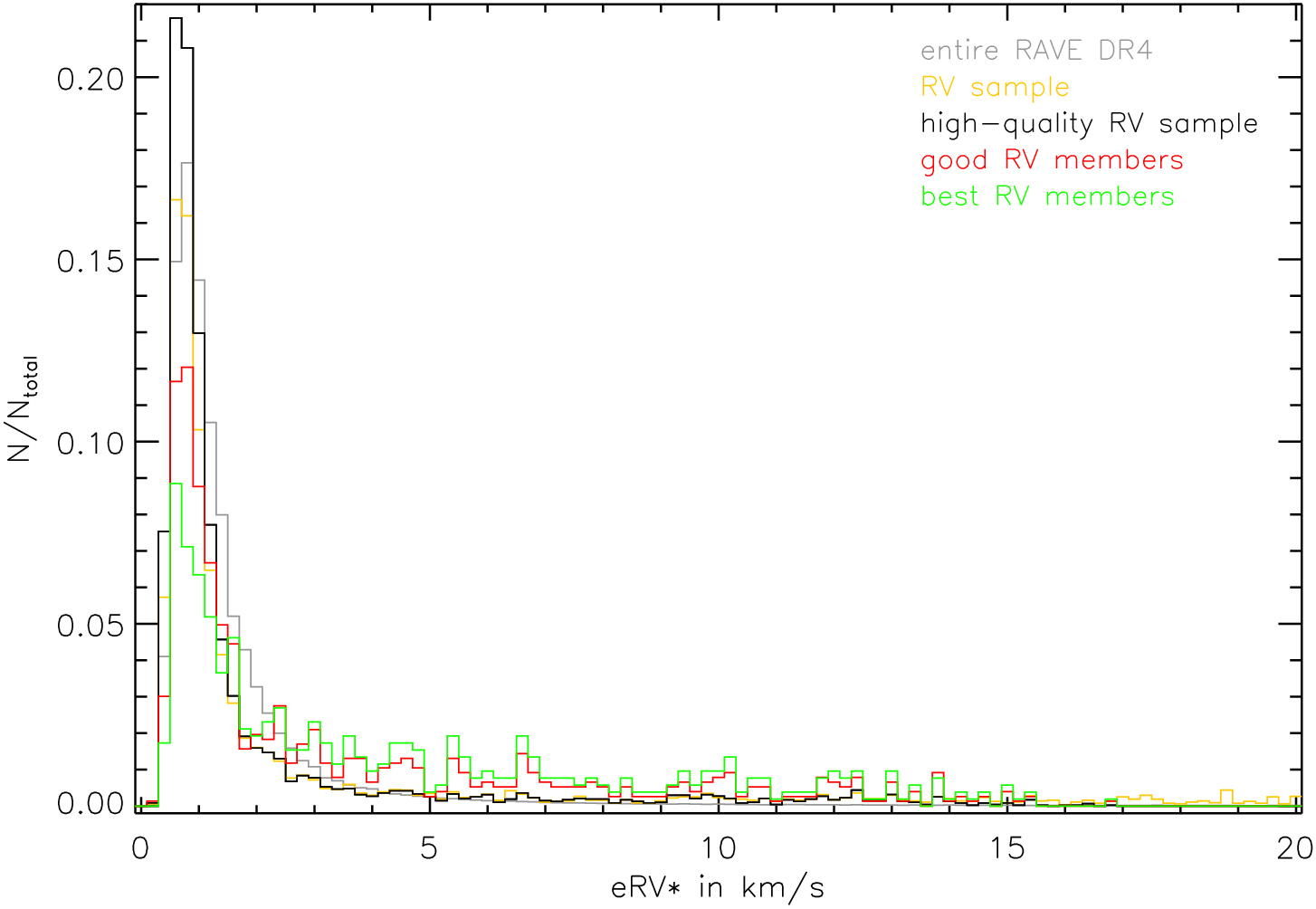}
\caption{Histograms for $eRV^*$ for the entire RAVE DR4 (grey), our RV sample (yellow), our high-quality RV sample (black), and our good (red) and best (green) RV members.}
\label{his-erv}
\end{center}
\end{figure}

To better characterise our working samples we checked the distribution of $eRV^*$ for our different samples (Fig. \ref{his-erv}). Since the size of each sample is different, we normalised each
histogram by the corresponding total number of measurements to make them comparable. As we expected, all histograms peak at about 1 km/s. However, $eRV^*$ below 1 km/s, as present in Fig.
\ref{his-erv}, are too optimistic, and especially for computing the $\overline{RV}$ we set all these very low $eRV^*$ to 1 km/s. Our good and best RV members show a significant fraction of
measurements with $eRV^* >$ 3 km/s and therefore do not reflect the quality of the entire RAVE survey; yet we have to identify the reason for this finding. 

First, we checked for a possible relation between the $eRV^*$ and RAVE observing date. In Tab. \ref{year_erv} we list the number of entries and $\epsilon RV$ in each observing year for our best RV
members and the entire RAVE DR4 for comparison. The majority of best RV members (394 out of 520 measurements) were observed in 2004, 2005, and 2010. The corresponding $\epsilon RV$ are about a factor
of 4 higher than the values of the remaining years. This is a specific feature of our OC member sample, since for the entire RAVE the $\epsilon RV$ are almost equal for all observing years. Although
we can now relate the less accurate RVs of our best RV members to certain RAVE observing years, we cannot sufficiently explain the difference in data quality between RAVE and our good and best RV
members.

\begin{table}[!ht]
\begin{center}
\caption{Comparison of $\epsilon RV$ between our best RV members and RAVE for each observing year.}
\label{year_erv}
\begin{tabular}{r | r r | r r}
\hline
             &  \multicolumn{2}{|c|}{best RV members}  & \multicolumn{2}{|c}{entire RAVE} \\ 
\hline
   Observing &      No. of & $\epsilon RV$ &       No. of & $\epsilon RV$ \\
        year &      entries &       in km/s &      entries &       in km/s \\
\hline\hline
 2003        &             0 &           --- &        19164 &          1.90 \\
 2004        &           109 &          4.51 &        28924 &          1.67 \\
 2005        &           104 &          4.20 &        30889 &          1.56 \\
 2006        &             9 &          1.64 &        78493 &          1.22 \\
 2007        &            18 &          0.88 &        53899 &          1.20 \\
 2008        &            18 &          1.13 &        60387 &          1.06 \\
 2009        &            15 &          1.11 &        75465 &          1.03 \\
 2010        &           181 &          4.47 &        59192 &          1.08 \\
 2011        &            20 &          0.87 &        50576 &          1.04 \\
 2012        &            46 &          1.66 &        25441 &          1.15 \\
 2013        &             0 &           --- &         1419 &          1.40 \\
\hline
total        &           520 &          3.03 &       483849 &          1.18 \\
\hline
\end{tabular}
\end{center}
\end{table}

\begin{figure}[!ht]
\begin{center}
\includegraphics[width=8.5cm]{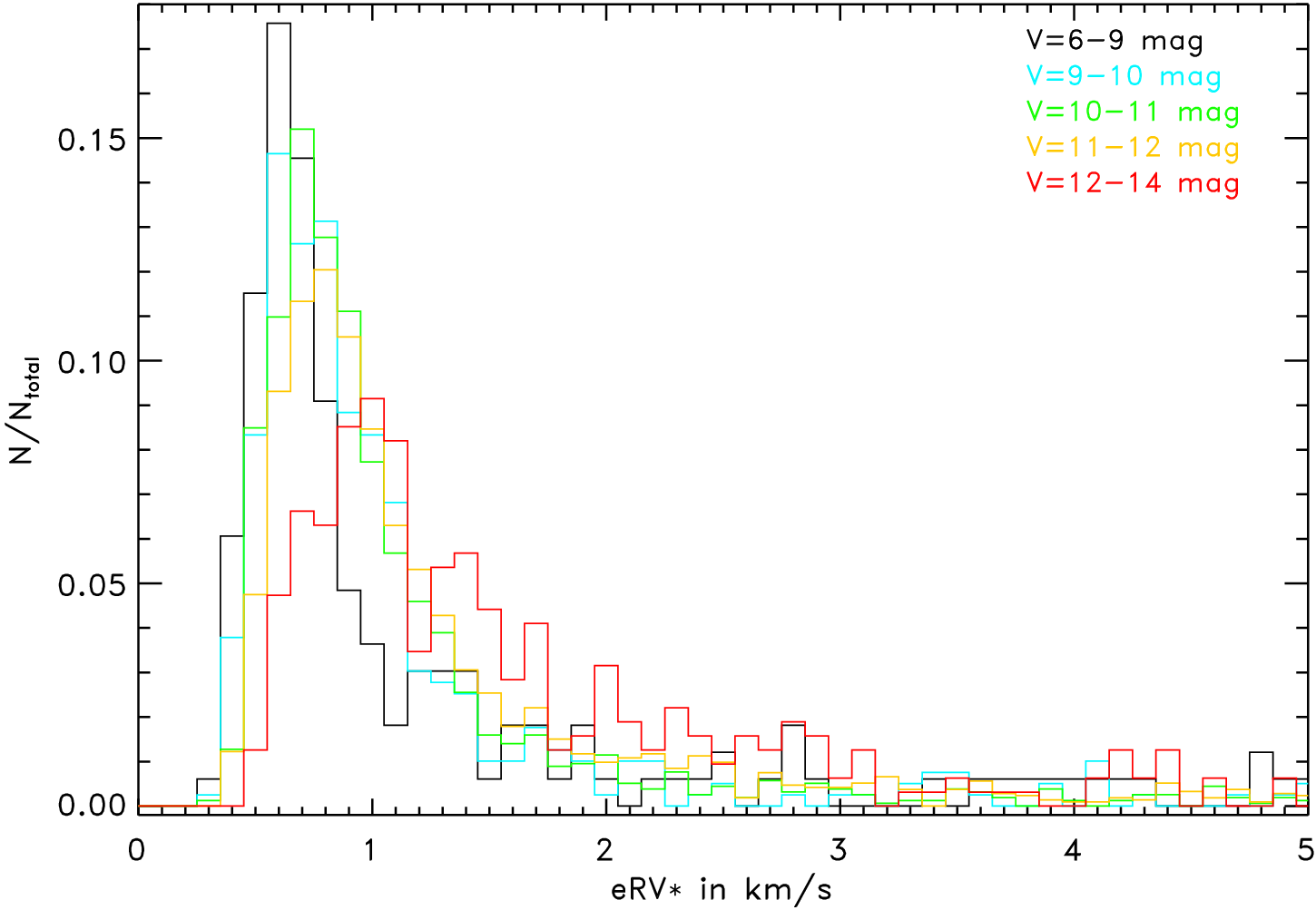}
\caption{Magnitude-dependent $eRV^*$ histograms for our high-quality RV sample. The $V_{Johnson}$ intervals are 6-9 mag (black), 9-10 mag (blue), 10-11 mag (green), 11-12 mag (yellow), and 12-14 mag
(red).}
\label{his-emag}
\end{center}
\end{figure}

To check for the degree of magnitude dependence in $eRV^*$, we show the magnitude-separated $eRV^*$ histograms for our high-quality RV sample in Fig. \ref{his-emag} and give the corresponding numbers
of measurements and $\epsilon RV$ in Tab. \ref{mag-erv}. For $8-12$ mag the $\epsilon RV$ are almost equal, only for the faintest magnitude interval the $\epsilon RV$ value is about 0.5 km/s higher,
as seen in Fig. \ref{his-emag} as well. Since the change in $eRV^*$ is only 0.5 km/s, the magnitude dependence can be considered negligible in our working sample.

Open clusters are relatively young objects and are expected to be dominated by dwarfs. In our samples we separated dwarfs from giants based on $\log{g}$ in RAVE DR4. We considered giants to have
$\log{g} <$ 3.75 dex and dwarfs to show $\log{g} \ge$ 3.75 dex. Objects with no $\log{g}$ were not included in this separation. The DR4 pipeline providing $\log{g}$, $T_{eff}$ and $[M/H]$ also list
flags indicating potential problems in the convergence of the algorithm. Targets indicated to not converge or that had to be rerun were excluded from the $\log{g}$ separation. Thus, the number of
dwarfs and giants in Tab. \ref{mag-erv} does not necessarily add up to the total number of measurements in the corresponding magnitude bin. \\
In Tab. \ref{mag-erv} we summarise the results for our high-quality RV sample and our good RV members. By total numbers the high-quality RV sample is dominated by giants with a giant-to-dwarf ratio
of 2.96, while the good RV members contain an almost equal number of dwarfs and giants, showing a ratio of 1.08. These numbers confirm our expectation that OCs contain a larger number of dwarfs and
that RAVE preferably observes giants.\\
Considering each magnitude interval, this becomes even more evident, because the number of good RV members that are dwarfs in $6 \le V_{Johnson} < 11$ mag is higher than the number of giants, and for
$11 \le V_{Johnson} \le 14$ mag the number of dwarfs and giants are almost equal for the good RV members. In all magnitude intervals the $\epsilon RV$ of our good RV members are higher than the
respective values in our high-quality RV sample, indicating a potential relation between stellar type and $eRV^*$.

\begin{table}[!ht]
\begin{center}
\caption{Number of entries, giant-to-dwarf ratios, and $\epsilon RV$ in magnitude intervals as shown in Fig. \ref{his-emag} for our high-quality RV sample and good RV members.}
\label{mag-erv}
\begin{tabular}{r | r r r | r r r}
\hline
$V_{Johnson}$ &\multicolumn{3}{c|}{high-quality RV sample} & \multicolumn{3}{c}{good RV members} \\
       in mag & No. &        G/D$^a$ & $\epsilon RV$ & No. &   G/D$^a$ & $\epsilon RV$ \\
\hline
    6-9 &       193 & 110/\ \ \ \ 78 &    0.95 &        34 & 10/\ \ 23 &    3.79 \\
   9-10 &       472 &    261/\ \ 186 &    1.01 &        49 & 18/\ \ 29 &    1.83 \\
  10-11 &      1582 &   1231/\ \ 243 &    0.92 &       136 & 51/\ \ 74 &    1.50 \\
  11-12 &      2170 &   1505/\ \ 477 &    1.03 &       419 &   224/150 &    1.45 \\
  12-14 &       350 &    175/\ \ 123 &    1.48 &       126 & 50/\ \ 52 &    2.63 \\
\hline
  total &      4768 &      3282/1108 &    1.00 &       764 &   353/328 &    1.73 \\
\hline
\multicolumn{7}{l}{\footnotesize $^a$G/D - giant-to-dwarf ratio.} \\
\end{tabular}
\end{center}
\end{table}

\begin{figure}[!ht]
\begin{center}
\includegraphics[width=8.5cm]{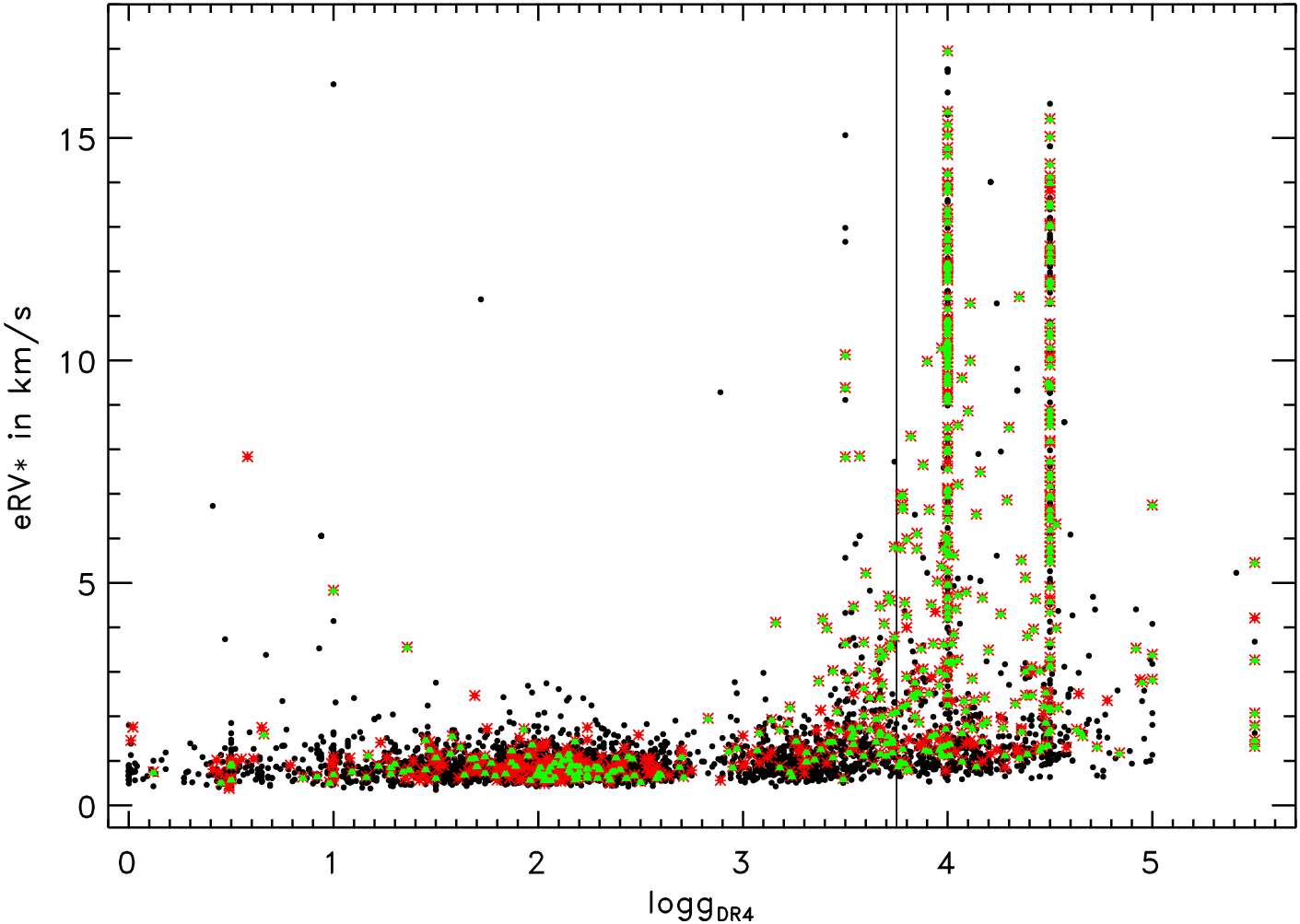}
\caption{Distribution of $eRV^*$ with respect to $\log{g}$. Symbol colour-coding is the same as in Fig. \ref{spatial}. Our giant/dwarf separation limit at $\log{g}=3.75$ is included as the black
solid line.}
\label{logg-erv}
\end{center}
\end{figure}

To investigate this aspect in more detail, we display the $eRV^*$ vs. $\log{g}$ diagram in Fig. \ref{logg-erv}. The pillar-like features in the $\log{g}$ distribution are due to the grid of synthetic
spectra used to derive stellar parameters in RAVE DR4 \citep[see][]{Kordopatis2011, Kordopatis2013}. We found that higher values of $\log{g}$ also show higher $eRV^*$. Potential reasons for this
dependence could be that dwarfs show fewer and weaker absorption lines, which are used to derive RV. For our good and best RV members the effect of higher $eRV^*$ with higher $\log{g}$ appears to be
stronger. Moreover, the location of our OCs in or near the Galactic disk might affect the quality of our working sample.

\begin{figure}[!ht]
\begin{center}
\includegraphics[width=8.5cm]{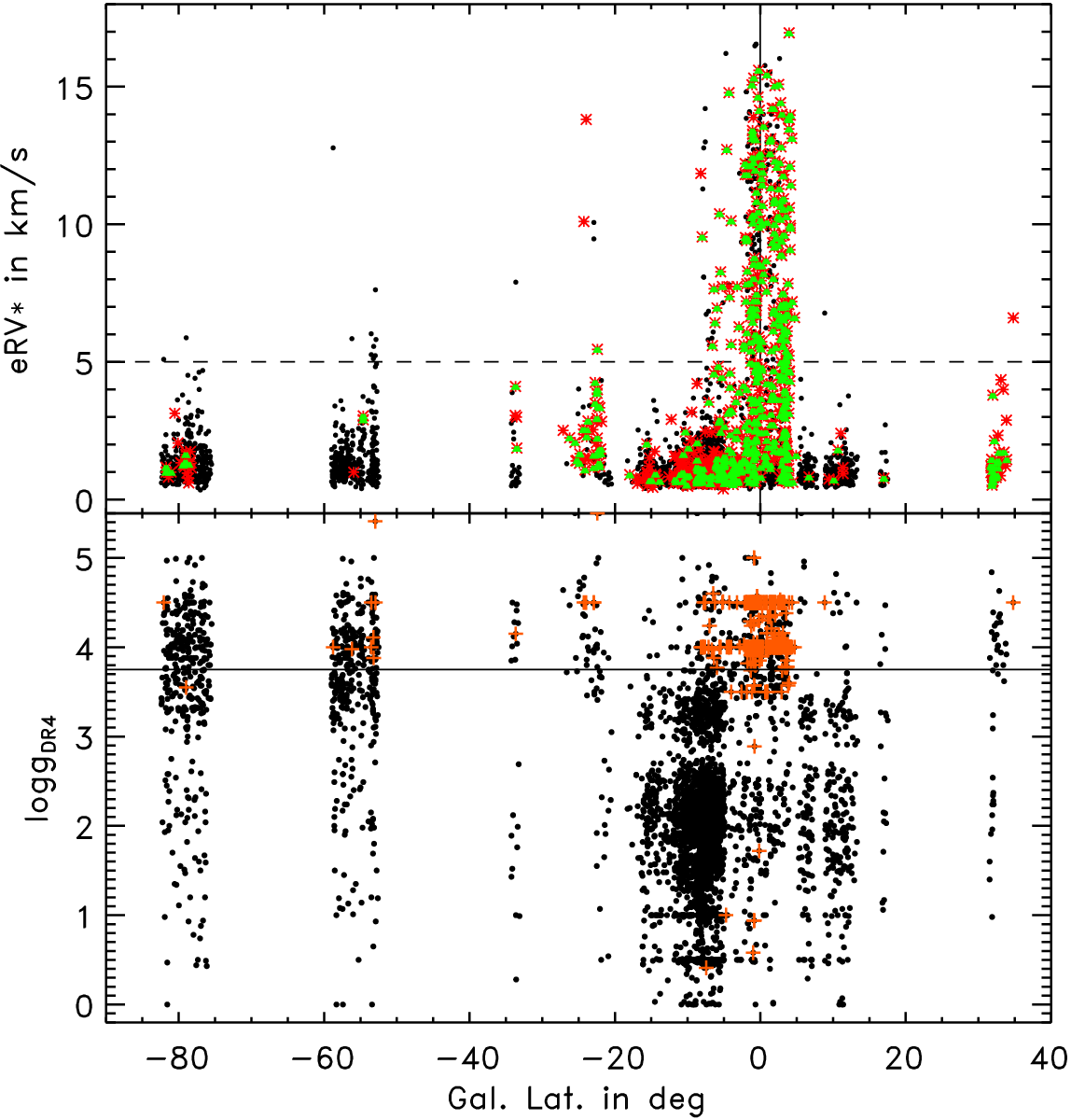}
\caption{Distribution of $eRV^*$ and $\log{g}$ with respect to $b$ along with the mid-plane and $\log{g}$ limit (3.75) overplotted as the black solid line in the upper and lower panel, respectively.
The symbol colour-coding is the same as in Fig. \ref{spatial}, and dark orange crosses highlight targets with $eRV^*>5$ km/s. This $eRV^*$ limit is displayed as the black dashed line.}
\label{glat-erv}
\end{center}
\end{figure}

Therefore, we present the $eRV^*$ distribution with respect to the Galactic latitude ($b$) in the upper panel of Fig. \ref{glat-erv}. One can see that almost all good and best RV members with
$eRV^*>5$ km/s are located very close to the Galactic plane. In the lower panel we show the $\log{g}$ vs. $b$ distribution and highlight all targets with $eRV^*>5$ km/s, which appear to be
predominantly dwarfs. This confirms that the higher $eRV^*$ for our good and best RV members are mainly caused by the higher percentage of dwarfs in our OC sample. The possible effect of undetected
binarity, extinction, or change in exposure time on $eRV^*$ we cannot study in detail with the data set used.

We can conclude that even though our OC sample in RAVE does not reflect the accuracy of the entire survey, the quality of our working sample is still sufficient for our purposes , which are
determining the average radial velocities ($\overline{RV}$) for open clusters.

\begin{table*}[!ht]
\begin{center}
\caption{Comparison of numbers and RV uncertainties between RAVE, CRVAD-2, and the resulting common sample. The $\epsilon RV$ values are the median of the RV uncertainties and $\sigma \Delta RV$
correspond to the standard deviation of the difference distribution.}
\label{cat-rv}
\begin{tabular}{l| r r | r r r r }
\hline
                               & \multicolumn{2}{|c|}{Catalogues} & \multicolumn{4}{|c}{OC sample} \\
\hline
                                &     entire &   high- &     RV & high-quality & good RV & best RV \\
                                &            & quality & sample &    RV sample & members & members \\
\hline\hline
--- RAVE ---                    &            &         &        &         &         &         \\
No. of entries                  &     483849 &  405944 &   6402 &    4768 &     764 &     520 \\
No. of clusters                 &        --- &     --- &    244 &     217 &     120 &     105 \\
$\epsilon RV$ in km/s           &       1.18 &    1.11 &   1.23 &    1.00 &    1.73 &    3.03 \\
\hline\hline
--- CRVAD-2 ---                 &            &         &        &         &         &         \\
No. of entries                  &      54907 &     --- &   6782 &     --- &    1586 &    1092 \\
No. of clusters                 &        650 &     --- &    595 &     --- &     318 &     306 \\
$\epsilon RV$ in km/s           &       0.86 &     --- &   3.60 &     --- &    3.70 &    3.70 \\
\hline\hline
--- common sample ---           &            &         &        &         &         &         \\
No. of entries                  &       2475 &    1774 &    531 &     262 &      51 &      32 \\
No. of clusters                 &        --- &     --- &    104 &      73 &      13 &       9 \\
$\epsilon RV_{RAVE}$ in km/s    &       1.23 &    1.02 &   6.06 &    1.45 &    2.04 &    2.28 \\
$\epsilon RV_{CRVAD-2}$ in km/s &       0.60 &    0.50 &   2.90 &    1.80 &    1.70 &    1.70 \\
$\sigma \Delta RV$ in km/s      &      90.66 &   22.65 &  81.21 &   38.20 &   22.75 &   21.02 \\
\hline
\end{tabular}
\end{center}
\end{table*}

\subsection{Radial velocity} \label{radvel}

\begin{figure}[!ht]
\begin{center}
\includegraphics[width=8.5cm]{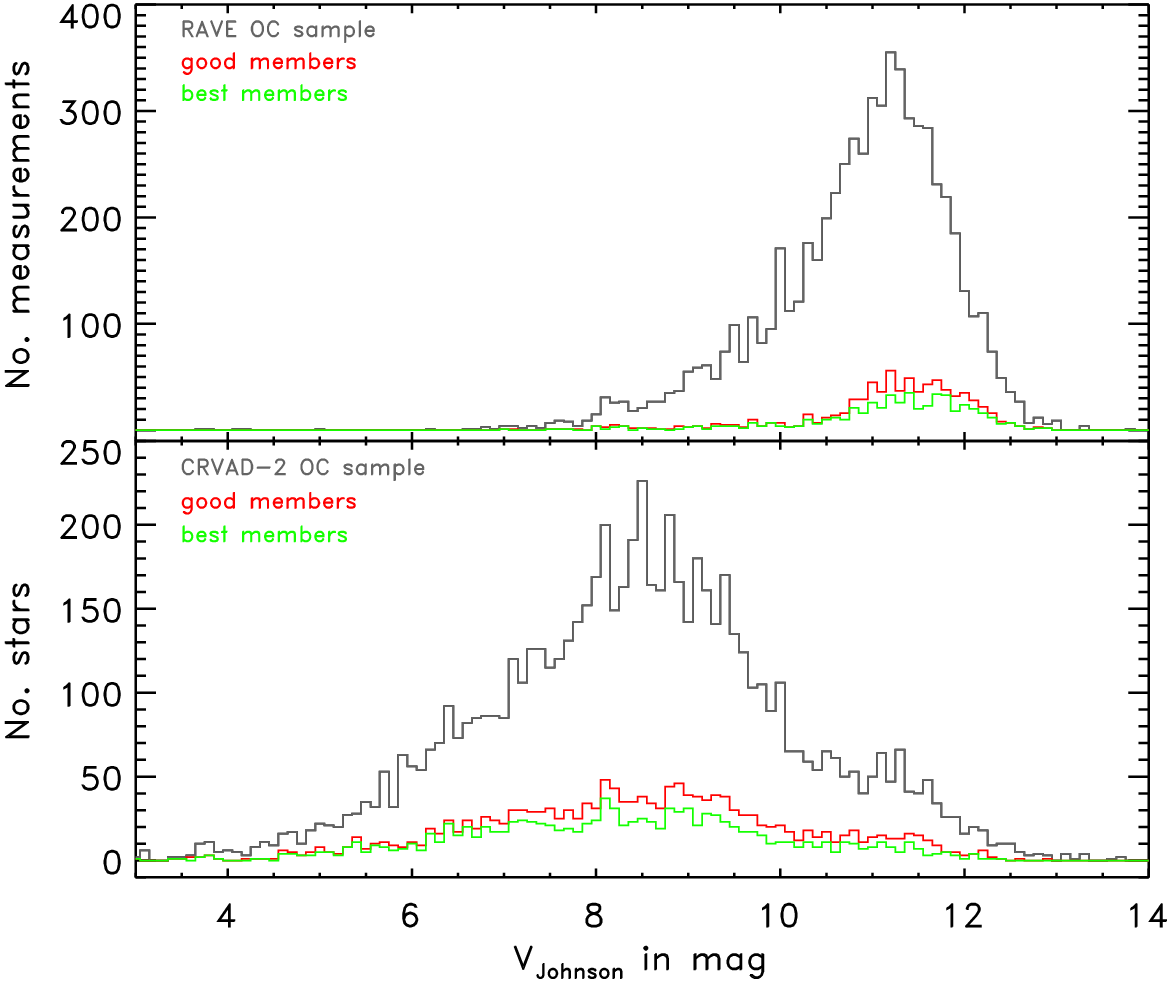}
\caption{$V_{Johnson}$ histograms in RAVE (upper panel) and CRVAD-2 (lower panel) for objects in OC areas (grey), as well as our good (red) and best (green) RV members.}
\label{his-mag}
\end{center}
\end{figure}

\begin{figure*}[!ht]
\begin{center}
\subfigure{\includegraphics[width=8.4cm]{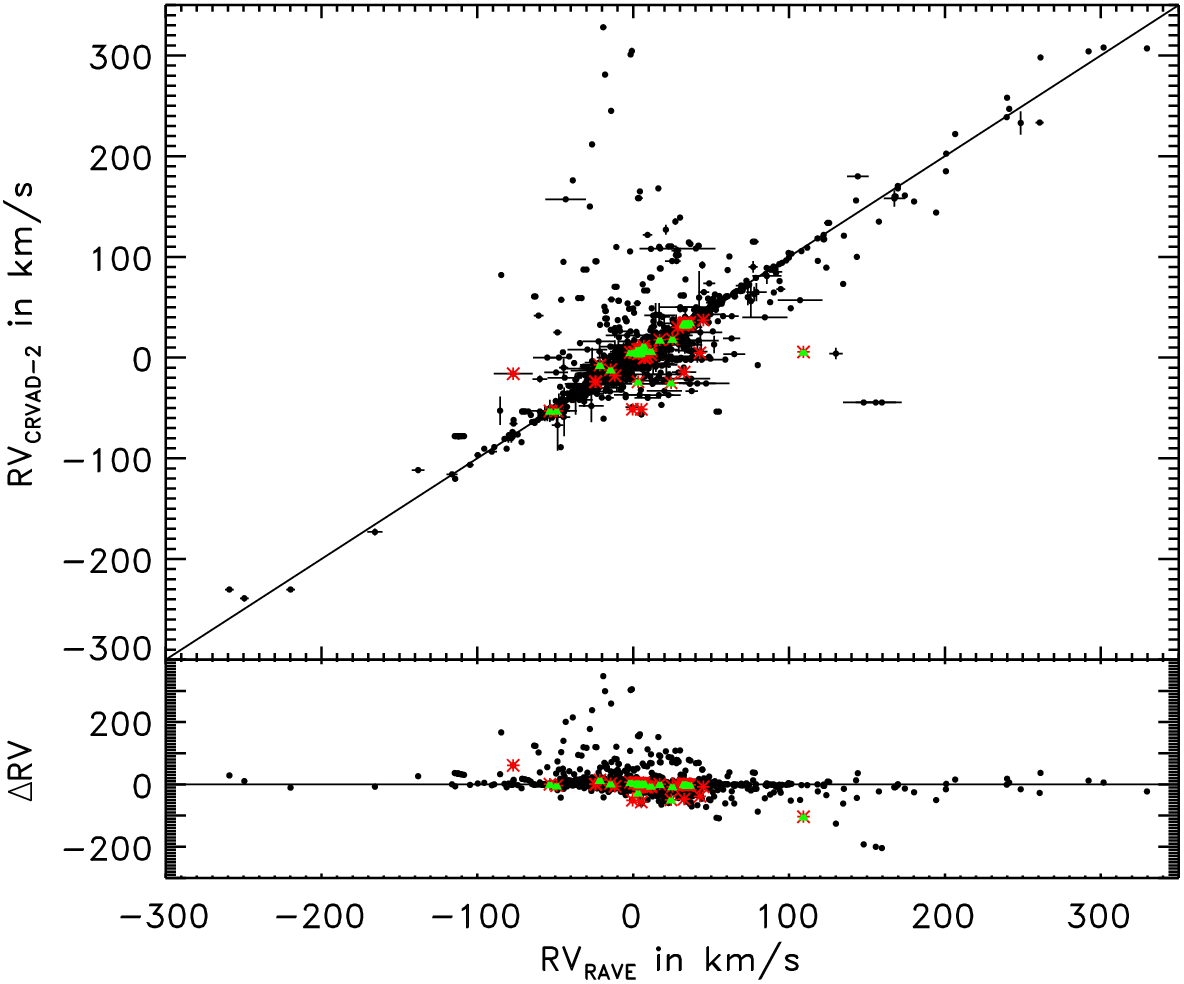}}
\subfigure{\includegraphics[width=8cm]{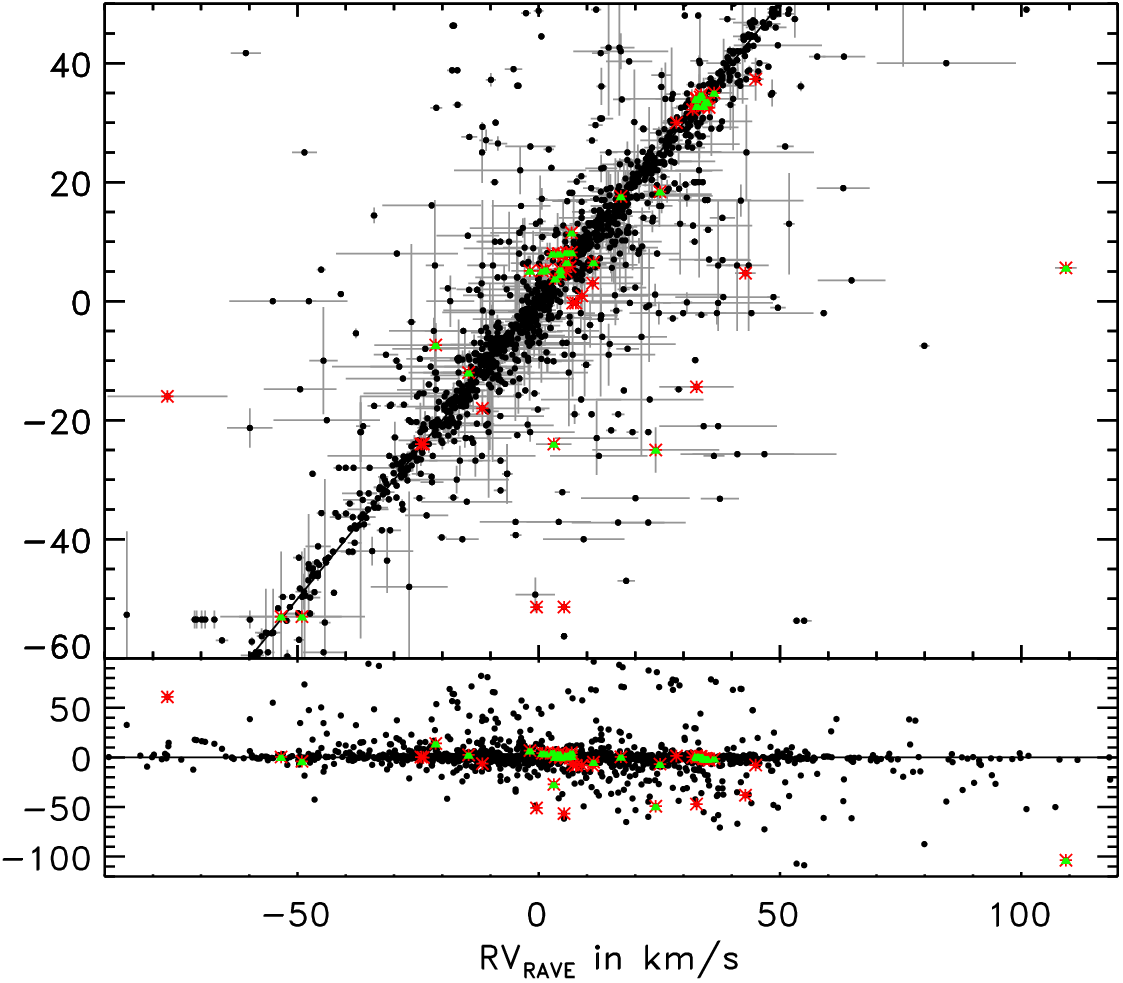}}
\caption{Upper panel: RV comparison between CRVAD-2 and RAVE. The black solid line refers to the one-to-one relation. Lower panel: Corresponding difference distribution along with the zero-difference
line (black solid line). Black dots show the high-quality common sample, while red asterisks and green triangles highlight good and best RV members in the common sample, respectively. The right panels
show the same diagrams enlarged to the RV range of our good and best RV members.}
\label{rv-comp}
\end{center}
\end{figure*}

\begin{figure}[!ht]
\begin{center}
\includegraphics[width=8cm]{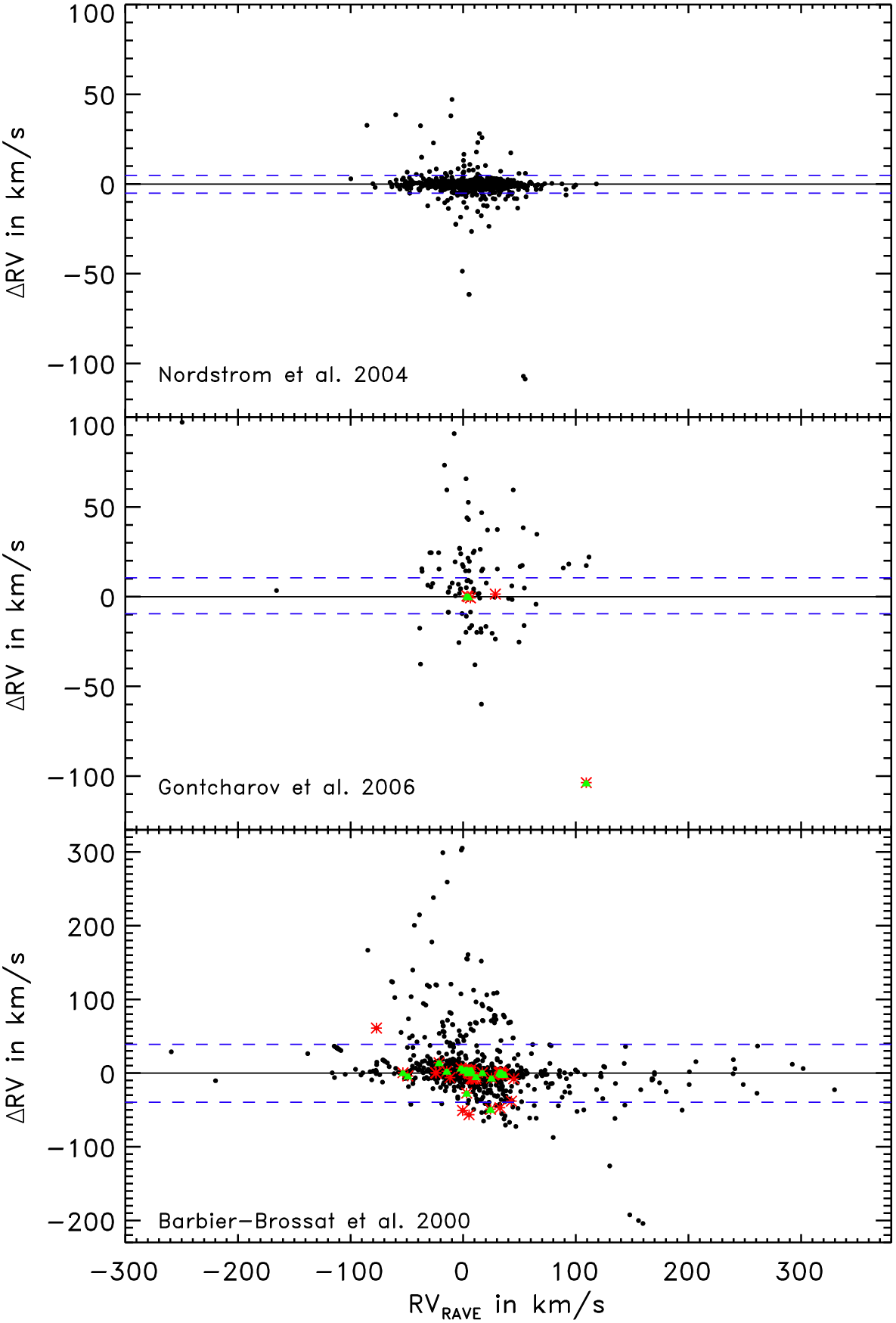}
\caption{Unclipped RV difference distributions between RAVE and \citet{Nordstrom2004} (upper panel), \citet{Gontcharov2006} (middle panel), and \citet{Barbier2000} (lower panel). The colour-coding is
the same as in Fig. \ref{rv-comp} and the blue dashed lines define the limits of the 3$\sigma$-clipped distributions.}
\label{rv-source}
\end{center}
\end{figure}

To better evaluate the RVs obtained by RAVE, we obtained reference values from CRVAD-2 and created a common sample for comparison via a cross-match based on coordinates with a matching radius of
3\arcsec. The numbers and $\epsilon RV$ for the two catalogues and the common sample are given in Tab. \ref{cat-rv}. The increase of $\epsilon RV$ after including membership probabilities, as stated
above, is a RAVE-specific characteristic, since it is only present in the RAVE data, but not in CRVAD-2. For the good and best OC members with RV, on the other hand, the $\epsilon RV$ are similar
in the two catalogues.\\
Interestingly, the common sample is very small (2500 listings) compared to the size of the two catalogues (RAVE: $\sim 460000$ entries and CRVAD-2: $\sim 55000$ stars) and only a very small fraction
of objects in each catalogue is located within OC regions (about 1.3\% in RAVE and about 12.3\% in CRVAD-2). One reason for the small overlap between CRVAD-2 and RAVE is that each catalogue has
different observing samples: RAVE is a southern-sky survey, while CRVAD-2 was an all-sky project.\\
Moreover, RAVE and CRVAD-2 cover different magnitude ranges shifted by almost 3 mag, as presented in Fig. \ref{his-mag}, also showing that RAVE only covers fainter OC members. Within OC areas, on the
other hand, the fraction of good and best members are comparably large, that is, in RAVE 12.3\% of objects in OC areas are good members and in CRVAD-2 the corresponding percentage is 23.4\%. This
indicates that the majority of objects in OC regions, included in each catalogue, are at least good members.\\
For the high-quality common sample we display the RV comparison between RAVE and CRVAD-2 source catalogues in Fig. \ref{rv-comp}, along with the corresponding difference distribution. The RV
differences were computed as $\Delta RV=RV_{CRVAD-2}-RV_{RAVE}$. Near $RV_{RAVE} = 0$ km/s we found several stars with intrinsically higher $RV_{CRVAD-2}$ than $RV_{RAVE}$. For our good and
best RV members this feature entirely disappears. In the difference distribution a slight negative slope is also visible in the high-quality sample. Our good and best RV members do not show this
slope distinctly, since only two stars show significant differences, which could be by chance. The remaining good and best members, except for the two deviating ones, show a spread in the difference
distribution of 20 km/s. Hence, our selected good and best RV members agree well with the reference values and show a sufficiently good quality to derive $\overline{RV}$ for OCs in RAVE.\\
Still, we have to understand the identified systematics of our high-quality sample (see Fig. \ref{rv-comp}). Accordingly, we investigated the major CRVAD-2 source catalogues, namely
\citet{Nordstrom2004}, \citet{Gontcharov2006}, and \citet{Barbier2000}. The results are presented visually in Fig. \ref{rv-source} and in numbers in Tab. \ref{sources}. The vast majority of CRVAD-2
values were obtained from \citet{Barbier2000} and \citet{Nordstrom2004}. The displayed difference distributions in Fig. \ref{rv-source} are relatively broad and might include several outliers.
Therefore, we applied a 3$\sigma$-clipping algorithm to identify the actual distribution characteristics and also included the results for the clipped distributions in Tab. \ref{sources} and Fig.
\ref{rv-source}.\\
In the difference distributions (clipped and unclipped) for reference values from \citet{Nordstrom2004} and \citet{Gontcharov2006} the standard deviations in the high-quality sample are considerably
lower than for the comparison with values from \citet{Barbier2000}. Therefore, the reference values from the first two catalogues seem to be more reliable. Moreover, the systematic effect near
$RV_{RAVE} = 0$ km/s is visible in all source catalogues, whereas the possible negative slope only appears in the comparison of our high-quality sample with values from \citet{Barbier2000}. Thus, we
can conclude that the trend is not a feature induced by the RAVE data but by the reference values from \citet{Barbier2000}.\\
Surprisingly, we found no good and best members in common with \citet{Nordstrom2004}. Moreover, the number of common good and best RV members with \citet{Gontcharov2006} is negligible, which in turn
makes the questionable values by \citet{Barbier2000} the dominant source for RV references. However, their values are the best RV references for OCs available, and since our good and best RV members
in RAVE show a better agreement with these references than the high-quality data, it indicates that our cuts are suitable for deriving reliable $\overline{RV}$ for our OC sample.

\begin{table}[!ht]
\begin{center}
\caption{Characteristics for the RV difference distributions between RAVE and the source catalogues in CRVAD-2 for the high-quality sample as well as for the good and best RV members in our common
sample.}
\label{sources}
\begin{tabular}{l| r r r r }
\hline
                    &    No. & $\epsilon RV$ & $\overline{\Delta RV}$ & $\sigma \Delta RV$ \\
\hline\hline
high-quality sample &        \multicolumn{4}{c}{before 3$\sigma$-clipping}        \\
\hline
Nordstr\"{o}m       &    825 &          0.40 &             -0.69 &          8.10  \\
Gontcharov          &     93 &          0.60 &             -1.86 &         12.71  \\
Barbier-Brossat     &    852 &          1.70 &              6.54 &         42.54  \\
\hline
                    &        \multicolumn{4}{c}{after 3$\sigma$-clipping}         \\
\hline
Nordstr\"{o}m       &    743 &          0.30 &             -0.36 &           1.78 \\
Gontcharov          &     89 &          0.60 &             -0.18 &           3.82 \\
Barbier-Brossat     &    728 &          1.70 &             -0.57 &          11.27 \\
\hline\hline
good RV members     &     \multicolumn{4}{c}{before 3$\sigma$-clipping}           \\
\hline
Nordstr\"{o}m       &     -- &           --- &               --- &            --- \\
Gontcharov          &      5 &          0.40 &            -20.50 &          46.50 \\
Barbier-Brossat     &     46 &          2.00 &             -4.77 &          18.93 \\
\hline
                    &     \multicolumn{4}{c}{after 3$\sigma$-clipping}            \\
\hline
Nordstr\"{o}m       &     -- &           --- &               --- &            --- \\
Gontcharov          &      4 &          1.30 &              0.29 &           0.90 \\
Barbier-Brossat     &     38 &          1.80 &             -0.66 &           4.04 \\
\hline\hline
best RV members     &      \multicolumn{4}{c}{before 3$\sigma$-clipping}          \\
\hline
Nordstr\"{o}m       &     -- &           --- &               --- &            --- \\
Gontcharov          &      3 &          0.40 &            -34.42 &          59.96 \\
Barbier-Brossat     &     29 &          1.80 &             -1.44 &          11.27 \\
\hline
                    &      \multicolumn{4}{c}{after 3$\sigma$-clipping}           \\
\hline
Nordstr\"{o}m       &     -- &           --- &               --- &            --- \\
Gontcharov          &      2 &          1.50 &              0.20 &           0.30 \\
Barbier-Brossat     &     26 &          1.70 &              0.79 &           3.12 \\
\hline
\end{tabular}
\end{center}
\end{table}

\subsection{Metallicity} \label{metal}

\begin{figure}[!ht]
\begin{center}
\includegraphics[width=8.5cm]{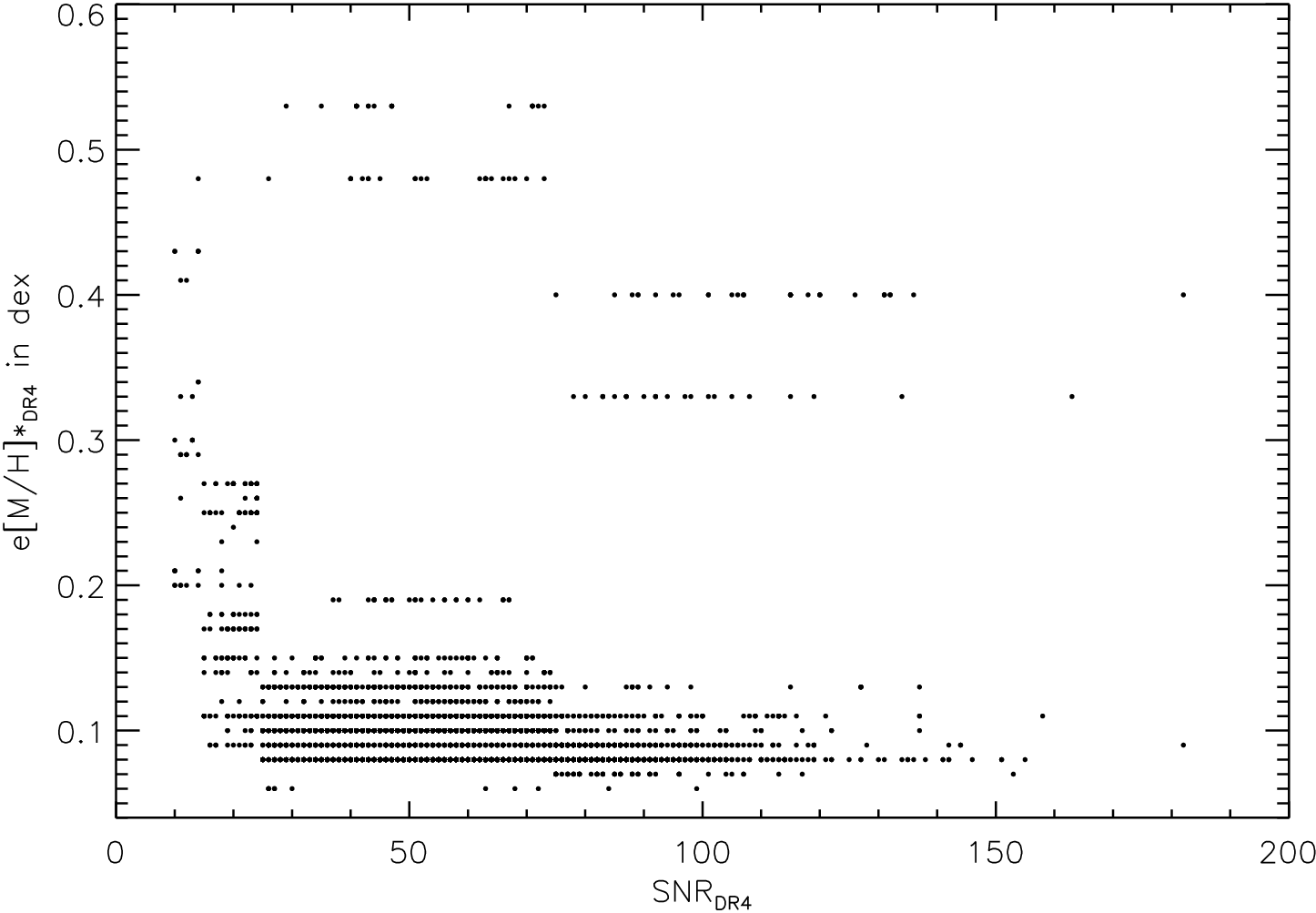}
\caption{Distribution of $e[M/H]^*$  with respect to $SNR$ for our high-quality RV sample.}
\label{SN-emet}
\end{center}
\end{figure}

We also aimed to provide mean metallicities ($\overline{[M/H]}$) for our RAVE clusters. Spectra of higher quality are typically needed for the metallicity determination and different template spectra
were used than for deriving RVs. In DR4 \citet{Kordopatis2013} applied several prior constraints, namely SNR $\ge$ 20, $v_{rot}<$ 100 km/s, $eRV^* <$ 8 km/s, $logg > 0.5$ and $T_{eff} > 3800$ K. This
resulted in a slightly smaller sample; 6209 out of the 6402 RAVE observations in OC regions are equipped with $[M/H]$ and we had to slightly adapt our quality constraints to conduct a reliable
metallicity study. In addition, the DR4 pipeline provides quality flags for the convergence of the stellar parameter algorithm used to derive $\log{g}$, $T_{eff}$, and $[M/H]$. Since the RV values
were derived by a different algorithm, we did not include them in our RV sample but have to do so now for our metallicity study. Objects with no converging algorithm or which had to be rerun by the
pipeline were excluded from our metallicity study on open clusters.\\
As noted by \citet{Kordopatis2013}, the internal metallicity uncertainties ($e[M/H]^*$) in RAVE DR4 were derived from different sets of synthetic spectra, leading to a discrete distribution (see Fig.
\ref{SN-emet}). These $e[M/H]^*$ might reflect model errors instead of realistic measurement uncertainties. Therefore, we preferred to evaluate the actual $[M/H]$ values and not the uncertainties to
define the adapted cuts for our metallicity study in open clusters.

\begin{figure}[!ht]
\begin{center}
\includegraphics[width=8.5cm]{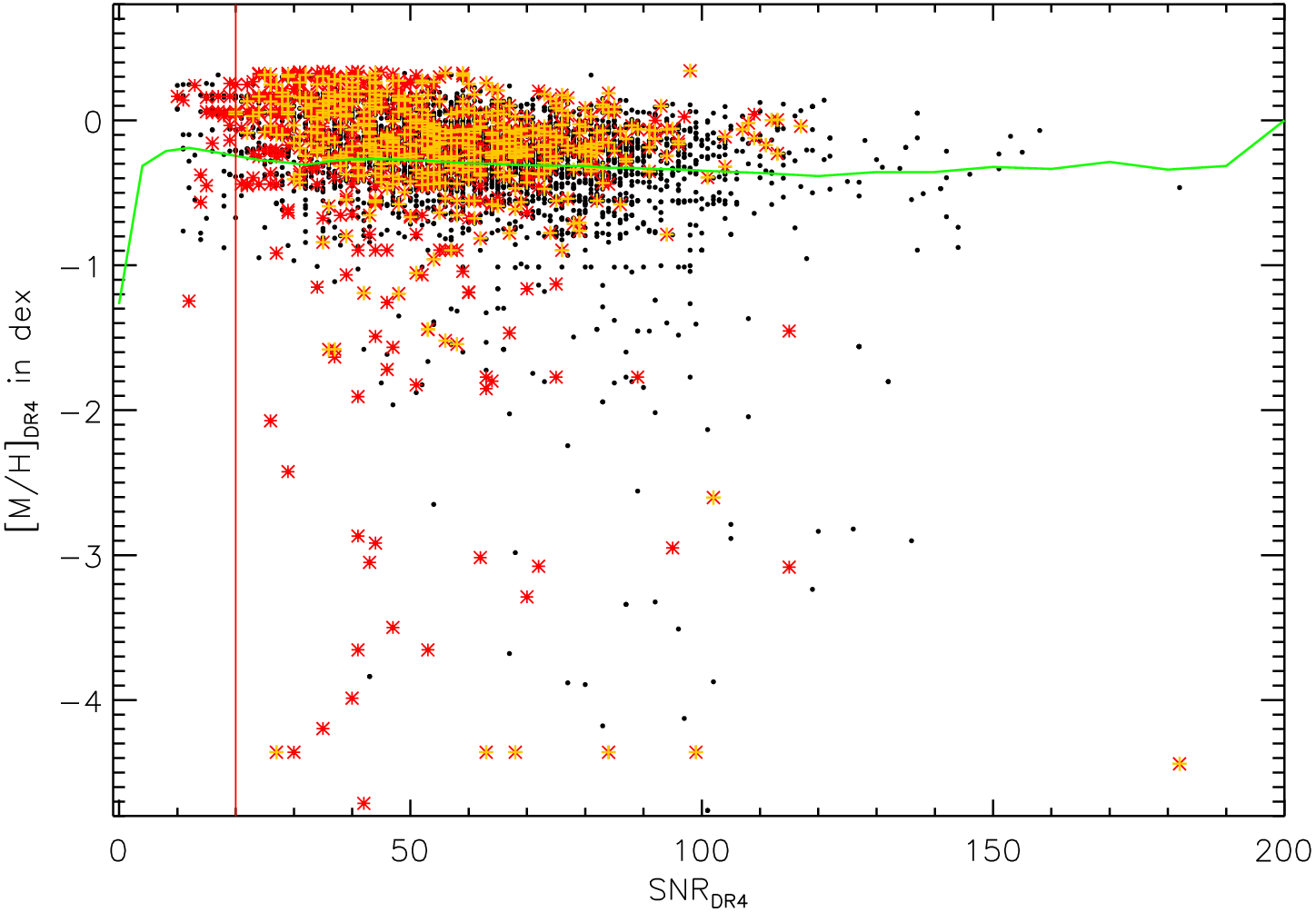}
\caption{$[M/H]$ distribution with respect to SNR for our high-quality RV sample (black dots). Red asterisks and orange crosses illustrate our good RV and $[M/H]$ members, respectively. The red and
green solid lines visualise our adapted cut at an SNR $\ge$ 20 and the overall trend for the entire RAVE DR4, respectively.}
\label{SN-met}
\end{center}
\end{figure}

In Fig. \ref{SN-met} we display the $[M/H]$ distribution with respect to SNR. To illustrate the overall trend in RAVE DR4 we calculated $\overline{[M/H]}$ in bins of 4 along SNR and changed the bin
size to 10 for SNR $\ge$ 100, to gain enough data points in each bin. This overall trend is quite flat and shows no specific correlation, not even for low SNR. Therefore, we simply adapted the same
cut as the RAVE DR4 pipeline at an SNR $\ge$ 20.

\begin{figure}[!ht]
\begin{center}
\includegraphics[width=8.5cm]{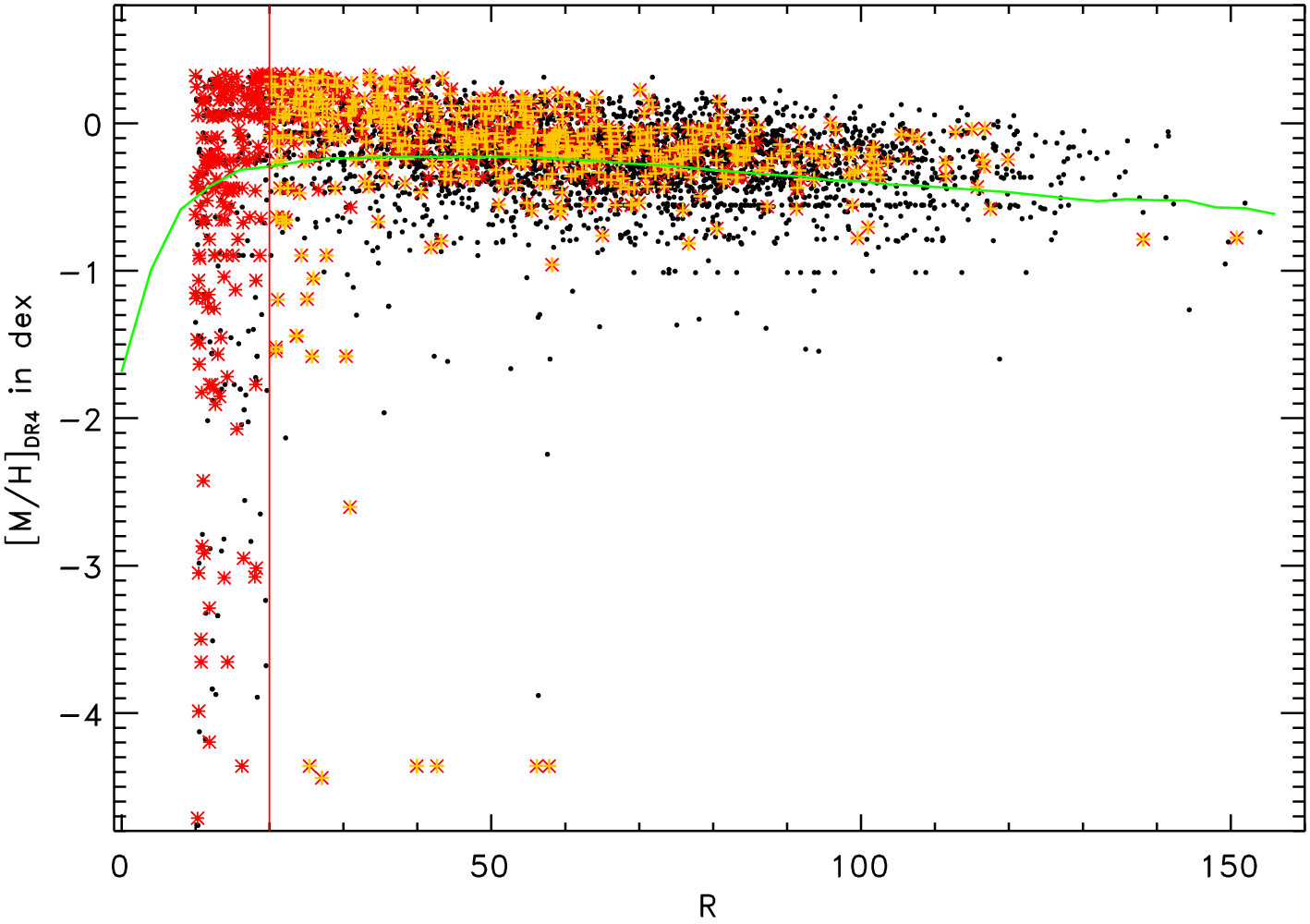}
\caption{$[M/H]$ distribution with respect to $R$. The symbol color-coding is the same as in Fig \ref{SN-met}. The red and green solid lines visualise our adapted cut at $R \ge$ 20 and the overall
trend for the entire RAVE DR4, respectively.}
\label{TDC-met}
\end{center}
\end{figure}

Moreover, we examined the $[M/H]$ distribution with respect to $R$ (Fig. \ref{TDC-met}) and computed the overall trend in RAVE DR4 as $\overline{[M/H]}$ in bins of 4 along $R$. This overall trend
indicates a slight correlation of $[M/H]$ with $R$, suggesting that the fewer lines in metal-poor stars lead to a better match of the observed to the template spectrum, at least for stars with $[M/H]
\ge -1$ dex. Because of this slope we cannot use the overall trend to evaluate the cut refinement in $R$. However, for $R \le$ 20 a non-negligible number of good RV members show unexpectedly low
$[M/H]$, and we chose the corresponding cut to $R \ge$ 20 for our metallicity study in Galactic open clusters.\\
We were unable to identify any dependencies of $[M/H]$ on corr\_RV and saw no need for additional changes of the constraints for our high-quality $[M/H]$ sample. Combined with the membership
probabilities ($P_{kin}$ and $P_{phot}\ge$ 14\% or $P_{kin}$ and $P_{phot}\ge$ 61\%), the new cuts define our good and best $[M/H]$ members, respectively. In Tab. \ref{tab_met} we summarise the
corresponding numbers of measurements, stars, and clusters for our metallicity study.

\begin{figure}[!ht]
\begin{center}
\includegraphics[width=8.5cm]{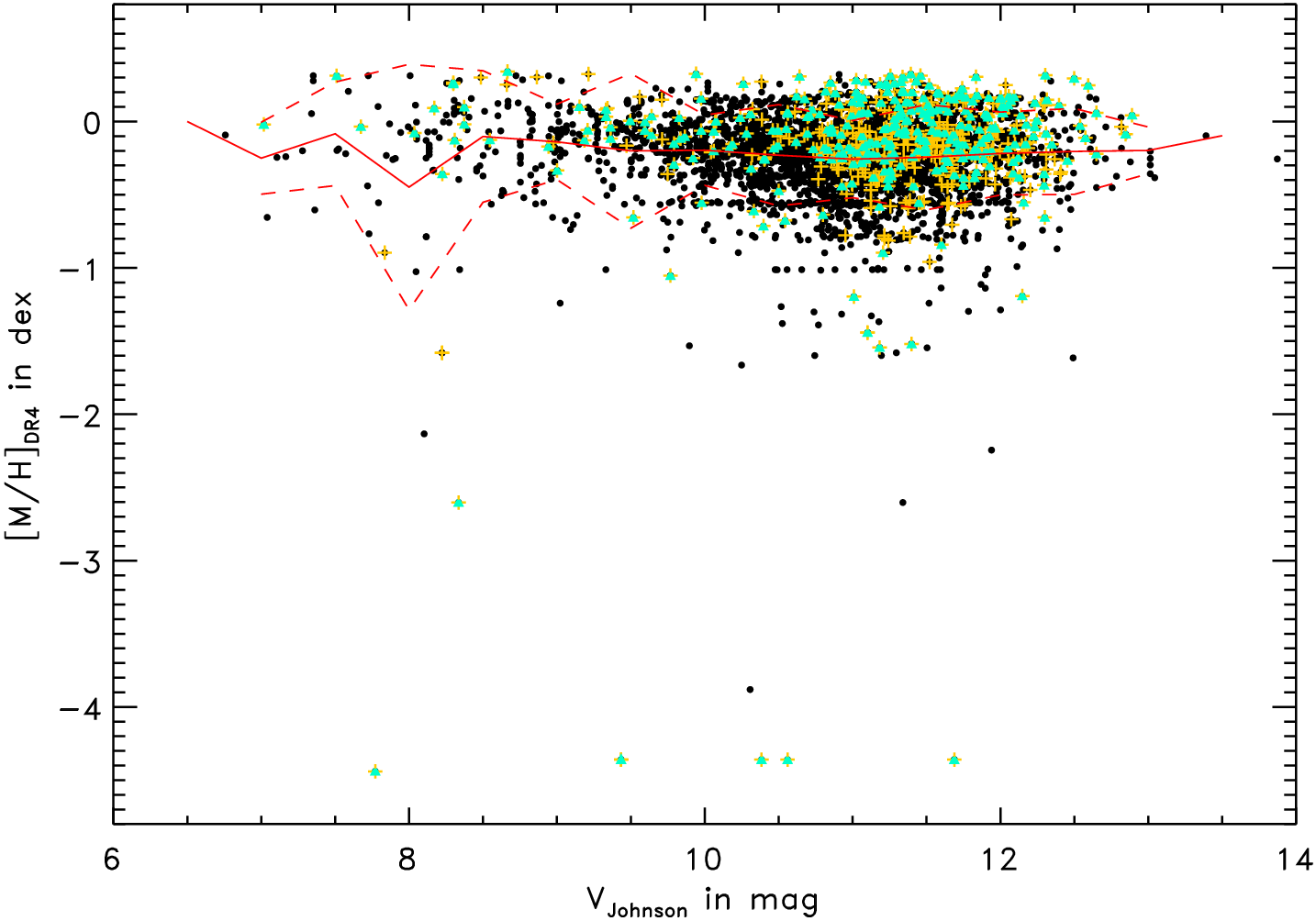}
\caption{$[M/H]$ distribution with respect to $V_{Johnson}$ for our high-quality $[M/H]$ sample (black dots). Orange crosses and turquoise triangles illustrate good and best $[M/H]$ members,
respectively. Red solid and dashed lines visualise $\overline{[M/H]}$ and $\sigma [M/H]$ for our high-quality $[M/H]$ sample, respectively.}
\label{met-vmag}
\end{center}
\end{figure}

Furthermore, we investigated a potential magnitude dependence of $[M/H]$, which might affect the reliability of our data (see Fig. \ref{met-vmag}). The few members at $[M/H]= - $4.36 dex show
obviously unrealistic values and were therefore not considered any further in our metallicity study of OCs. To identify a possible dependence more clearly, we computed the unweighted
$\overline{[M/H]}$ and $\sigma [M/H]$ of our high-quality $[M/H]$ sample in bins of 0.5 mag along $V_{Johnson}$. Both show a very flat behaviour and the variations at brighter magnitudes are most
likely due to small number statistics and are not representative for the overall trend. Hence, we were unable to identify any considerable magnitude dependence of metallicities in RAVE, confirming our
sample to provide reliable results.

Since CSOCA does not provide any metallicity data, no reference values for individual cluster members were available. For cluster mean metallicities, on the other hand, we found reference values in
DAML, which we discuss in more detail in Sect. \ref{cl-met}.

\begin{table*}[!ht]
\begin{center}
\caption{Numbers for our different $[M/H]$ samples in RAVE and OC areas.}
\label{tab_met}
\begin{tabular}{l| r r | r r r r}
\hline
             & \multicolumn{2}{|c|}{RAVE DR4} & \multicolumn{4}{|c}{OC sample} \\
\hline
Number of    & entire & high-quality & $[M/H]$ &   high-quality & good $[M/H]$ & best $[M/H]$ \\
             &   RAVE &      in RAVE &  sample & $[M/H]$ sample &      members &      members \\
\hline\hline
Measurements & 451474 &       354906 &    6209 &           3947 &          517 &          308 \\
Stars        & 405176 &       322843 &    4785 &           3485 &          455 &          265 \\
Clusters     &    --- &          --- &     244 &            192 &           94 &           77 \\
\hline
\end{tabular}
\end{center}
\end{table*}

\section{Mean values for our Galactic open clusters}  \label{cluster}

\subsection{Radial velocity} \label{clu_rv}

First of all, we cleaned each OC from outliers by applying a $3\sigma$-clipping algorithm to obtain the most representative $\overline{RV}$. Then we determined $\overline{RV}$ for in total 110 OCs and
summarise the results in Tab. \ref{tab1} along with catalogue identifiers, that is, COCD number (Seq) and Name. In addition, we provide two kinds of reference values. On the one hand, we computed
$\overline{RV}$ in CRVAD-2, and on the other hand we list values from CRVOCA \citep{Kharchenko2007}. We prefer to use their computed $\overline{RV}$ and only where no calculated $\overline{RV}$ were
available we give literature values. For 37 OCs we provide $\overline{RV}$ for the first time.

\begin{eqnarray}
 \overline{RV}            & = & \frac{\sum\limits_{i} RV_i \cdot g_i}{\sum\limits_{i} g_i} \label{rv-mean} \\ \nonumber
\\
 \sigma \overline{RV}     & = & \sqrt{\frac{n}{n-1} \cdot \frac{\sum\limits_{i} g_i \cdot (RV_i - \overline{RV})^2}{\sum\limits_{i} g_i}} \label{rv-scat} \\ \nonumber
 \\
\textrm{e}{\overline{RV}} & = & \frac{\sigma \overline{RV}}{\sqrt{n}} \label{rv-err} \\ \nonumber
 \\
\overline{eRV^*}          & = & \frac{\sum\limits_{i} eRV^*_i \cdot (P_{kin, i} \cdot P_{phot, i})}{\sum\limits_{i} (P_{kin, i} \cdot P_{phot, i})}, \label{erv-mean}  \\ \nonumber
 \\ \nonumber
\textrm{with the weights} &g_i& \textrm{defined as}  \\ \nonumber
 \\
    g_i                   & = & \frac{P_{kin, i} \cdot P_{phot, i}}{(eRV^*_i)^2}. \label{rv-weight}
\end{eqnarray}

The $\overline{RV}$ from RAVE and CRVAD-2 were primarily derived from best RV or 1$\sigma$-members, respectively. Only where just one or no most probable member was available we included good RV or
2$\sigma$-members as well to compute the $\overline{RV}$ in RAVE and CRVAD-2, respectively. The corresponding numbers are also included in Tab. \ref{tab1}. CRVOCA includes $\overline{RV}$ based on
3$\sigma$-members, while the $\overline{RV}$ references computed in this work consider at worst 2$\sigma$-members to reduce the field star contamination. A comparison between the reference catalogues
yielded a very good agreement, as expected, indicating that in CRVOCA as well the field star contamination can be considered to be relatively low and the values as suitable references.\\
The provided $\overline{RV}$ in RAVE and CRVAD-2 were calculated as weighted mean considering individual $eRV^*$ and membership probabilities $P_{kin}$ and $P_{phot}$ (Eq. \ref{rv-mean}). As mentioned
above, we considered all $eRV^*<1$ km/s to be too optimistic and replaced them with 1 km/s, which is also reflected in Tab. \ref{tab1}. We also give typical RV uncertainties in OCs
($\overline{eRV^*}$), computed as weighted mean from the individual $eRV^*$ of the members (Eq. \ref{erv-mean}), including only OC membership probabilities as weights. The weighted standard deviation
($\sigma \overline{RV}$; Eq. \ref{rv-scat}) and uncertainty of $\overline{RV}$ (e$\overline{RV}$; Eq. \ref{rv-err}) could only be computed for OCs with at least two individual measurements. For
clusters with only one representative we do not provide $\sigma \overline{RV}$ and assume $\overline{eRV^*}=eRV^*$.

\begin{figure}[!ht]
\begin{center}
\includegraphics[width=8.5cm]{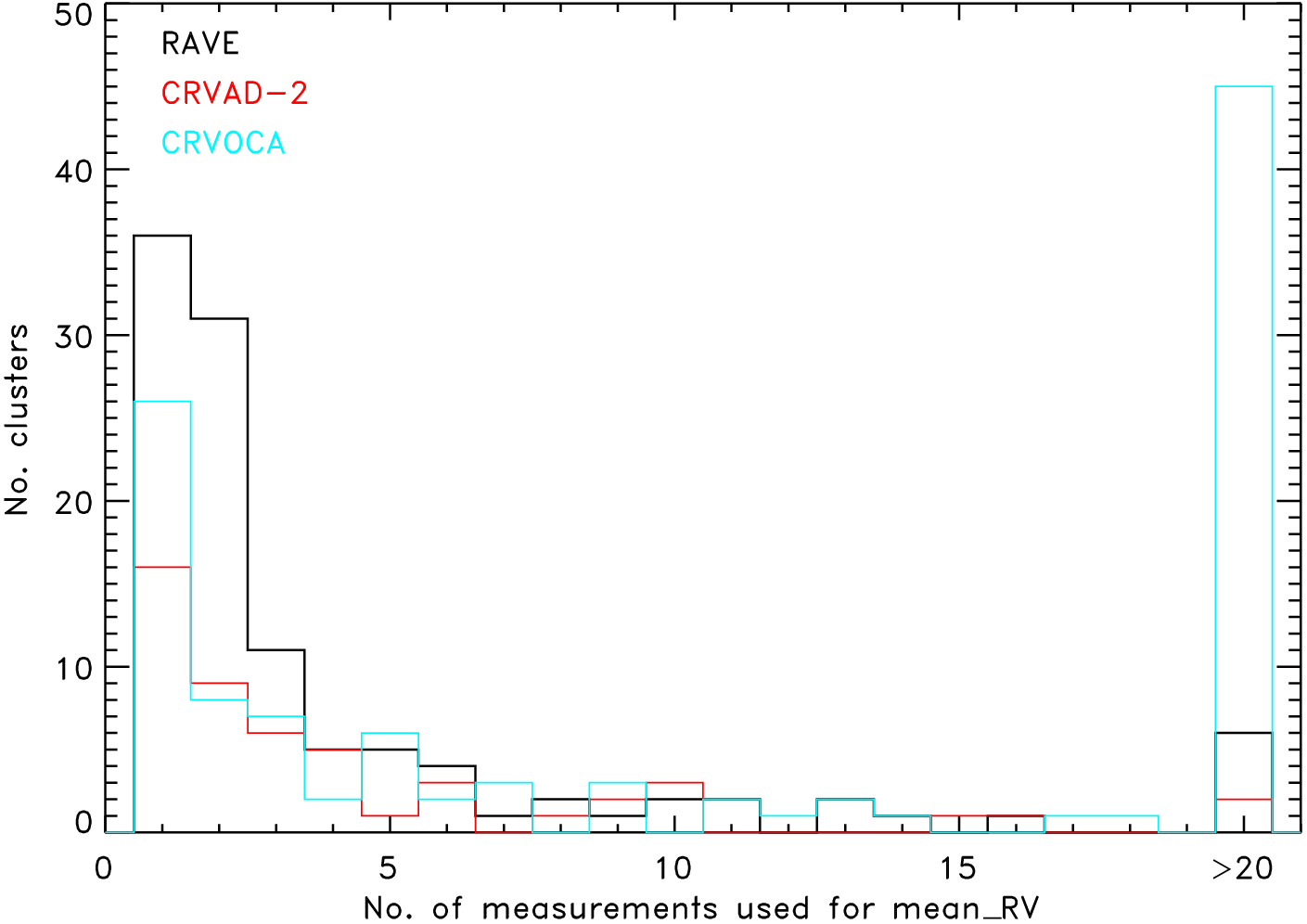}
\caption{Histogram for the number of measurements or stars used to derive $\overline{RV}$ in RAVE (black) and CRVAD-2 (red), respectively. The cyan histogram shows the number histogram for CRVOCA.}
\label{num-rv}
\end{center}
\end{figure}

In Fig.\ref{num-rv} we show the histograms for the total number of measurements and stars used to obtain the RAVE based and reference $\overline{RV}$, respectively. We only included OCs observed in
RAVE. The vast majority of $\overline{RV}$ in all catalogues are based on fewer than six individual RV measurements and only a few OCs show $\overline{RV}$ derived from more than 20 individual RV
measurements in either data set. CRVOCA shows the largest number of OCs with more than 20 individual RV values, since they used stars with lower membership probability than we did. Considering the
different numbers of OCs covered by the catalogues, the distributions for the number of individual measurements show a very similar shape. This indicates that the resulting $\overline{RV}$ are of
similar quality, as expected.

\begin{figure*}[!ht]
\begin{center}
\rotatebox{180}{\includegraphics[width=17cm]{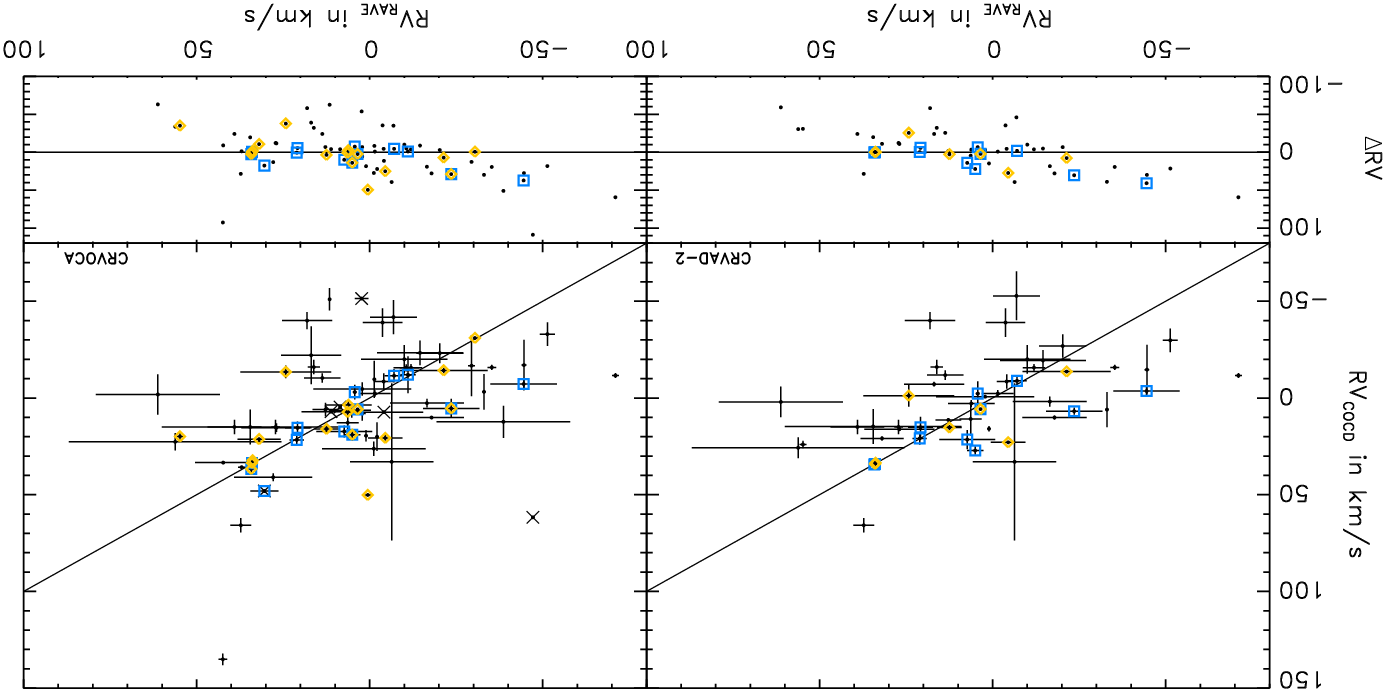}}
\caption{Upper panels: $\overline{RV}$ comparison between RAVE and reference values from CRVAD-2 (left) and CRVOCA (right). The black line shows the one-to-one relations. Lower panels:
Corresponding difference distributions with the zero-difference lines included as black solid lines. Blue squares and yellow diamonds illustrate clusters with $\ge$ 10 individual RVs in RAVE and the
reference catalogue, respectively. Black crosses indicate missing e$\overline{RV}$ information in CRVOCA.}
\label{clu-rv}
\end{center}
\end{figure*}

Fig. \ref{clu-rv} illustrates a visual comparison between our RAVE results and available references. The error bars represent the e$\overline{RV}$ in each catalogue. The RV difference ($\Delta
\overline{RV}$) is defined as $\Delta \overline{RV} =\overline{RV_{Ref}}-\overline{RV_{RAVE}}$, where $\overline{RV_{Ref}}$ are the reference values obtained from CRVAD-2 or CRVOCA for the
corresponding panel.  The differences between RAVE results and reference values for our OCs (Fig. \ref{clu-rv}) appear to be larger than for the individual stars (Fig. \ref{rv-comp}). One can see a
negative slope in the difference distribution, which is mainly caused by two OCs with very large differences and cannot be verified to be statistically significant. Contributing factors to the
apparently larger RV differences are the different OC members targeted by either survey and the potential systematics induced by the reference values from \citet{Barbier2000}. In general, cluster
$\overline{RVs}$ derived from only up to five individual measurements have to be considered with caution in all data sets used in the presented project, that is, RAVE, CRVAD-2, and CRVOCA.\\
OCs with more than ten individual measurements in RAVE, on the other hand, show a very good agreement, except for three. The three exceptions (Platais 8, Sco-OB 4, and Sgr-OB 7; left panel of Fig.
\ref{clu-rv}) are all associations, which naturally show an intrinsically higher velocity dispersion, because they are not as tightly bound as open clusters. Since the membership selection is partly
based on kinematics, it might be possible that for associations as well mistaken membership can contribute to the larger differences, in particular because different objects were targeted by RAVE and
CRVAD-2. CRVAD-2 references with more than ten individual RV measurements also show a good agreement, except for two actual open clusters: NGC 2516 and Collinder 228. In CRVOCA even better measured
OCs show relatively large differences to the RAVE results. Thus, the field star contamination in CRVOCA is not negligible, though we stated it to be relatively low. Furthermore, we can conclude that
RAVE provides more reliable $\overline{RV}$ than CRVAD-2.

\begin{figure}[!ht]
\begin{center}
\includegraphics[width=8.5cm]{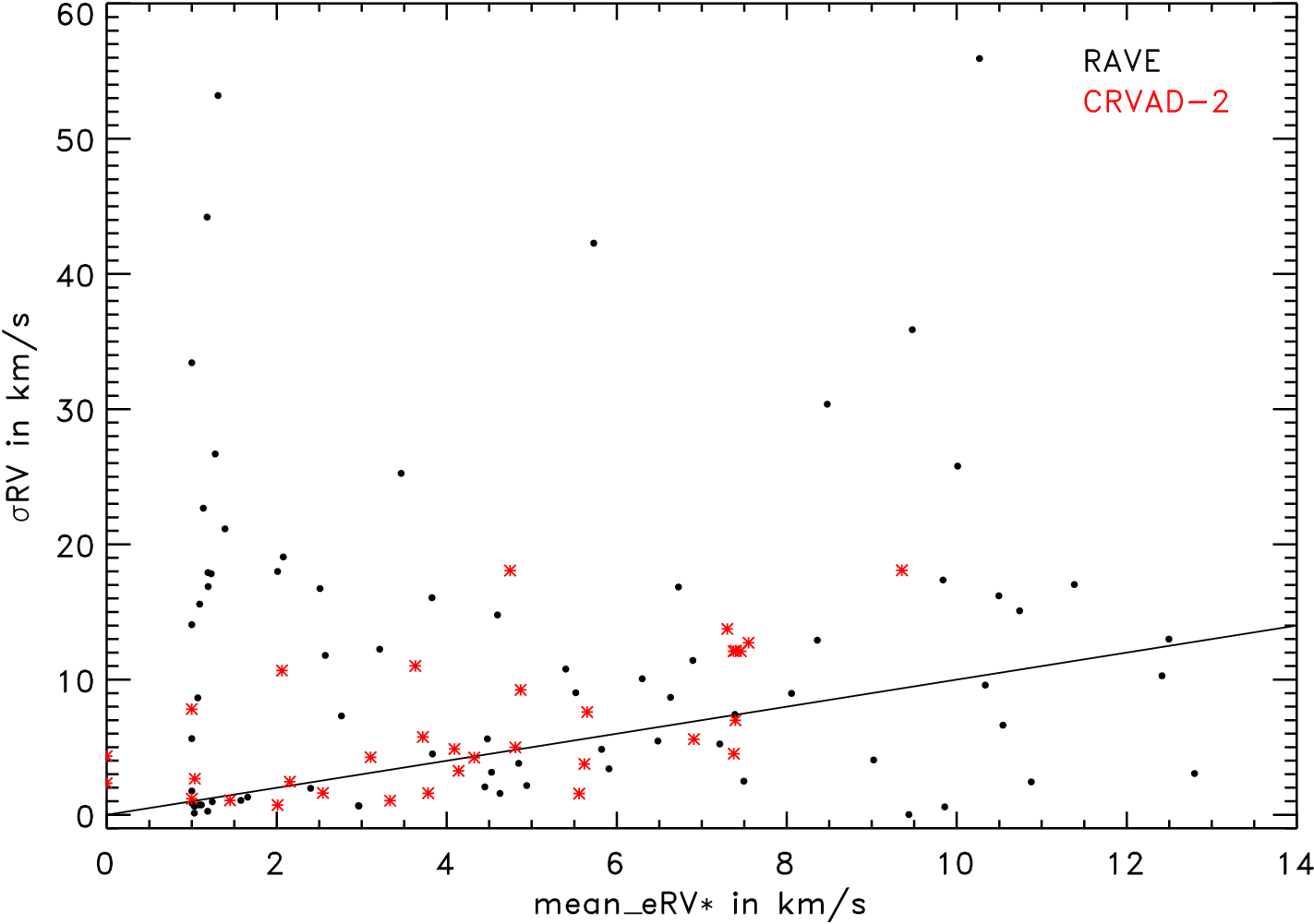}
\caption{Comparison of $\sigma\overline{RV}$ to $\overline{eRV^*}$ in CRVAD-2 (red asterisks) and RAVE (black dots) for OCs observed by RAVE. The black solid line represents the one-to-one relation.}
\label{err-scat}
\end{center}
\end{figure}

In addition, we compared $\sigma\overline{RV}$ and $\overline{eRV^*}$ in RAVE and CRVAD-2 (Fig. \ref{err-scat}). In both catalogues only very few OCs show $\sigma \overline{RV}$ similar to
$\overline{eRV^*}$, the majority show higher $\sigma\overline{RV}$, and in certain cases they are about a factor of 5-10 higher than $\overline{eRV^*}$. There are several possible reasons, namely
small number statistics, partly mistaken membership, or undetected binarity. Due to the first aspect, the $\sigma \overline{RV}$ have to be considered with care and cannot be regarded in any way
representative for the internal cluster velocity dispersion. The aspect of binarity in our OCs is discussed in Sect. \ref{binary}. Partly mistaken membership might be minimised when updated membership
probabilities from the Milky Way Star Cluster (MWSC) survey \citep{Kharchenko2012} become available.\\
Moreover, it would be a great improvement to also include RVs as criteria for OC membership, but this is only reasonable when RV data are available for all stars in OC areas. The CRVAD-2
$\sigma\overline{RV}$ are well below 20 km/s, whereas the RAVE values reach up to 60 km/s. Most likely, this is due to the different targets included to compute $\overline{RV}$ for the two catalogues
(see Sect. \ref{radvel}).

\subsection{Binarity fraction} \label{binary}

Above we pointed out that undetected binaries can have a significant influence on the accuracy of our $\overline{RV}$ results. For a detailed study multiple epochs for each member would be needed. We
examined our best RV members in RAVE for multiple epochs and only identified 76 out of 443 stars, where each object is only provided with two measurements. This is by far not enough for a deep binary
study based on RAVE data. Hence, we have to work with limited sources of information to give an approximate idea on the binary fraction in our sample.\\
In a first step we checked the duplicity flags in CSOCA and found 14 stars indicated as potential or confirmed binaries among our 443 best RV members. Secondly, we cross-matched our best RV members
with the list of SB1 \citep{Matijevic2011} and SB2 \citep{Matijevic2010} binaries in RAVE and found no common object. This is not surprising, since we rejected objects with bad spectral flags from
\citet{Matijevic2012}. If we only consider the cuts SNR $\ge$ 10, $R \ge$ 10, and $|corr\_RV| \le$ 9 km/s in RAVE along with $P_{kin}$ and $P_{phot}\ge$ 61\%, we find 11 SB2 binaries in 4 OCs.
However, all these numbers are far below the 6\% binary fraction suggested by \citet{Matijevic2011}.\\
Moreover, we provide a rough estimate on the binary fraction based on RAVE data using a very simple approach, namely that the large scatter in Fig. \ref{clu-rv} and the high $\sigma \overline{RV}$ are
mainly caused by undetected binarity. For each cluster we first computed the difference between individual RVs and $\overline{RV}$. Then we compared these differences with 3$\overline{eRV^*}$,
defining our assumed velocity dispersion. This analysis can only be made for OCs with at least two individual measurements, which reduces the number of clusters considered to 76. We assumed members
exceeding the 3$\overline{eRV^*}$ limit to be potential binaries and calculated the binary fraction with respect to the total number of RAVE measurements in the corresponding OC. The results are
summarised in Tab. \ref{tab-bin}. 

\begin{table}[!ht]
\begin{center}
\caption{Results for our rough binary fraction estimate in OCs with at least two RV measurements in RAVE.}
\label{tab-bin}
\begin{tabular}{l| r r r r | r }
\hline
binary fraction &   0\% & $\le$25\% & 25-50\% & $\ge$50\% & total \\
\hline\hline
No. of OCs      &    41 &         9 &       7 &        17 &    74 \\
Proportion (\%) &  55.4 &      12.2 &     9.5 &      23.0 &   --- \\
\hline
\end{tabular}
\end{center}
\end{table}

About half of our OCs with at least two RV measurements show no binarity and another 23\% show a very high estimated binary fraction ($\ge$50\%). This effect is most likely due to small number
statistics, where the binary fraction can change fast from 0\% to more than 50\% if just one more star is outside the defined 3$\overline{eRV^*}$ limit. Therefore, the listed numbers can at most be
considered as lower limits. In Tab. \ref{tab1} about 45.9\% of OCs with at least two RV measurements show $\sigma \overline{RV} \ge10 $ km/s, which is similar to the 44.7\% of OCs with non-zero binary
fraction. This verifies that undetected binaries are a dominant effect that induces unexpectedly high $\sigma \overline{RV}$ for our OCs.

\subsection{Metallicity} \label{cl-met}

Because of the more stringent requirements in our $[M/H]$ study, we were able to determine $\overline{[M/H]}$ for only 81 of our 110 OCs with $\overline{RV}$ in RAVE. Because we strictly distinguished
between iron abundances and overall metallicities in DAML (see Sect. \ref{DAML}), we obtained reference $\overline{[M/H]}$ for only 12 OCs. Hence, for 69 clusters we present $\overline{[M/H]}$ for the
first time. The results are summarised in Tab. \ref{tab2} along with the cluster identifiers (COCD number and cluster name). Our metallicity results were primarily obtained from best $[M/H]$ member
measurements after cleaning each OC from outliers by applying a 3$\sigma$-clipping algorithm. Only where no or just one best $[M/H]$ member measurement was available we included good $[M/H]$ member
measurements as well. The number of best and additional good $[M/H]$ member measurements are also included in Tab. \ref{tab2}. We computed the $\overline{[M/H]}$ as weighted mean with respect to the
membership probabilities (Eq. \ref{met-mean}), since the listed $e[M/H]^*$ show a very discrete distribution and might not reflect realistic measurements errors (see Sect. \ref{metal}). For OCs with
at least two individual $[M/H]$ measurements we computed weighted standard deviations ($\sigma \overline{[M/H]}$; Eq. \ref{met-scat}) and uncertainties of $\overline{[M/H]}$ (e$\overline{[M/H]}$; Eq.
\ref{emet-mean}).

\begin{eqnarray}
 \overline{[M/H]}            & = & \frac{\sum\limits_{i} [M/H]_i \cdot w_i}{\sum\limits_{i} w_i} \label{met-mean} \\ \nonumber
\\
 \sigma \overline{[M/H]}     & = & \sqrt{\frac{n}{n-1} \cdot \frac{\sum\limits_{i} w_i \cdot ([M/H]_i - \overline{[M/H]})^2}{\sum\limits_{i} w_i}} \label{met-scat}\\ \nonumber
 \\
\textrm{e}{\overline{[M/H]}} & = & \frac{\sigma \overline{[M/H]}}{\sqrt{n}}, \label{emet-mean}\\ \nonumber
\\ \nonumber
\textrm{with the weights} &w_i& \textrm{defined as} \\ \nonumber
\\
 w_i & = & P_{kin, i} \cdot P_{phot, i}.
\end{eqnarray}

\begin{figure}[!ht]
\begin{center}
\includegraphics[width=8.5cm]{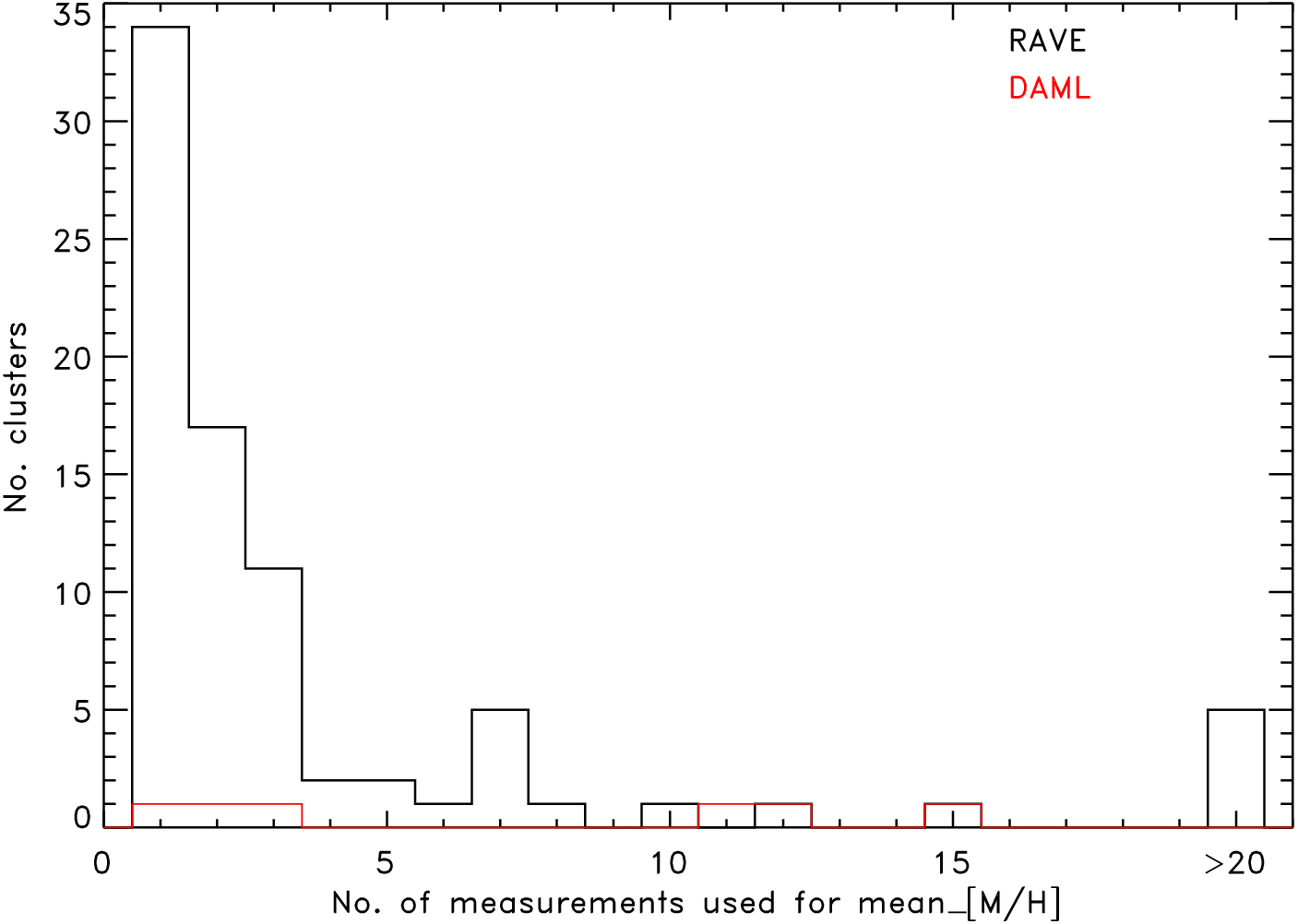}
\caption{Histogram for the number of measurements or stars used to obtain $\overline{[M/H]}$ in RAVE (black) and DAML (red), respectively.}
\label{num-met}
\end{center}
\end{figure}

In Fig. \ref{num-met} we display the histograms for the number of measurements and stars used to obtain $\overline{[M/H]}$ in RAVE and DAML, respectively. Again we only included OCs with $[M/H]$ data
available in RAVE. As expected, the vast majority of OCs are covered by fewer than six individual $[M/H]$ measurements and small number statistics might affect our results. The number of references is
too small to conclude about the shape of the number distribution.

\begin{figure}[!ht]
\begin{center}
\includegraphics[width=8.5cm]{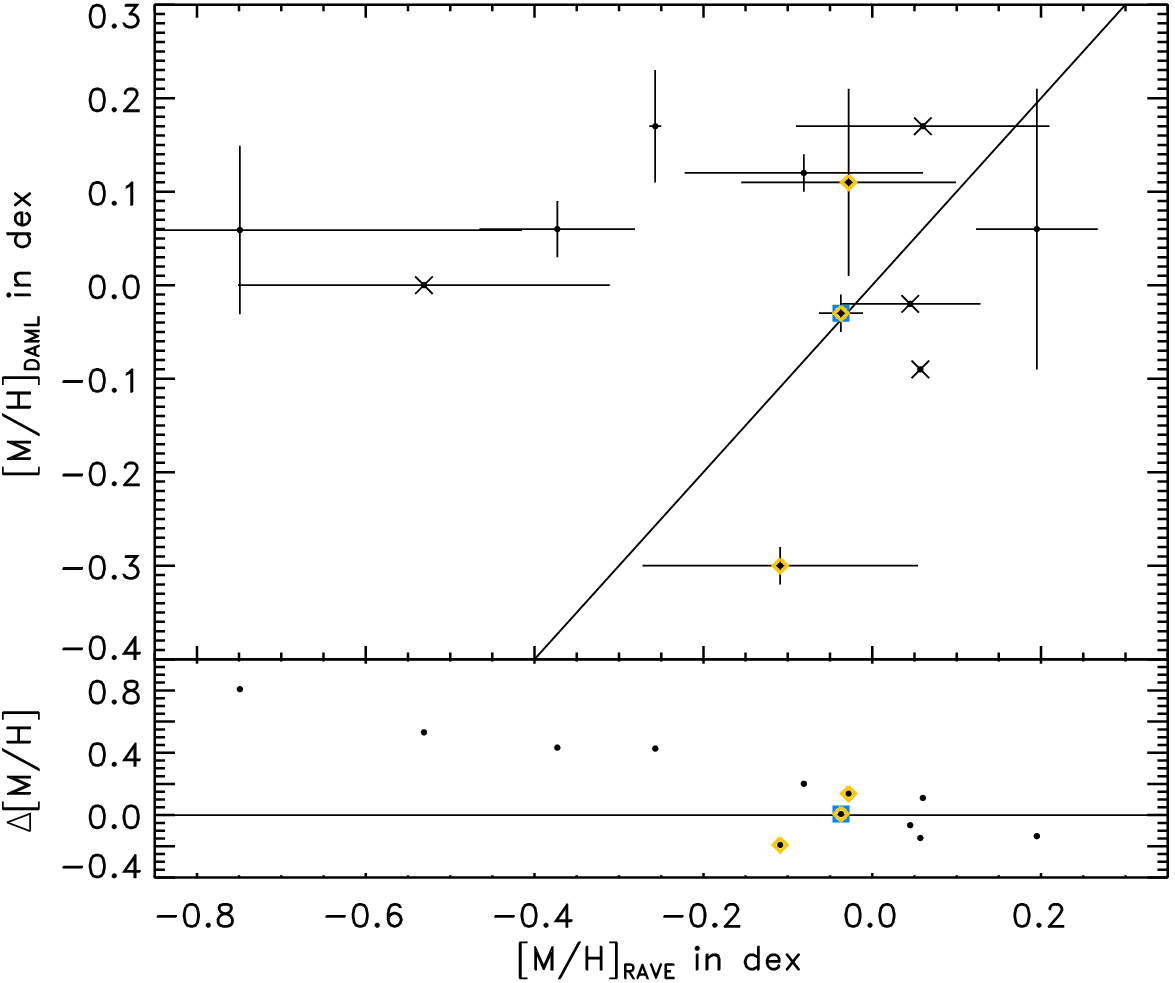}
\caption{$\overline{[M/H]}$ comparison (upper panel) and difference distribution (lower panel) between RAVE DR4 and DAML, along with the one-to-one relation and zero-difference line (black solid
lines). Blue squares and yellow diamonds highlight OCs with $\ge$10 individual $[M/H]$ measurements in RAVE DR4 and DAML, respectively. Black crosses indicate e$\overline{[M/H]}$ missing in one or
both catalogues.}
\label{met-comp}
\end{center}
\end{figure}

From Fig. \ref{met-comp} one can see that the majority of OCs in RAVE, except for four, agree very well with the values from DAML within the uncertainties. We define the differences between the
catalogues as $\Delta \overline{[M/H]} = \overline{[M/H]_{DAML}} - \overline{[M/H]_{RAVE}}$ and they appear to be similar to the uncertainties. Only the Pleiades (Melotte 22) are covered by more than
ten individual measurements in RAVE and agree very well. In addition to the Pleiades, DAML lists two more clusters with $\overline{[M/H]}$ based on more than ten values, namely NGC 2422 and NGC
2354.\\
Our metallicity study in RAVE can only give a rough idea on the $[M/H]$ behaviour of the Galactic OC system. The typical uncertainties of $\overline{[M/H]}$ and individual members, obtained from the
pipeline, are about 0.1 dex and reflect only internal errors. When including external errors as well, the typical errors are about 0.3 dex \citep{Boeche2011}. The RAVE $[M/H]$ accuracy is apparently
not high enough to carry out a detailed metallicity study within OCs.

\begin{figure}[!ht]
\begin{center}
\includegraphics[width=8.5cm]{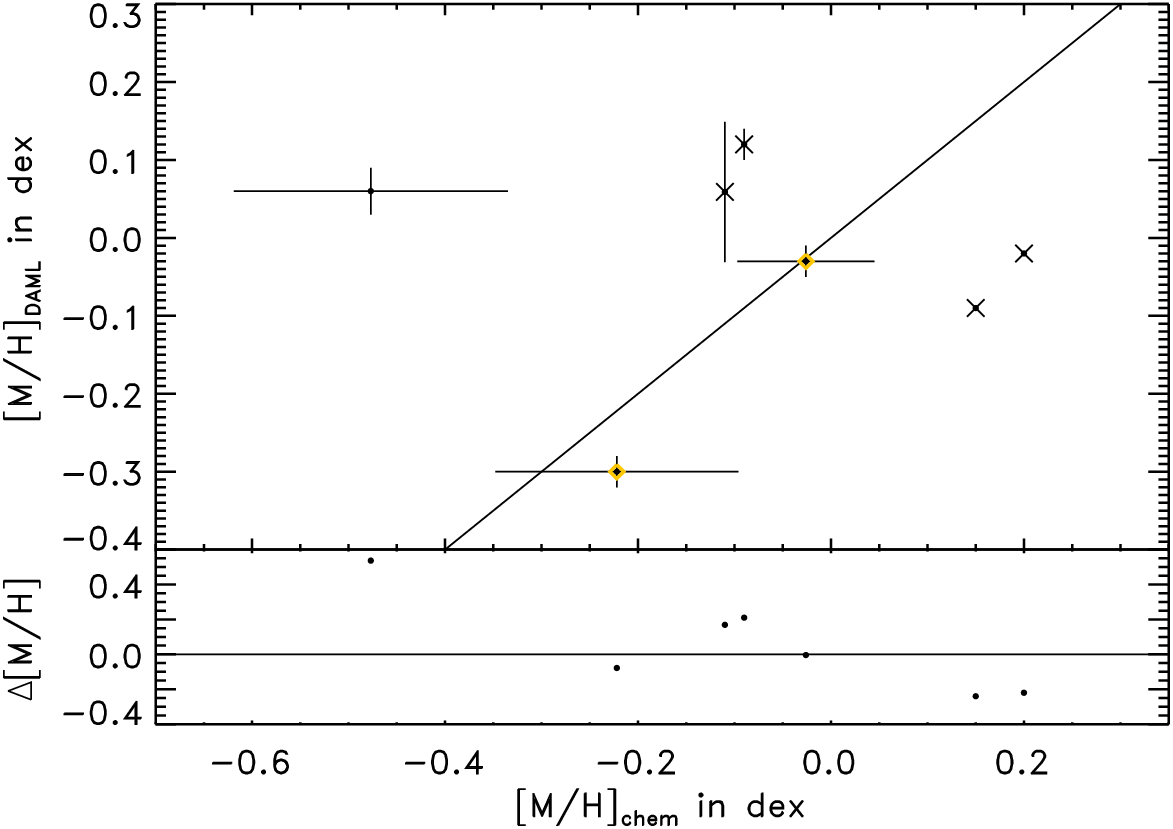}
\caption{$\overline{[M/H]}$ comparison (upper panel) and difference distribution (lower panel) between the results from the RAVE chemical pipeline \citep{Boeche2011} and DAML, along with the
one-to-one relation and zero-difference line (black solid lines). Yellow diamonds highlight OCs with $\ge$10 individual $[M/H]$ measurements in DAML. Black crosses indicate e$\overline{[M/H]}$ missing
in one or both catalogues.}
\label{chem-comp}
\end{center}
\end{figure}

A brief look at the difference distribution might suggest a negative slope with increasing metallicities. This apparent slope is primarily caused by four clusters, which are metal poor in RAVE. If we
eliminate them, the distribution is consistent with not showing any trend and is centred around zero. In Tab. \ref{tab2} we found ten clusters and associations with $\overline{[M/H]}$ below $-0.5$
dex. This contradicts our expectation that open clusters and associations in the solar neighbourhood have about solar metallicity. Except for one OC with three best $[M/H]$ member measurements, the
$\overline{[M/H]}$ values for all metal-poor OCs are based on either one best $[M/H]$ member or mainly on good $[M/H]$ members. Therefore, mistaken membership in combination with small number
statistics can be one reason for very low $\overline{[M/H]}$.\\
However, this would not explain the amount of very metal poor OCs we found in our sample, since our membership selection used a uniform algorithm on homogeneous spatial, photometric, and kinematic
information. These unexpectedly metal-poor OCs could also indicate that the RAVE DR4 pipeline might underestimate the corresponding metallicities for certain spectra. This is supported by our finding
that three out of the 23 individual $[M/H]$ measurements of Pleiades best members show values of $-4.36$ dex, which we excluded when we computed $\overline{[M/H]}$.\\
To verify this hypothesis we analysed the results of the chemical pipeline implemented for RAVE by \citet{Boeche2011}. These authors employed slightly more stringent quality constraints (SNR $\ge20$,
$v_{rot}<50$ km/s and $4000<T_{eff}<7000$ K). It also has to be noted that the chemical pipeline does not cover the very metal-poor end, which the DR4 pipeline does, since either the data quality is
too low or the spectral characteristics are not covered by the data grid used in the chemical pipeline. Hence, the chemical pipeline provides $\overline{[M/H]}$ for only 52 OCs with typically fewer
individual measurements after applying our quality requirements on this data set. We included these additional results in Tab. \ref{tab2} along with the number of good and best member measurements in
this data set and show a comparison to our reference $\overline{[M/H]}$ in Fig. \ref{chem-comp}.

The two RAVE metallicity sets, DR4 and the chemical pipeline, agree well with the references from DAML in the range $-0.5<\overline{[M/H]}<0.5$. However, the chemical pipeline does not provide any
very metal-poor values for targets that match our quality requirements, and such stars are simply not listed in the resulting data table. This might indicate that the apparently very metal poor
stars in DR4 suffer from lower data quality. Future investigation will show whether all these very metal-poor OCs simply arise from mistaken membership combined with low number statistics or if
potentially underestimated metallicities in RAVE DR4 might also play a role.

\section{Summary and discussion}  \label{disc}

Current compilations and catalogues of Galactic open clusters significantly lack spectroscopic information, such as RVs and abundances. The RAVE survey allows us to fill in some of the missing data.
Our project is based on the most homogeneous OC catalogue by \citet{Kharchenko2005a, Kharchenko2005b} (COCD) and the corresponding stellar catalogue (CSOCA).\\
Via a cross-match we identified OC members in RAVE DR4, with a bias towards fainter stars. For the cleaned working sample we provided new RV and $[M/H]$ data. Interestingly, our OC members in RAVE do
not represent the accuracy of the entire survey. We showed that this is most likely due to the higher percentage of dwarfs in our OC sample. Still, the data quality is sufficient for determining
$\overline{RV}$ and $\overline{[M/H]}$ for Galactic open clusters, since the selected members agree well with previous RV data in OCs.\\
We were able to derive $\overline{RV}$ for 110 OCs, including new data for 37 open clusters. $\overline{[M/H]}$ we derived for only 81 OCs, due to more stringent constraints for our metallicity
sample. For 69 of these OCs we presented metallicities for the first time. The $\overline{RV}$ sample agrees better with the reference values than the $\overline{[M/H]}$ based on RAVE DR4. The
relatively large spread in both comparison distributions is most likely caused by different stellar samples for each OC in RAVE and the reference catalogue, partly mistaken OC membership, or
undetected binarity. Partly mistaken membership may be minimised when the updated membership probabilities from the Milky Way Star Cluster (MWSC) survey \citep{Kharchenko2012} become available.
Furthermore, most of our results are based on only a few individual measurements, which in general makes them less robust against the effects mentioned. All these clusters in RAVE and the reference
catalogues have to be considered with caution.\\
Studies by \citet{Kouwenhoven2008}, \citet{Geller2008, Geller2010}, and \citet{Gieles2010} also indicate that binarity may significantly affect the internal velocity dispersion of open clusters.
Although we cannot consider our $\sigma \overline{RV}$ to be representative for the internal cluster velocity dispersion, we come to the same conclusion based on a rough estimate on binarity in the
considered OCs, yielding a similar number of OCs with potential binaries present and OCs with unusually high $\sigma \overline{RV}$.\\
Our $\sigma \overline{RV}$ results are of sufficient quality to derive reliable 3D-kinematics for the Galactic OC system. Combined with previous RV data on OCs this enabled us to re-evaluate the open
cluster groups and complexes, proposed by \citet{Piskunov2006}. The additional abundance data obtained by RAVE may only give us a rough idea on the $[M/H]$ behaviour of the Galactic OC system. We
found ten OCs with $\overline{[M/H]}<-0.5$ dex, which are too metal poor considering that they are located in the solar neighbourhood. Hence, the DR4 metallicities presented in this work have to be
considered with care.\\
Based on inter-cluster differences we can draw conclusions on potential formation scenarios of the re-investigated open cluster groupings. For a very detailed picture high-resolution results would be
necessary, which was previously suggested by \citet{Carrera2007} and \citet{Carrera2012}. In a second paper (Conrad et al. in prep.) we will present more results of our ongoing project on the OC
groups and complexes.
\vspace*{0.8cm}

\begin{acknowledgements}
This work was supported by DFG grant RO 528/10-1, and RFBR grant 10-02-91338, and by Sonderforschungsbereich SFB 881 “The MilkyWay System” (subproject B5) of the German Research Foundation (DFG).
Funding for RAVE has been provided by: the Australian Astronomical Observatory; the Leibniz-Institut f\"ur Astrophysik Potsdam (AIP); the Australian National University; the Australian Research
Council; the French National Research Agency; the German Research Foundation; the European Research Council (ERC-StG 240271 Galactica); the Istituto Nazionale di Astrofisica at Padova; The Johns
Hopkins University; the National Science Foundation of the USA (AST-0908326); the W. M. Keck foundation; the Macquarie University; the Netherlands Research School for Astronomy; the Natural Sciences
and Engineering Research Council of Canada; the Slovenian Research Agency; the Swiss National Science Foundation; the Science \& Technology Facilities Council of the UK; Opticon; Strasbourg
Observatory; and the Universities of Groningen, Heidelberg and Sydney. The RAVE web site is at http://www.rave-survey.org
\end{acknowledgements}

\bibliographystyle{aa}
\bibliography{paper_ref}

\begin{thebibliography}{60}
\expandafter\ifx\csname natexlab\endcsname\relax\def\natexlab#1{#1}\fi

\bibitem[{{Alessi} {et~al.}(2003){Alessi}, {Moitinho}, \& {Dias}}]{Alessi2003}
{Alessi}, B.~S., {Moitinho}, A., \& {Dias}, W.~S. 2003, \aap, 410, 565

\bibitem[{{Barbier-Brossat} \& {Figon}(2000)}]{Barbier2000}
{Barbier-Brossat}, M. \& {Figon}, P. 2000, \aaps, 142, 217

\bibitem[{{Bastian} \& {R{\"o}ser}(1993)}]{PPM1993}
{Bastian}, U. \& {R{\"o}ser}, S. 1993, {PPM Star Catalogue. Positions and
  proper motions of 197179 stars south of -2.5 degrees declination for equinox
  and epoch J2000.0. Vol. III: Zones -00{$^{\circ}$} to -20{$^{\circ}$}. Vol.
  IV: Zones -30{$^{\circ}$} to -80{$^{\circ}$}.}

\bibitem[{{Bica} {et~al.}(2003{\natexlab{a}}){Bica}, {Dutra}, \&
  {Barbuy}}]{Bica2003a}
{Bica}, E., {Dutra}, C.~M., \& {Barbuy}, B. 2003{\natexlab{a}}, \aap, 397, 177

\bibitem[{{Bica} {et~al.}(2003{\natexlab{b}}){Bica}, {Dutra}, {Soares}, \&
  {Barbuy}}]{Bica2003b}
{Bica}, E., {Dutra}, C.~M., {Soares}, J., \& {Barbuy}, B. 2003{\natexlab{b}},
  \aap, 404, 223

\bibitem[{{Boeche}\noopsort{a} {et~al.}(2011){Boeche}\noopsort{a}, {Siebert},
  {Williams}, {de Jong}, {Steinmetz}, {Fulbright}, {Ruchti}, {Bienaym{\'e}},
  {Bland-Hawthorn}, {Campbell}, {Freeman}, {Gibson}, {Gilmore}, {Grebel},
  {Helmi}, {Munari}, {Navarro}, {Parker}, {Reid}, {Seabroke}, {Siviero},
  {Watson}, {Wyse}, \& {Zwitter}}]{Boeche2011}
{Boeche}\noopsort{a}, C., {Siebert}, A., {Williams}, M., {et~al.} 2011, \aj,
  142, 193

\bibitem[{{Carrera\noopsort{a}} {et~al.}(2007){Carrera\noopsort{a}}, {Gallart},
  {Pancino}, \& {Zinn}}]{Carrera2007}
{Carrera\noopsort{a}}, R., {Gallart}, C., {Pancino}, E., \& {Zinn}, R. 2007,
  \aj, 134, 1298

\bibitem[{{Carrera\noopsort{b}}(2012)}]{Carrera2012}
{Carrera\noopsort{b}}, R. 2012, \aap, 544, A109

\bibitem[{{Clari{\'a}} {et~al.}(1999){Clari{\'a}}, {Mermilliod}, \&
  {Piatti}}]{Claria1999}
{Clari{\'a}}, J.~J., {Mermilliod}, J.-C., \& {Piatti}, A.~E. 1999, \aaps, 134,
  301

\bibitem[{{Cutri} {et~al.}(2003){Cutri}, {Skrutskie}, {van Dyk}, {Beichman},
  {Carpenter}, {Chester}, {Cambresy}, {Evans}, {Fowler}, {Gizis}, {Howard},
  {Huchra}, {Jarrett}, {Kopan}, {Kirkpatrick}, {Light}, {Marsh}, {McCallon},
  {Schneider}, {Stiening}, {Sykes}, {Weinberg}, {Wheaton}, {Wheelock}, \&
  {Zacarias}}]{MASS2003}
{Cutri}, R.~M., {Skrutskie}, M.~F., {van Dyk}, S., {et~al.} 2003, VizieR Online
  Data Catalog, 2246, 0

\bibitem[{{Dias} {et~al.}(2002){Dias}, {Alessi}, {Moitinho}, \&
  {L{\'e}pine}}]{Dias2002}
{Dias}, W.~S., {Alessi}, B.~S., {Moitinho}, A., \& {L{\'e}pine}, J.~R.~D. 2002,
  \aap, 389, 871

\bibitem[{{Dufolt} {et~al.}(1995){Dufolt}, {Figon}, \&
  {Meyssonier}}]{Dufolt1995}
{Dufolt}, M., {Figon}, P., \& {Meyssonier}, N. 1995, \aap Suppl., 114

\bibitem[{{Dutra} {et~al.}(2003){Dutra}, {Bica}, {Soares}, \&
  {Barbuy}}]{Dutra2003}
{Dutra}, C.~M., {Bica}, E., {Soares}, J., \& {Barbuy}, B. 2003, \aap, 400, 533

\bibitem[{{Epchtein} {et~al.}(1997){Epchtein}, {de Batz}, {Capoani},
  {Chevallier}, {Copet}, {Fouqu{\'e}}, {Lacombe}, {Le Bertre}, {Pau}, {Rouan},
  {Ruphy}, {Simon}, {Tiph{\`e}ne}, {Burton}, {Bertin}, {Deul}, {Habing},
  {Borsenberger}, {Dennefeld}, {Guglielmo}, {Loup}, {Mamon}, {Ng}, {Omont},
  {Provost}, {Renault}, {Tanguy}, {Kimeswenger}, {Kienel}, {Garzon}, {Persi},
  {Ferrari-Toniolo}, {Robin}, {Paturel}, {Vauglin}, {Forveille}, {Delfosse},
  {Hron}, {Schultheis}, {Appenzeller}, {Wagner}, {Balazs}, {Holl},
  {L{\'e}pine}, {Boscolo}, {Picazzio}, {Duc}, \& {Mennessier}}]{DENIS1997}
{Epchtein}, N., {de Batz}, B., {Capoani}, L., {et~al.} 1997, The Messenger, 87,
  27

\bibitem[{{Fabricius}(1993)}]{CMC1993}
{Fabricius}, C. 1993, Bulletin d'Information du Centre de Donnees Stellaires,
  42, 5

\bibitem[{{Famaey} {et~al.}(2005){Famaey}, {Jorissen}, {Luri}, {Mayor}, {Udry},
  {Dejonghe}, \& {Turon}}]{Famaey2005}
{Famaey}, B., {Jorissen}, A., {Luri}, X., {et~al.} 2005, \aap, 430, 165

\bibitem[{{Froebrich} {et~al.}(2007){Froebrich}, {Scholz}, \&
  {Raftery}}]{Froebrich2007}
{Froebrich}, D., {Scholz}, A., \& {Raftery}, C.~L. 2007, \mnras, 374, 399

\bibitem[{{Geller}\noopsort{a} {et~al.}(2008){Geller}\noopsort{a}, {Mathieu},
  {Harris}, \& {McClure}}]{Geller2008}
{Geller}\noopsort{a}, A.~M., {Mathieu}, R.~D., {Harris}, H.~C., \& {McClure},
  R.~D. 2008, \aj, 135, 2264

\bibitem[{{Geller}\noopsort{b} {et~al.}(2010){Geller}\noopsort{b}, {Mathieu},
  {Braden}, {Meibom}, {Platais}, \& {Dolan}}]{Geller2010}
{Geller}\noopsort{b}, A.~M., {Mathieu}, R.~D., {Braden}, E.~K., {et~al.} 2010,
  \aj, 139, 1383

\bibitem[{{Gieles} {et~al.}(2010){Gieles}, {Sana}, \& {Portegies
  Zwart}}]{Gieles2010}
{Gieles}, M., {Sana}, H., \& {Portegies Zwart}, S.~F. 2010, \mnras, 402, 1750

\bibitem[{{Gontcharov}(2006)}]{Gontcharov2006}
{Gontcharov}, G.~A. 2006, Astronomy Letters, 32, 759

\bibitem[{{Gratton}(2000)}]{Gratton2000}
{Gratton}, R. 2000, in Astronomical Society of the Pacific Conference Series,
  Vol. 198, Stellar Clusters and Associations: Convection, Rotation, and
  Dynamos, ed. R.~{Pallavicini}, G.~{Micela}, \& S.~{Sciortino}, 225

\bibitem[{{H{\o}g} {et~al.}(1997){H{\o}g}, {B{\"a}ssgen}, {Bastian}, {Egret},
  {Fabricius}, {Gro{\ss}mann}, {Halbwachs}, {Makarov}, {Perryman},
  {Schwekendiek}, {Wagner}, \& {Wicenec}}]{TYCHO1997}
{H{\o}g}, E., {B{\"a}ssgen}, G., {Bastian}, U., {et~al.} 1997, \aap, 323, L57

\bibitem[{{H{\o}g} {et~al.}(2000){H{\o}g}, {Fabricius}, {Makarov}, {Urban},
  {Corbin}, {Wycoff}, {Bastian}, {Schwekendiek}, \& {Wicenec}}]{TYCHO2000}
{H{\o}g}, E., {Fabricius}, C., {Makarov}, V.~V., {et~al.} 2000, \aap, 355, L27

\bibitem[{{Kazarovets} {et~al.}(1998){Kazarovets}, {Samus}, \&
  {Durlevich}}]{NSV1998}
{Kazarovets}, E.~V., {Samus}, N.~N., \& {Durlevich}, O.~V. 1998, Information
  Bulletin on Variable Stars, 4655, 1

\bibitem[{{Kharchenko\noopsort{a}}(2001)}]{Kharchenko2001}
{Kharchenko\noopsort{a}}, N.~V. 2001, Kinematika i Fizika Nebesnykh Tel, 17,
  409

\bibitem[{{Kharchenko\noopsort{b}} {et~al.}(2004a){Kharchenko\noopsort{b}},
  {Piskunov}, \& {Scholz}}]{Kharchenko2004a}
{Kharchenko\noopsort{b}}, N.~V., {Piskunov}, A.~E., \& {Scholz}, R.-D. 2004a,
  Astronomische Nachrichten, 325, 439

\bibitem[{{Kharchenko\noopsort{c}} {et~al.}(2004b){Kharchenko\noopsort{c}},
  {Piskunov}, {R{\"o}ser}, {Schilbach}, \& {Scholz}}]{Kharchenko2004b}
{Kharchenko\noopsort{c}}, N.~V., {Piskunov}, A.~E., {R{\"o}ser}, S.,
  {Schilbach}, E., \& {Scholz}, R.-D. 2004b, Astronomische Nachrichten, 325,
  740

\bibitem[{{Kharchenko\noopsort{d}} {et~al.}(2005a){Kharchenko\noopsort{d}},
  {Piskunov}, {R{\"o}ser}, {Schilbach}, \& {Scholz}}]{Kharchenko2005a}
{Kharchenko\noopsort{d}}, N.~V., {Piskunov}, A.~E., {R{\"o}ser}, S.,
  {Schilbach}, E., \& {Scholz}, R.-D. 2005a, \aap, 438, 1163

\bibitem[{{Kharchenko\noopsort{e}} {et~al.}(2005b){Kharchenko\noopsort{e}},
  {Piskunov}, {R{\"o}ser}, {Schilbach}, \& {Scholz}}]{Kharchenko2005b}
{Kharchenko\noopsort{e}}, N.~V., {Piskunov}, A.~E., {R{\"o}ser}, S.,
  {Schilbach}, E., \& {Scholz}, R.-D. 2005b, \aap, 440, 403

\bibitem[{{Kharchenko\noopsort{f}} {et~al.}(2007){Kharchenko\noopsort{f}},
  {Scholz}, {Piskunov}, {R{\"o}ser}, \& {Schilbach}}]{Kharchenko2007}
{Kharchenko\noopsort{f}}, N.~V., {Scholz}, R.-D., {Piskunov}, A.~E.,
  {R{\"o}ser}, S., \& {Schilbach}, E. 2007, Astronomische Nachrichten, 328, 889

\bibitem[{{Kharchenko\noopsort{g}} {et~al.}(2012){Kharchenko\noopsort{g}},
  {Piskunov}, {Schilbach}, {R{\"o}ser}, \& {Scholz}}]{Kharchenko2012}
{Kharchenko\noopsort{g}}, N.~V., {Piskunov}, A.~E., {Schilbach}, E.,
  {R{\"o}ser}, S., \& {Scholz}, R.-D. 2012, \aap, 543, A156

\bibitem[{{Kordopatis}\noopsort{a} {et~al.}(2011){Kordopatis}\noopsort{a},
  {Recio-Blanco}, {de Laverny}, {Bijaoui}, {Hill}, {Gilmore}, {Wyse}, \&
  {Ordenovic}}]{Kordopatis2011}
{Kordopatis}\noopsort{a}, G., {Recio-Blanco}, A., {de Laverny}, P., {et~al.}
  2011, \aap, 535, A106

\bibitem[{{Kordopatis}\noopsort{b} \& {the RAVE
  Collaboration}(2013)}]{Kordopatis2013}
{Kordopatis}\noopsort{b}, G. \& {the RAVE Collaboration}. 2013, in prep.

\bibitem[{{Kouwenhoven} \& {de Grijs}(2008)}]{Kouwenhoven2008}
{Kouwenhoven}, M.~B.~N. \& {de Grijs}, R. 2008, \aap, 480, 103

\bibitem[{{Lada\noopsort{a}} \& {Lada}(2003)}]{Lada2003}
{Lada\noopsort{a}}, C.~J. \& {Lada}, E.~A. 2003, \araa, 41, 57

\bibitem[{{Lada\noopsort{b}}(2006)}]{Lada2006}
{Lada\noopsort{b}}, C.~J. 2006, \apjl, 640, L63

\bibitem[{{Lyng\aa}(1987)}]{Lynga1987}
{Lyng\aa}, G. 1987, Catalogue of open cluster data, Fifth edition

\bibitem[{{Margheim} {et~al.}(2000){Margheim}, {King}, {Deliyannis}, \&
  {Platais}}]{Margheim2000}
{Margheim}, S.~J., {King}, J.~R., {Deliyannis}, C.~P., \& {Platais}, I. 2000,
  in Bulletin of the American Astronomical Society, Vol.~32, American
  Astronomical Society Meeting Abstracts \#196, 742

\bibitem[{{Matijevi{\v c}}\noopsort{a} {et~al.}(2010){Matijevi{\v
  c}}\noopsort{a}, {Zwitter}, {Munari}, {Bienaym{\'e}}, {Binney},
  {Bland-Hawthorn}, {Boeche}, {Campbell}, {Freeman}, {Gibson}, {Gilmore},
  {Grebel}, {Helmi}, {Navarro}, {Parker}, {Seabroke}, {Siebert}, {Siviero},
  {Steinmetz}, {Watson}, {Williams}, \& {Wyse}}]{Matijevic2010}
{Matijevi{\v c}}\noopsort{a}, G., {Zwitter}, T., {Munari}, U., {et~al.} 2010,
  \aj, 140, 184

\bibitem[{{Matijevi{\v c}}\noopsort{b} {et~al.}(2011){Matijevi{\v
  c}}\noopsort{b}, {Zwitter}, {Bienaym{\'e}}, {Bland-Hawthorn}, {Freeman},
  {Gilmore}, {Grebel}, {Helmi}, {Munari}, {Navarro}, {Parker}, {Reid},
  {Seabroke}, {Siebert}, {Siviero}, {Steinmetz}, {Watson}, {Williams}, \&
  {Wyse}}]{Matijevic2011}
{Matijevi{\v c}}\noopsort{b}, G., {Zwitter}, T., {Bienaym{\'e}}, O., {et~al.}
  2011, \aj, 141, 200

\bibitem[{{Matijevi{\v c}}\noopsort{c} {et~al.}(2012){Matijevi{\v
  c}}\noopsort{c}, {Zwitter}, {Bienaym{\'e}}, {Bland-Hawthorn}, {Boeche},
  {Freeman}, {Gibson}, {Gilmore}, {Grebel}, {Helmi}, {Munari}, {Navarro},
  {Parker}, {Reid}, {Seabroke}, {Siebert}, {Siviero}, {Steinmetz}, {Watson},
  {Williams}, \& {Wyse}}]{Matijevic2012}
{Matijevi{\v c}}\noopsort{c}, G., {Zwitter}, T., {Bienaym{\'e}}, O., {et~al.}
  2012, \apjs, 200, 14

\bibitem[{{Melnik} \& {Efremov}(1995)}]{Melnik1995}
{Melnik}, A.~M. \& {Efremov}, Y.~N. 1995, Astronomy Letters, 21, 10

\bibitem[{{Mermilliod}(1988)}]{WEBDA1988}
{Mermilliod}, J.~C. 1988, Bulletin d'Information du Centre de Donnees
  Stellaires, 35, 77

\bibitem[{{Munari} {et~al.}(2005){Munari}, {Sordo}, {Castelli}, \&
  {Zwitter}}]{Munari2005}
{Munari}, U., {Sordo}, R., {Castelli}, F., \& {Zwitter}, T. 2005, \aap, 442,
  1127

\bibitem[{{Netopil} {et~al.}(2012){Netopil}, {Paunzen}, \&
  {St{\"u}tz}}]{WEBDA2012}
{Netopil}, M., {Paunzen}, E., \& {St{\"u}tz}, C. 2012, {in: Developments of the
  Open Cluster Database WEBDA}, ed. A.~{Moitinho} \& J.~{Alves}, 53

\bibitem[{{Nissen}(1988)}]{Nissen1988}
{Nissen}, P.~E. 1988, \aap, 199, 146

\bibitem[{{Nordstr{\"o}m} {et~al.}(2004){Nordstr{\"o}m}, {Mayor}, {Andersen},
  {Holmberg}, {Pont}, {Jorgensen}, {Olsen}, {Udry}, \&
  {Mowlavi}}]{Nordstrom2004}
{Nordstr{\"o}m}, B., {Mayor}, M., {Andersen}, J., {et~al.} 2004, \aap, 419

\bibitem[{{Perryman} {et~al.}(1997){Perryman}, {Lindegren}, {Kovalevsky},
  {Hoeg}, {Bastian}, {Bernacca}, {Cr{\'e}z{\'e}}, {Donati}, {Grenon},
  {Grewing}, {van Leeuwen}, {van der Marel}, {Mignard}, {Murray}, {Le Poole},
  {Schrijver}, {Turon}, {Arenou}, {Froeschl{\'e}}, \& {Petersen}}]{HIP1997}
{Perryman}, M.~A.~C., {Lindegren}, L., {Kovalevsky}, J., {et~al.} 1997, \aap,
  323, L49

\bibitem[{{Piatti} {et~al.}(1995){Piatti}, {Claria}, \& {Abadi}}]{Piatti1995}
{Piatti}, A.~E., {Claria}, J.~J., \& {Abadi}, M.~G. 1995, \aj, 110, 2813

\bibitem[{{Piskunov} {et~al.}(2006){Piskunov}, {Kharchenko}, {R{\"o}ser},
  {Schilbach}, \& {Scholz}}]{Piskunov2006}
{Piskunov}, A.~E., {Kharchenko}, N.~V., {R{\"o}ser}, S., {Schilbach}, E., \&
  {Scholz}, R.-D. 2006, \aap, 445, 545

\bibitem[{{Platais} {et~al.}(1998){Platais}, {Kozhurina-Platais}, \& {van
  Leeuwen}}]{Platais1998}
{Platais}, I., {Kozhurina-Platais}, V., \& {van Leeuwen}, F. 1998, \aj, 116,
  2423

\bibitem[{{P{\"o}hnl} \& {Paunzen}(2010)}]{Pohnl2010}
{P{\"o}hnl}, H. \& {Paunzen}, E. 2010, \aap, 514, A81

\bibitem[{{R{\"o}ser} \& {Bastian}(1991)}]{PPM1991}
{R{\"o}ser}, S. \& {Bastian}, U. 1991, {PPM Star Catalogue. Positions and
  proper motions of 181731 stars north of -2.5 degrees declination for equinox
  and epoch J2000.0. Vol. I: Zones +80{$^{\circ}$} to +30{$^{\circ}$}. Vol. II:
  Zones +20{$^{\circ}$} to -0{$^{\circ}$}.}

\bibitem[{{Ruprecht} {et~al.}(1981){Ruprecht}, {Balazs}, \&
  {White}}]{Ruprecht1981}
{Ruprecht}, J., {Balazs}, B.~A., \& {White}, R.~E. 1981, {"Catalogue of star
  clusters and associations"}

\bibitem[{{Samus} {et~al.}(1997){Samus}, {Durlevich}, \&
  {Kazarovets}}]{GCVS1997}
{Samus}, N.~N., {Durlevich}, O.~V., \& {Kazarovets}, R.~V. 1997, Baltic
  Astronomy, 6, 296

\bibitem[{{Siebert} {et~al.}(2011){Siebert}, {Williams}, {Siviero}, {Reid},
  {Boeche}, {Steinmetz}, {Fulbright}, {Munari}, {Zwitter}, {Watson}, {Wyse},
  {de Jong}, {Enke}, {Anguiano}, {Burton}, {Cass}, {Fiegert}, {Hartley},
  {Ritter}, {Russel}, {Stupar}, {Bienaym{\'e}}, {Freeman}, {Gilmore}, {Grebel},
  {Helmi}, {Navarro}, {Binney}, {Bland-Hawthorn}, {Campbell}, {Famaey},
  {Gerhard}, {Gibson}, {Matijevi{\v c}}, {Parker}, {Seabroke}, {Sharma},
  {Smith}, \& {Wylie-de Boer}}]{RAVE3}
{Siebert}, A., {Williams}, M.~E.~K., {Siviero}, A., {et~al.} 2011, \aj, 141,
  187

\bibitem[{{Steinmetz} {et~al.}(2006){Steinmetz}, {Zwitter}, {Siebert},
  {Watson}, {Freeman}, {Munari}, {Campbell}, {Williams}, {Seabroke}, {Wyse},
  {Parker}, {Bienaym{\'e}}, {R{\"o}ser}, {Gibson}, {Gilmore}, {Grebel},
  {Helmi}, {Navarro}, {Burton}, {Cass}, {Dawe}, {Fiegert}, {Hartley},
  {Russell}, {Saunders}, {Enke}, {Bailin}, {Binney}, {Bland-Hawthorn},
  {Boeche}, {Dehnen}, {Eisenstein}, {Evans}, {Fiorucci}, {Fulbright},
  {Gerhard}, {Jauregi}, {Kelz}, {Mijovi{\'c}}, {Minchev}, {Parmentier},
  {Pe{\~n}arrubia}, {Quillen}, {Read}, {Ruchti}, {Scholz}, {Siviero}, {Smith},
  {Sordo}, {Veltz}, {Vidrih}, {von Berlepsch}, {Boyle}, \& {Schilbach}}]{RAVE1}
{Steinmetz}, M., {Zwitter}, T., {Siebert}, A., {et~al.} 2006, \aj, 132, 1645

\bibitem[{{Twarog} {et~al.}(1997){Twarog}, {Ashman}, \&
  {Anthony-Twarog}}]{Twarog1997}
{Twarog}, B.~A., {Ashman}, K.~M., \& {Anthony-Twarog}, B.~J. 1997, \aj, 114,
  2556

\bibitem[{{Zwitter} {et~al.}(2008){Zwitter}, {Siebert}, {Munari}, {Freeman},
  {Siviero}, {Watson}, {Fulbright}, {Wyse}, {Campbell}, {Seabroke}, {Williams},
  {Steinmetz}, {Bienaym{\'e}}, {Gilmore}, {Grebel}, {Helmi}, {Navarro},
  {Anguiano}, {Boeche}, {Burton}, {Cass}, {Dawe}, {Fiegert}, {Hartley},
  {Russell}, {Veltz}, {Bailin}, {Binney}, {Bland-Hawthorn}, {Brown}, {Dehnen},
  {Evans}, {Re Fiorentin}, {Fiorucci}, {Gerhard}, {Gibson}, {Kelz}, {Kujken},
  {Matijevi{\v c}}, {Minchev}, {Parker}, {Pe{\~n}arrubia}, {Quillen}, {Read},
  {Reid}, {R{\"o}ser}, {Ruchti}, {Scholz}, {Smith}, {Sordo}, {Tolstoi},
  {Tomasella}, {Vidrih}, \& {Wylie-de Boer}}]{RAVE2}
{Zwitter}, T., {Siebert}, A., {Munari}, U., {et~al.} 2008, \aj, 136, 421

\end{thebibliography}

\clearpage

\onecolumn
\begin{landscape}
\begin{center}
\begin{longtable}{r l | r r r r r | r r r r r | r r r}
\multicolumn{15}{l}{\textbf{Tab. \ref{tab1}.} Results from our radial velocity study on open clusters in RAVE, along with reference values from CRVAD-2 and CRVOCA (see Sect. \ref{clu_rv}).}
\vspace*{0.3cm}
\label{tab1}
\\
\hline
    &      &             \multicolumn{5}{|c|}{RAVE}            &            \multicolumn{5}{|c|}{CRVAD-2}         &         \multicolumn{3}{|c}{CRVOCA}         \\
\hline
Seq & Name & $\overline{RV}$  & e$\overline{RV}$ & $\sigma\overline{RV}$ & $\overline{eRV^*}$ &      No. of & $\overline{RV}$ & e$\overline{RV}$ & $\sigma\overline{RV}$ & $\overline{eRV^*}$ &     
     No. of & $\overline{RV}$ & e$\overline{RV}$ &   No. of \\
    &      &             km/s &             km/s &                  km/s &               km/s & entries$^a$ &            km/s &             km/s &                  km/s &               km/s &
entries$^a$ &            km/s &             km/s & stars$^b$ \\
\hline\hline
\endfirsthead

\multicolumn{11}{l}{\textbf{\tablename\ \thetable{} continued}}
\vspace*{0.3cm}
\\ 
\hline
    &      &             \multicolumn{5}{|c|}{RAVE}            &            \multicolumn{5}{|c|}{CRVAD-2}         &         \multicolumn{3}{|c}{CRVOCA}         \\
\hline
Seq & Name & $\overline{RV}$  & e$\overline{RV}$ & $\sigma\overline{RV}$ & $\overline{eRV^*}$ &      No. of & $\overline{RV}$ & e$\overline{RV}$ & $\sigma\overline{RV}$ & $\overline{eRV^*}$ &     
     No. of & $\overline{RV}$ & e$\overline{RV}$ &   No. of \\
    &      &             km/s &             km/s &                  km/s &               km/s & entries$^a$ &            km/s &             km/s &                  km/s &               km/s &
entries$^a$ &            km/s &             km/s & stars$^b$ \\
\hline\hline
\endhead

\hline
\endfoot

\hline
\multicolumn{15}{l}{\footnotesize  $^a$The numbers in brackets are the numbers of additional good-member measurements used to compute the $\overline{RV}$.}\\
\multicolumn{15}{l}{\footnotesize  $^b$ ``$-1$'' indicates clusters where only one star was considered as representative with $P_{kin}>1\%$.}\\
\endlastfoot

    3 &  Blanco 1       &    6.168 &    6.737 &   17.825 &   1.229 &  7 (\ \ --) &     2.843 &    1.761 &    4.981 &    4.807 &  8 (\ \ --) &   3.580 &    2.360 &  13 \\
   44 &  Alessi 13      &    1.053 &    0.458 &    0.647 &   2.967 &  2 (\ \ --) &    15.916 &    1.353 &    2.344 &    0.000 &  1  (\ \ 2) &  19.530 &    3.000 &   3 \\
   47 &  Melotte 22     &    3.503 &    0.391 &    1.955 &   2.399 & 25 (\ \ --) &     5.766 &    0.241 &    1.616 &    2.543 & 45 (\ \ --) &   5.900 &    0.450 & 106 \\
   65 &  NGC 1901       &   -1.354 &    0.478 &    0.676 &   2.962 &  2 (\ \ --) &       --- &      --- &      --- &      --- & -- (\ \ --) &  -9.630 &    9.630 &   3 \\
   77 &  Collinder 70   &   54.817 &    1.000 &      --- &   1.000 &  1 (\ \ --) &    24.038 &    1.917 &    5.750 &    3.719 &  9 (\ \ --) &  19.870 &    2.190 &  23 \\
  127 &  NGC 2354       &   42.320 &    8.117 &   14.060 &   1.000 &  3 (\ \ --) &       --- &      --- &      --- &      --- & -- (\ \ --) &  33.400 &    0.270 &   6 \\
  129 &  Alessi 3       &   -2.143 &    0.531 &    1.302 &   1.658 &  1  (\ \ 5) &       --- &      --- &      --- &      --- & -- (\ \ --) &  20.000 &    7.400 &  -1 \\
  133 &  Collinder 135  &   26.957 &   10.140 &   22.673 &   1.136 &  5 (\ \ --) &    16.173 &    0.782 &    1.564 &    5.558 &  4 (\ \ --) &  15.350 &    2.200 &   4 \\
  142 &  Bochum 5       &   27.881 &   11.293 &   25.251 &   3.464 &  5 (\ \ --) &       --- &      --- &      --- &      --- & -- (\ \ --) &  41.000 &    2.000 &   1 \\
  147 &  NGC 2422       &   34.212 &    0.872 &    3.145 &   4.527 & 13 (\ \ --) &    34.408 &    1.862 &    5.587 &    6.908 &  9 (\ \ --) &  36.720 &    2.920 &  13 \\
  148 &  NGC 2423       &   21.069 &    2.041 &   12.244 &   3.211 & 36 (\ \ --) &    20.913 &    3.190 &    7.813 &    1.000 &  6 (\ \ --) &  21.670 &    2.540 &   9 \\
  149 &  Ruprecht 26    &   39.079 &    7.765 &   17.362 &   9.839 &  5 (\ \ --) &    15.000 &    3.700 &      --- &    3.700 &  1 (\ \ --) &  15.000 &    3.700 &   1 \\
  150 &  Melotte 71     &    0.545 &    1.000 &      --- &   1.000 &  1 (\ \ --) &       --- &      --- &      --- &      --- & -- (\ \ --) &  50.140 &    0.140 &  11 \\
  152 &  NGC 2428       &   46.665 &    2.727 &    5.454 &   6.484 &  4 (\ \ --) &       --- &      --- &      --- &      --- & -- (\ \ --) &     --- &      --- &  -- \\
  153 &  NGC 2430       &   31.337 &   11.831 &   16.732 &   2.509 &  2 (\ \ --) &       --- &      --- &      --- &      --- & -- (\ \ --) &     --- &      --- &  -- \\
  158 &  Ruprecht 151   &   31.294 &    4.841 &   16.056 &   3.826 & 11 (\ \ --) &       --- &      --- &      --- &      --- & -- (\ \ --) &     --- &      --- &  -- \\
  159 &  NGC 2451A      &   56.185 &   30.712 &   53.195 &   1.309 & --  (\ \ 3) &    25.721 &    5.332 &   10.664 &    2.062 &  4 (\ \ --) &  22.630 &    4.490 &   9 \\
  160 &  NGC 2437       &   30.432 &    4.048 &   16.194 &  10.496 & 16 (\ \ --) &       --- &      --- &      --- &      --- & -- (\ \ --) &  48.090 &    0.000 &   1 \\
  163 &  NGC 2447       &   31.959 &    6.222 &   10.777 &   5.400 &  3 (\ \ --) &    20.925 &    1.088 &    2.665 &    1.036 &  6 (\ \ --) &  21.310 &    0.670 &  11 \\
  164 &  NGC 2448       &   27.195 &    1.245 &    2.157 &   4.941 & --  (\ \ 3) &    15.000 &    3.700 &      --- &    3.700 &  1 (\ \ --) &  15.000 &    3.700 &   1 \\
  166 &  Haffner 16     &   37.246 &    3.031 &      --- &   3.031 &  1 (\ \ --) &    65.800 &    3.700 &      --- &    3.700 &  1 (\ \ --) &  65.800 &    3.700 &   1 \\
  167 &  NGC 2477       &    6.374 &    0.184 &    0.261 &   1.188 &  2 (\ \ --) &       --- &      --- &      --- &      --- & -- (\ \ --) &   7.320 &    0.130 &  49 \\
  171 &  NGC 2482       &   42.155 &    0.910 &    1.577 &   4.626 &  3 (\ \ --) &       --- &      --- &      --- &      --- & -- (\ \ --) &     --- &      --- &  -- \\
  175 &  NGC 2516       &   -4.492 &    4.994 &    8.649 &   1.071 & --  (\ \ 3) &    22.937 &    0.330 &    1.043 &    3.334 & 10 (\ \ --) &  20.670 &    1.450 &  23 \\
  180 &  NGC 2527       &   42.425 &    1.268 &    3.803 &   4.846 &  9 (\ \ --) &       --- &      --- &      --- &      --- & -- (\ \ --) & 135.100 &    3.000 &  -1 \\
  182 &  Vel OB2        &   20.840 &    3.598 &   16.878 &   1.193 & --     (22) &    15.163 &    5.370 &    7.594 &    5.652 &  2 (\ \ --) &  15.450 &    3.200 &   6 \\
  191 &  Haffner 26     &   62.377 &   12.076 &      --- &  12.076 &  1 (\ \ --) &       --- &      --- &      --- &      --- & -- (\ \ --) &     --- &      --- &  -- \\
  192 &  NGC 2567       &   37.020 &    1.000 &      --- &   1.000 &  1 (\ \ --) &       --- &      --- &      --- &      --- & -- (\ \ --) &  35.790 &    0.090 &   1 \\
  193 &  NGC 2571       &   44.021 &   10.280 &      --- &  10.280 &  1 (\ \ --) &       --- &      --- &      --- &      --- & -- (\ \ --) &     --- &      --- &  -- \\
  201 &  NGC 2632       &   34.001 &    0.285 &    1.065 &   1.578 & 14 (\ \ --) &    34.117 &    0.425 &    2.442 &    2.154 & 33 (\ \ --) &  33.660 &    1.180 &  73 \\
  202 &  IC 2391        &   12.487 &    3.533 &      --- &   3.533 &  1 (\ \ --) &    15.318 &    1.336 &    4.225 &    4.323 & 10 (\ \ --) &  16.040 &    2.530 &  18 \\
  212 &  NGC 2682       &   33.806 &    0.301 &    0.852 &   1.004 &  8 (\ \ --) &    33.555 &    0.307 &    1.190 &    1.000 & 15 (\ \ --) &  32.300 &    1.100 &  33 \\
  216 &  Platais 8      &    7.349 &    8.046 &   26.687 &   1.276 & 11 (\ \ --) &    21.495 &    4.935 &    6.980 &    7.396 &  2 (\ \ --) &  17.320 &    3.090 &   5 \\
  223 &  Turner 5       &   -1.195 &   14.952 &   21.145 &   1.391 &  2 (\ \ --) &       --- &      --- &      --- &      --- & -- (\ \ --) &  26.200 &    3.700 &  -1 \\
  226 &  Ruprecht 80    &   45.749 &    1.000 &      --- &   1.000 &  1 (\ \ --) &       --- &      --- &      --- &      --- & -- (\ \ --) &     --- &      --- &  -- \\
  230 &  NGC 3036       &   11.777 &    1.000 &      --- &   1.000 &  1 (\ \ --) &       --- &      --- &      --- &      --- & -- (\ \ --) &     --- &      --- &  -- \\
  243 &  IC 2581        &    2.165 &   14.146 &      --- &  14.146 &  1 (\ \ --) &    -0.658 &    0.510 &    0.721 &    2.010 &  2 (\ \ --) &  -4.640 &    3.510 &   5 \\
  248 &  Collinder 223  &    8.553 &    0.087 &    0.124 &   1.031 &  2 (\ \ --) &       --- &      --- &      --- &      --- & -- (\ \ --) &   4.700 &    0.000 &   1 \\
  249 &  Ruprecht 90    &   18.079 &    7.269 &   10.280 &  12.415 &  2 (\ \ --) &   -40.000 &    4.400 &      --- &    4.400 &  1 (\ \ --) & -40.000 &    4.400 &   1 \\
  251 &  Loden 153      &   13.698 &    5.247 &    7.420 &   7.390 &  2 (\ \ --) &   -11.824 &    2.502 &    4.334 &    0.000 &  1  (\ \ 2) & -10.330 &    2.400 &   3 \\
  254 &  NGC 3324       &   -4.042 &    2.864 &    4.051 &   9.023 &  2 (\ \ --) &    -8.500 &    3.700 &      --- &    3.700 &  1 (\ \ --) &  -8.500 &    4.200 &   1 \\
  255 &  vdBergh 99     &   12.790 &    3.684 &    9.025 &   5.518 &  6 (\ \ --) &    11.340 &    0.751 &    1.062 &    1.450 &  2 (\ \ --) &   6.000 &    3.490 &   4 \\
  258 &  Melotte 101    &   20.265 &   10.383 &      --- &  10.383 &  1 (\ \ --) &       --- &      --- &      --- &      --- & -- (\ \ --) &     --- &      --- &  -- \\
  259 &  IC 2602        &    5.042 &    2.365 &   19.066 &   2.077 & 65 (\ \ --) &    27.163 &    2.127 &    4.255 &    3.104 &  4 (\ \ --) &  19.020 &    2.930 &  12 \\
  260 &  Bochum 10      &   -1.485 &    4.592 &      --- &   4.592 &  1 (\ \ --) &    -2.300 &    1.800 &      --- &    1.800 &  1 (\ \ --) &  -2.300 &    1.800 &   1 \\
  261 &  Alessi 5       &    6.339 &    3.183 &    4.501 &   3.832 &  2 (\ \ --) &    10.867 &    8.585 &   12.141 &    7.400 &  2 (\ \ --) &  12.800 &    8.800 &   2 \\
  262 &  Collinder 228  &   24.240 &   13.119 &      --- &  13.119 &  1 (\ \ --) &    -1.161 &    5.712 &   18.062 &    4.746 & 10 (\ \ --) & -13.460 &    3.270 &  24 \\
  266 &  Ruprecht 91    &   -4.115 &   11.325 &      --- &  11.325 &  1 (\ \ --) &       --- &      --- &      --- &      --- & -- (\ \ --) &   7.300 &    0.000 &   1 \\
  267 &  Loden 189      &   10.493 &    4.499 &   10.060 &   6.300 &  5 (\ \ --) &       --- &      --- &      --- &      --- & -- (\ \ --) &     --- &      --- &  -- \\
  268 &  Ruprecht 92    &  -10.582 &    3.703 &    5.237 &   7.212 &  2 (\ \ --) &       --- &      --- &      --- &      --- & -- (\ \ --) & -15.600 &    1.800 &  -1 \\
  270 &  Trumpler 17    &   25.506 &    1.719 &    2.431 &  10.877 &  2 (\ \ --) &       --- &      --- &      --- &      --- & -- (\ \ --) &     --- &      --- &  -- \\
  271 &  Collinder 236  &   14.140 &   11.531 &   25.785 &  10.011 &  5 (\ \ --) &       --- &      --- &      --- &      --- & -- (\ \ --) &     --- &      --- &  -- \\
  273 &  NGC 3496       &    2.321 &    1.962 &    3.398 &   5.909 &  1  (\ \ 2) &       --- &      --- &      --- &      --- & -- (\ \ --) & -51.400 &    0.000 &   1 \\
  274 &  Pismis 17      &  -47.146 &    0.517 &    0.731 &   1.086 &  2 (\ \ --) &       --- &      --- &      --- &      --- & -- (\ \ --) &  61.700 &    0.000 &   1 \\
  275 &  Ruprecht 93    &    5.072 &    3.502 &      --- &   3.502 &  1 (\ \ --) &       --- &      --- &      --- &      --- & -- (\ \ --) &     --- &      --- &  -- \\
  276 &  NGC 3532       &    4.311 &    0.572 &    2.061 &   4.448 & 13 (\ \ --) &    -2.252 &    6.355 &   11.008 &    3.630 &  3 (\ \ --) &  -3.040 &    3.810 &   5 \\
  283 &  Trumpler 18    &  -10.016 &   12.474 &      --- &  12.474 &  1 (\ \ --) &   -20.000 &    7.400 &      --- &    7.400 &  1 (\ \ --) & -20.000 &    7.400 &   1 \\
  287 &  NGC 3680       &    2.154 &    1.000 &      --- &   1.000 &  1 (\ \ --) &       --- &      --- &      --- &      --- & -- (\ \ --) &   8.000 &    3.670 &   7 \\
  298 &  Loden 481      &   16.915 &    8.672 &      --- &   8.672 &  1 (\ \ --) &    -7.100 &    1.200 &      --- &    1.200 &  1 (\ \ --) & -22.050 &   14.950 &   2 \\
  304 &  ESO 130-06     &   -1.227 &    3.424 &    4.842 &   5.822 &  2 (\ \ --) &       --- &      --- &      --- &      --- & -- (\ \ --) &     --- &      --- &  -- \\
  305 &  Loden 565      &  -17.872 &    9.328 &      --- &   9.328 &  1 (\ \ --) &    10.100 &    1.000 &      --- &    1.000 &  1 (\ \ --) &  10.100 &    0.200 &   1 \\
  306 &  ESO 130-08     &  -38.522 &    6.346 &    8.974 &   8.057 &  2 (\ \ --) &       --- &      --- &      --- &      --- & -- (\ \ --) &     --- &      --- &  -- \\
  309 &  NGC 4349       &  -11.929 &    3.239 &    5.611 &   4.478 &  3 (\ \ --) &   -15.700 &    1.800 &      --- &    1.800 &  1 (\ \ --) & -15.690 &    1.780 &   1 \\
  310 &  Collinder 258  &  -14.349 &    9.186 &   12.992 &  12.496 &  2 (\ \ --) &       --- &      --- &      --- &      --- & -- (\ \ --) &     --- &      --- &  -- \\
  322 &  Stock 16       &   -6.875 &    6.780 &    9.589 &  10.336 &  2 (\ \ --) &   -52.787 &   12.634 &   21.883 &   37.024 &  3 (\ \ --) & -41.800 &    8.790 &   5 \\
  324 &  Loden 821      &   -3.718 &    5.705 &   11.410 &   6.896 &  4 (\ \ --) &   -39.000 &    7.400 &      --- &    7.400 &  1 (\ \ --) & -39.000 &    7.400 &   1 \\
  327 &  Basel 18       &  -21.989 &    3.021 &      --- &   3.021 &  1 (\ \ --) &       --- &      --- &      --- &      --- & -- (\ \ --) &     --- &      --- &  -- \\
  328 &  Hogg 16        &  -51.341 &    2.159 &    3.053 &  12.798 &  2 (\ \ --) &   -29.793 &    6.047 &   12.093 &    7.460 &  4 (\ \ --) & -33.000 &    6.160 &   5 \\
  333 &  Loden 915      &  -21.605 &   11.760 &      --- &  11.760 &  1 (\ \ --) &       --- &      --- &      --- &      --- & -- (\ \ --) &     --- &      --- &  -- \\
  335 &  Loden 1010     &  -18.238 &    3.313 &    6.627 &  10.544 &  4 (\ \ --) &       --- &      --- &      --- &      --- & -- (\ \ --) &     --- &      --- &  -- \\
  336 &  NGC 5281       &  -15.912 &   13.073 &      --- &  13.073 &  1 (\ \ --) &       --- &      --- &      --- &      --- & -- (\ \ --) &     --- &      --- &  -- \\
  337 &  Platais 12     &  -11.048 &    2.228 &   14.777 &   4.597 & 44 (\ \ --) &       --- &      --- &      --- &      --- & -- (\ \ --) & -11.990 &    9.510 &   2 \\
  338 &  NGC 5316       &    9.983 &    5.271 &   12.911 &   8.360 &  6 (\ \ --) &       --- &      --- &      --- &      --- & -- (\ \ --) &     --- &      --- &  -- \\
  339 &  Loden 995      &   -6.064 &    1.224 &      --- &   1.224 &  1 (\ \ --) &       --- &      --- &      --- &      --- & -- (\ \ --) &     --- &      --- &  -- \\
  343 &  Ruprecht 110   &  -36.646 &    3.981 &    5.631 &   1.000 &  2 (\ \ --) &       --- &      --- &      --- &      --- & -- (\ \ --) &     --- &      --- &  -- \\
  356 &  Alessi 6       &   19.950 &    1.000 &      --- &   1.000 &  1 (\ \ --) &       --- &      --- &      --- &      --- & -- (\ \ --) &     --- &      --- &  -- \\
  367 &  Nor OB5        &  -20.248 &    6.805 &   11.787 &   2.571 &  1  (\ \ 2) &   -26.863 &    6.048 &   12.097 &    7.375 &  4 (\ \ --) & -23.140 &    5.110 &   7 \\
  382 &  NGC 6204       &   11.580 &    0.411 &    0.581 &   9.859 &  2 (\ \ --) &       --- &      --- &      --- &      --- & -- (\ \ --) & -51.000 &    5.800 &   5 \\
  387 &  NGC 6242       &   39.324 &   39.549 &   55.931 &  10.268 &  2 (\ \ --) &       --- &      --- &      --- &      --- & -- (\ \ --) &     --- &      --- &  -- \\
  390 &  NGC 6250       &  -16.754 &    0.010 &    0.015 &   9.437 &  2 (\ \ --) &       --- &      --- &      --- &      --- & -- (\ \ --) &     --- &      --- &  -- \\
  393 &  Sco OB4        &  -23.559 &    8.136 &   42.275 &   5.729 & 27 (\ \ --) &     6.853 &    1.843 &    4.514 &    7.378 &  6 (\ \ --) &   5.390 &    4.950 &  17 \\
  396 &  vdBergh 221    &    5.196 &    5.566 &      --- &   5.566 &  1 (\ \ --) &       --- &      --- &      --- &      --- & -- (\ \ --) &   6.330 &    3.840 &   3 \\
  397 &  IC 4651        &  -30.406 &    0.312 &    0.624 &   1.031 &  4 (\ \ --) &       --- &      --- &      --- &      --- & -- (\ \ --) & -31.000 &    0.200 &  14 \\
  399 &  Antalova 1     &   -6.347 &   12.041 &   17.029 &  11.384 &  2 (\ \ --) &    33.000 &   40.600 &      --- &   40.600 &  1 (\ \ --) &  33.000 &   40.600 &   1 \\
  402 &  NGC 6383       &  -16.537 &   10.667 &   15.085 &  10.740 &  2 (\ \ --) &     1.881 &    2.659 &    3.760 &    5.620 &  2 (\ \ --) &   2.670 &    2.170 &   3 \\
  403 &  Trumpler 27    &  -35.281 &    1.241 &    1.755 &   1.000 &  2 (\ \ --) &   -15.800 &    1.300 &      --- &    1.300 &  1 (\ \ --) & -15.800 &    1.300 &   1 \\
  404 &  Trumpler 28    &  -32.768 &   11.912 &   16.846 &   6.726 &  2 (\ \ --) &       --- &      --- &      --- &      --- & -- (\ \ --) &     --- &      --- &  -- \\
  405 &  ESO 139-13     &  -33.871 &    1.000 &      --- &   1.000 &  1 (\ \ --) &       --- &      --- &      --- &      --- & -- (\ \ --) &     --- &      --- &  -- \\
  408 &  NGC 6405       &   -7.049 &    2.746 &    8.683 &   6.633 & 10 (\ \ --) &    -8.857 &    2.295 &    3.245 &    4.139 &  2 (\ \ --) & -11.500 &    3.500 &   2 \\
  410 &  Alessi 9       &  -71.015 &    1.000 &      --- &   1.000 &  1 (\ \ --) &   -11.600 &    1.200 &      --- &    1.200 &  1 (\ \ --) & -11.600 &    1.200 &   1 \\
  411 &  NGC 6416       &  -11.201 &   12.724 &   17.994 &   2.010 &  2 (\ \ --) &       --- &      --- &      --- &      --- & -- (\ \ --) &     --- &      --- &  -- \\
  413 &  NGC 6425       &    5.439 &   12.130 &      --- &  12.130 &  1 (\ \ --) &       --- &      --- &      --- &      --- & -- (\ \ --) &     --- &      --- &  -- \\
  418 &  Sco OB5        &  -10.453 &    6.425 &      --- &   6.425 &  1 (\ \ --) &       --- &      --- &      --- &      --- & -- (\ \ --) &     --- &      --- &  -- \\
  420 &  NGC 6475       &  -21.377 &   12.720 &      --- &  12.720 &  1 (\ \ --) &   -13.665 &    0.400 &    1.600 &    3.782 & 16 (\ \ --) & -14.210 &    1.390 &  31 \\
  429 &  NGC 6546       &  -29.427 &    1.000 &      --- &   1.000 &  1 (\ \ --) &       --- &      --- &      --- &      --- & -- (\ \ --) & -16.670 &   15.770 &   3 \\
  430 &  vdBergh 113    &   16.148 &    1.755 &    2.482 &   7.495 &  2 (\ \ --) &   -16.000 &    3.700 &      --- &    3.700 &  1 (\ \ --) & -16.000 &    3.700 &   1 \\
  435 &  Sgr OB7        &  -44.478 &    9.602 &   30.365 &   8.477 & 10 (\ \ --) &    -3.576 &    2.176 &    4.867 &    4.091 &  5 (\ \ --) &  -7.190 &    3.140 &   9 \\
  436 &  Markarian 38   &  -33.031 &    0.515 &    0.728 &   1.114 &  2 (\ \ --) &     6.000 &    9.100 &      --- &    9.100 &  1 (\ \ --) &  -3.200 &    9.200 &   2 \\
  444 &  NGC 6618       &  -44.577 &    0.682 &    0.965 &   1.242 &  2 (\ \ --) &   -14.662 &   12.788 &   18.085 &    9.355 &  2 (\ \ --) & -17.000 &   13.000 &   2 \\
  445 &  Trumpler 33    &   -8.544 &    7.741 &      --- &   7.741 &  1 (\ \ --) &       --- &      --- &      --- &      --- & -- (\ \ --) &     --- &      --- &  -- \\
  449 &  IC 4725        &   61.206 &   17.938 &   35.876 &   9.477 &  4 (\ \ --) &     2.051 &    7.935 &   13.745 &    7.301 &  3 (\ \ --) &  -1.810 &   10.460 &   7 \\
  452 &  Ruprecht 145   &   -8.357 &    1.000 &      --- &   1.000 &  1 (\ \ --) &       --- &      --- &      --- &      --- & -- (\ \ --) &     --- &      --- &  -- \\
 1033 &  ASCC 33        &   34.522 &   25.521 &   44.203 &   1.181 & --  (\ \ 3) &    14.692 &    8.995 &   12.720 &    7.551 &  2 (\ \ --) &  15.000 &    9.000 &   2 \\
 1057 &  ASCC 57        &    2.463 &    7.312 &   17.911 &   1.191 &  6 (\ \ --) &       --- &      --- &      --- &      --- & -- (\ \ --) &     --- &      --- &  -- \\
 1078 &  ASCC 78        &  -14.558 &    5.508 &   15.578 &   1.093 &  8 (\ \ --) &       --- &      --- &      --- &      --- & -- (\ \ --) &     --- &      --- &  -- \\
 1085 &  ASCC 85        &   11.160 &    8.549 &      --- &   8.549 &  1 (\ \ --) &       --- &      --- &      --- &      --- & -- (\ \ --) &   7.090 &    0.000 &   1 \\
 1089 &  Alessi 24      &  -38.648 &   19.301 &   33.430 &   1.000 &  3 (\ \ --) &       --- &      --- &      --- &      --- & -- (\ \ --) &  12.300 &    8.300 &   2 \\
 1091 &  ASCC 91        &  -14.513 &    1.000 &      --- &   1.000 &  1 (\ \ --) &       --- &      --- &      --- &      --- & -- (\ \ --) &     --- &      --- &  -- \\
 1093 &  ASCC 93        &  -14.545 &   12.428 &      --- &  12.428 &  1 (\ \ --) &   -19.371 &    5.334 &    9.238 &    4.869 &  3 (\ \ --) & -23.330 &    6.360 &   3 \\
 1097 &  Alessi 40      &  -16.312 &    5.168 &    7.308 &   2.761 &  2 (\ \ --) &       --- &      --- &      --- &      --- & -- (\ \ --) &     --- &      --- &  -- \\

\end{longtable}
\end{center}
\end{landscape}
\normalsize
\clearpage

\begin{landscape}
\begin{center}
\begin{longtable}{r l | r r r r r | r r r | r r r l l}
\multicolumn{15}{l}{\textbf{Tab. \ref{tab2}.} $\overline{[M/H]}$ for our OCs in RAVE, along with uncertainties, number of individual $[M/H]$ measurements, and reference values from DAML (see Sect.
\ref{cl-met}).}
\vspace*{0.3cm}
\label{tab2}
\\
\hline
    &       &               \multicolumn{5}{|c|}{RAVE DR4}            &           \multicolumn{3}{|c|}{RAVE chem. pipeline}         &                    \multicolumn{5}{|c}{DAML}        \\
\hline
Seq & Name  &  $\overline{[M/H]}$ & e$\overline{[M/H]}$ & $\sigma \overline{[M/H]}$ & $\overline{e[M/H]^*}$ &      No. of &  $\overline{[M/H]}$ & e$\overline{[M/H]}$ &      No. of & 
$\overline{[M/H]}$ & e$\overline{[M/H]}$ &   No. & Tech.& Ref.\\
    &       &                 dex &                 dex &                       dex &                   dex & entries$^a$ &                 dex &                 dex & entries$^a$ &                
               dex &                 dex & stars &      &     \\ 
\hline \hline
\endfirsthead

\multicolumn{15}{l}{\textbf{\tablename\ \thetable{} continued}} 
\vspace*{0.3cm}
\\
\hline
    &       &               \multicolumn{5}{|c|}{RAVE DR4}            &           \multicolumn{3}{|c|}{RAVE chem. pipeline}         &                    \multicolumn{5}{|c}{DAML}           \\
\hline
Seq & Name  &  $\overline{[M/H]}$ & e$\overline{[M/H]}$ & $\sigma \overline{[M/H]}$ & $\overline{e[M/H]^*}$ &      No. of &  $\overline{[M/H]}$ & e$\overline{[M/H]}$ &      No. of & 
$\overline{[M/H]}$ & e$\overline{[M/H]}$ &   No. & Tech.& Ref.\\
    &       &                 dex &                 dex &                       dex &                   dex & entries$^a$ &                 dex &                 dex & entries$^a$ &                
               dex &                 dex & stars &      &     \\  
\hline \hline
\endhead

\hline
\endfoot

\hline
\multicolumn{15}{l}{\footnotesize $^a$The numbers in brackets are the numbers of additional good-member measurements used to compute $\overline{[M/H]}$. }\\
\multicolumn{2}{l}{\footnotesize Technical abbreviations: } & \multicolumn{13}{l}{\footnotesize DDO: DDO photometry; DTR: recalibrated values from \citet{Piatti1995}; REC: recalibrated values from
\citet{Gratton2000}; }\\
\multicolumn{2}{l}{                                       } & \multicolumn{13}{l}{\footnotesize STO: Stromgren uvby photometry; UBV: UBV photometry; SPE: high, moderate, or low-resolution
spectroscopy}\\
\multicolumn{2}{l}{\footnotesize Reference abbreviations: } & \multicolumn{13}{l}{\footnotesize Cl99: \citet{Claria1999}; G00: \citet{Gratton2000}; Ma00: \citet{Margheim2000}; N88: \citet{Nissen1988};
Pi95: \citet{Piatti1995}; }\\
\multicolumn{2}{l}{                                       } & \multicolumn{13}{l}{\footnotesize Po10: \citet{Pohnl2010}; T97: \citet{Twarog1997}}\\
\vspace*{-2cm}
\endlastfoot
\normalsize
    3  & Blanco 1            &  -0.188 &  0.082 &  0.216 &  0.098 &   7 (\ \ --) & -0.148 &  0.092 &  6 (\ \ --) &     --- &    --- & -- & ---  & ---  \\
   44  & Alessi 13           &   0.060 &  0.150 &    --- &  0.150 &   1 (\ \ --) &    --- &    --- & -- (\ \ --) &   0.170 &    --- & -- & CMD  & Po10 \\
   47  & Melotte 22          &  -0.037 &  0.026 &  0.116 &  0.116 &  20 (\ \ --) & -0.026 &  0.071 &  6 (\ \ --) &  -0.030 &  0.020 & 15 & REC  & G00  \\
   65  & NGC 1901            &  -0.018 &  0.092 &  0.160 &  0.110 &   1  (\ \ 2) &    --- &    --- & -- (\ \ --) &     --- &    --- & -- & ---  & ---  \\
   77  & Collinder 70        &   0.144 &  0.080 &    --- &  0.080 &   1 (\ \ --) &  0.150 &    --- &  1 (\ \ --) &     --- &    --- & -- & ---  & ---  \\
  127  & NGC 2354            &  -0.109 &  0.163 &  0.282 &  0.094 &   3 (\ \ --) & -0.222 &  0.126 &  3 (\ \ --) &  -0.300 &  0.020 & 12 & DDO  & Cl99 \\
  129  & Alessi 3            &  -0.275 &  0.065 &  0.160 &  0.111 &   1  (\ \ 5) & -0.387 &  0.098 & --  (\ \ 5) &     --- &    --- & -- & ---  & ---  \\
  133  & Collinder 135       &  -0.219 &  0.093 &  0.209 &  0.096 &   5 (\ \ --) & -0.249 &  0.134 &  5 (\ \ --) &     --- &    --- & -- & ---  & ---  \\
  142  & Bochum 5            &  -0.171 &  0.096 &  0.193 &  0.105 &   4 (\ \ --) & -0.205 &  0.064 &  3 (\ \ --) &     --- &    --- & -- & ---  & ---  \\
  147  & NGC 2422            &  -0.028 &  0.127 &  0.336 &  0.117 &   7 (\ \ --) &    --- &    --- & -- (\ \ --) &   0.110 &  0.100 & 11 & STO  & N88  \\
  148  & NGC 2423            &   0.067 &  0.037 &  0.190 &  0.105 &  27 (\ \ --) & -0.083 &  0.044 &  9 (\ \ --) &     --- &    --- & -- & ---  & ---  \\
  149  & Ruprecht 26         &   0.313 &  0.110 &    --- &  0.110 &   1 (\ \ --) &    --- &    --- & -- (\ \ --) &     --- &    --- & -- & ---  & ---  \\
  150  & Melotte 71          &  -0.220 &  0.100 &    --- &  0.100 &   1 (\ \ --) & -0.200 &    --- &  1 (\ \ --) &     --- &    --- & -- & ---  & ---  \\
  152  & NGC 2428            &  -0.145 &  0.095 &  0.135 &  0.105 &   2 (\ \ --) & -0.151 &  0.070 &  2 (\ \ --) &     --- &    --- & -- & ---  & ---  \\
  153  & NGC 2430            &   0.130 &  0.151 &  0.213 &  0.110 &   2 (\ \ --) & -0.080 &    --- &  1 (\ \ --) &     --- &    --- & -- & ---  & ---  \\
  158  & Ruprecht 151        &  -0.102 &  0.077 &  0.203 &  0.102 &   7 (\ \ --) & -0.207 &  0.015 &  3 (\ \ --) &     --- &    --- & -- & ---  & ---  \\
  159  & NGC 2451A           &  -0.531 &  0.220 &  0.381 &  0.101 &  --  (\ \ 3) &    --- &    --- & -- (\ \ --) &   0.000 &    --- & -- & ---  & Ma00 \\
  160  & NGC 2437            &  -0.749 &  0.334 &  0.579 &  0.139 &   3 (\ \ --) & -0.110 &    --- &  1 (\ \ --) &   0.059 &  0.090 &  1 & DTR  & T97  \\
  163  & NGC 2447            &  -0.096 &  0.147 &  0.209 &  0.110 &   2 (\ \ --) &    --- &    --- & -- (\ \ --) &     --- &    --- & -- & ---  & ---  \\
  166  & Haffner 16          &  -0.111 &  0.090 &    --- &  0.090 &   1 (\ \ --) &    --- &    --- & -- (\ \ --) &     --- &    --- & -- & ---  & ---  \\
  167  & NGC 2477            &  -0.192 &  0.002 &  0.003 &  0.074 &   2 (\ \ --) & -0.148 &  0.111 &  2 (\ \ --) &     --- &    --- & -- & ---  & ---  \\
  171  & NGC 2482            &  -0.081 &  0.141 &  0.199 &  0.105 &   2 (\ \ --) & -0.090 &    --- &  1 (\ \ --) &   0.120 &  0.020 &  3 & DDO  & T97  \\
  175  & NGC 2516            &  -0.373 &  0.092 &  0.159 &  0.098 &  --  (\ \ 3) & -0.477 &  0.142 & --  (\ \ 3) &   0.060 &  0.030 &  2 & DDO  & T97  \\
  180  & NGC 2527            &   0.057 &  0.000 &  0.001 &  0.110 &   2 (\ \ --) &  0.150 &    --- &  1 (\ \ --) &  -0.090 &    --- & -- & DDO  & Pi95 \\
  182  & Vel OB2             &  -0.288 &  0.045 &  0.207 &  0.093 &  --     (21) & -0.279 &  0.056 & --     (18) &     --- &    --- & -- & ---  & ---  \\
  192  & NGC 2567            &  -0.075 &  0.090 &    --- &  0.090 &   1 (\ \ --) & -0.040 &    --- &  1 (\ \ --) &     --- &    --- & -- & ---  & ---  \\
  201  & NGC 2632            &   0.101 &  0.029 &  0.100 &  0.108 &  12 (\ \ --) &  0.090 &  0.033 &  9 (\ \ --) &     --- &    --- & -- & ---  & ---  \\
  202  & IC 2391             &  -0.155 &  0.100 &    --- &  0.100 &   1 (\ \ --) &    --- &    --- & -- (\ \ --) &     --- &    --- & -- & ---  & ---  \\
  212  & NGC 2682            &  -0.102 &  0.036 &  0.102 &  0.081 &   8 (\ \ --) & -0.161 &  0.058 &  8 (\ \ --) &     --- &    --- & -- & ---  & ---  \\
  216  & Platais 8           &  -0.296 &  0.082 &  0.260 &  0.096 &  10 (\ \ --) & -0.209 &  0.082 &  8 (\ \ --) &     --- &    --- & -- & ---  & ---  \\
  223  & Turner 5            &  -0.210 &  0.063 &  0.110 &  0.098 &  --  (\ \ 3) & -0.046 &  0.255 & --  (\ \ 3) &     --- &    --- & -- & ---  & ---  \\
  226  & Ruprecht 80         &  -0.373 &  0.100 &    --- &  0.100 &   1 (\ \ --) & -0.460 &    --- &  1 (\ \ --) &     --- &    --- & -- & ---  & ---  \\
  230  & NGC 3036            &  -0.714 &  0.090 &    --- &  0.090 &   1 (\ \ --) &    --- &    --- & -- (\ \ --) &     --- &    --- & -- & ---  & ---  \\
  248  & Collinder 223       &  -0.217 &  0.003 &  0.004 &  0.100 &   2 (\ \ --) & -0.215 &  0.005 &  2 (\ \ --) &     --- &    --- & -- & ---  & ---  \\
  251  & Loden 153           &   0.191 &  0.130 &    --- &  0.130 &   1 (\ \ --) &    --- &    --- & -- (\ \ --) &     --- &    --- & -- & ---  & ---  \\
  254  & NGC 3324            &  -0.474 &  0.140 &    --- &  0.140 &   1 (\ \ --) &    --- &    --- & -- (\ \ --) &     --- &    --- & -- & ---  & ---  \\
  255  & vdBergh-Hagen 99    &   0.094 &  0.117 &  0.203 &  0.113 &   3 (\ \ --) & -0.290 &    --- &  1 (\ \ --) &     --- &    --- & -- & ---  & ---  \\
  259  & IC 2602             &  -0.091 &  0.026 &  0.187 &  0.101 &  50 (\ \ --) & -0.100 &  0.032 & 37 (\ \ --) &     --- &    --- & -- & ---  & ---  \\
  261  & Alessi 5            &  -0.382 &  0.100 &    --- &  0.100 &   1 (\ \ --) & -0.490 &    --- &  1 (\ \ --) &     --- &    --- & -- & ---  & ---  \\
  267  & Loden 189           &   0.202 &  0.069 &  0.098 &  0.112 &   2 (\ \ --) &  0.100 &    --- &  1 (\ \ --) &     --- &    --- & -- & ---  & ---  \\
  268  & Ruprecht 92         &   0.201 &  0.130 &    --- &  0.130 &   1 (\ \ --) &    --- &    --- & -- (\ \ --) &     --- &    --- & -- & ---  & ---  \\
  271  & Collinder 236       &   0.042 &  0.130 &    --- &  0.130 &   1 (\ \ --) &    --- &    --- & -- (\ \ --) &     --- &    --- & -- & ---  & ---  \\
  274  & Pismis 17           &  -0.145 &  0.127 &  0.284 &  0.113 &   5 (\ \ --) & -0.330 &  0.000 &  2 (\ \ --) &     --- &    --- & -- & ---  & ---  \\
  275  & Ruprecht 93         &   0.153 &  0.130 &    --- &  0.130 &   1 (\ \ --) &    --- &    --- & -- (\ \ --) &     --- &    --- & -- & ---  & ---  \\
  276  & NGC 3532            &  -0.021 &  0.057 &  0.150 &  0.112 &   7 (\ \ --) & -0.000 &    --- &  1 (\ \ --) &     --- &    --- & -- & ---  & ---  \\
  287  & NGC 3680            &  -0.167 &  0.080 &    --- &  0.080 &   1 (\ \ --) & -0.230 &    --- &  1 (\ \ --) &     --- &    --- & -- & ---  & ---  \\
  304  & ESO 130-06          &  -1.520 &  0.190 &    --- &  0.190 &   1 (\ \ --) &    --- &    --- & -- (\ \ --) &     --- &    --- & -- & ---  & ---  \\
  306  & ESO 130-08          &  -0.250 &  0.140 &    --- &  0.140 &   1 (\ \ --) &    --- &    --- & -- (\ \ --) &     --- &    --- & -- & ---  & ---  \\
  309  & NGC 4349            &  -0.021 &  0.090 &    --- &  0.090 &   1 (\ \ --) &    --- &    --- & -- (\ \ --) &     --- &    --- & -- & ---  & ---  \\
  324  & Loden 821           &  -0.099 &  0.130 &    --- &  0.130 &   1 (\ \ --) &    --- &    --- & -- (\ \ --) &     --- &    --- & -- & ---  & ---  \\
  327  & Basel 18            &   0.043 &  0.270 &    --- &  0.270 &   1 (\ \ --) &    --- &    --- & -- (\ \ --) &     --- &    --- & -- & ---  & ---  \\
  337  & Platais 12          &  -0.007 &  0.054 &  0.241 &  0.128 &  20 (\ \ --) &  0.027 &  0.117 &  4 (\ \ --) &     --- &    --- & -- & ---  & ---  \\
  338  & NGC 5316            &   0.045 &  0.083 &  0.118 &  0.130 &   2 (\ \ --) &  0.200 &    --- &  1 (\ \ --) &  -0.020 &    --- & -- & DDO  & Pi95 \\
  339  & Loden 995           &  -0.131 &  0.090 &    --- &  0.090 &   1 (\ \ --) & -0.200 &    --- &  1 (\ \ --) &     --- &    --- & -- & ---  & ---  \\
  343  & Ruprecht 110        &  -0.359 &  0.193 &  0.273 &  0.085 &   2 (\ \ --) & -0.120 &    --- &  1 (\ \ --) &     --- &    --- & -- & ---  & ---  \\
  356  & Alessi 6            &  -0.154 &  0.080 &    --- &  0.080 &   1 (\ \ --) & -0.160 &    --- &  1 (\ \ --) &     --- &    --- & -- & ---  & ---  \\
  367  & Nor OB5             &  -2.063 &  0.361 &  0.625 &  0.095 &   1  (\ \ 2) &    --- &    --- & -- (\ \ --) &     --- &    --- & -- & ---  & ---  \\
  382  & NGC 6204            &  -1.053 &  0.150 &    --- &  0.150 &   1 (\ \ --) &    --- &    --- & -- (\ \ --) &     --- &    --- & -- & ---  & ---  \\
  393  & Sco OB4             &  -0.086 &  0.067 &  0.258 &  0.129 &  15 (\ \ --) & -0.180 &  0.129 &  4 (\ \ --) &     --- &    --- & -- & ---  & ---  \\
  397  & IC 4651             &  -0.128 &  0.029 &  0.059 &  0.082 &   4 (\ \ --) & -0.097 &  0.041 &  4 (\ \ --) &     --- &    --- & -- & ---  & ---  \\
  399  & Antalova 1          &  -0.655 &  0.190 &    --- &  0.190 &   1 (\ \ --) &    --- &    --- & -- (\ \ --) &     --- &    --- & -- & ---  & ---  \\
  403  & Trumpler 27         &  -0.193 &  0.013 &  0.019 &  0.090 &   2 (\ \ --) & -0.080 &  0.170 &  2 (\ \ --) &     --- &    --- & -- & ---  & ---  \\
  404  & Trumpler 28         &   0.326 &  0.110 &    --- &  0.110 &   1 (\ \ --) &    --- &    --- & -- (\ \ --) &     --- &    --- & -- & ---  & ---  \\
  405  & ESO 139-13          &  -0.320 &  0.080 &    --- &  0.080 &   1 (\ \ --) & -0.380 &    --- &  1 (\ \ --) &     --- &    --- & -- & ---  & ---  \\
  408  & NGC 6405            &   0.195 &  0.072 &  0.124 &  0.110 &   3 (\ \ --) &    --- &    --- & -- (\ \ --) &   0.060 &  0.150 & -- & UBV  & G00  \\
  410  & Alessi 9            &  -0.580 &  0.090 &    --- &  0.090 &   1 (\ \ --) &    --- &    --- & -- (\ \ --) &     --- &    --- & -- & ---  & ---  \\
  411  & NGC 6416            &  -0.613 &  0.556 &  0.786 &  0.095 &   2 (\ \ --) & -0.080 &    --- &  1 (\ \ --) &     --- &    --- & -- & ---  & ---  \\
  429  & NGC 6546            &  -0.334 &  0.080 &    --- &  0.080 &   1 (\ \ --) & -0.440 &    --- &  1 (\ \ --) &     --- &    --- & -- & ---  & ---  \\
  430  & vdBergh 113         &  -0.358 &  0.090 &    --- &  0.090 &   1 (\ \ --) &    --- &    --- & -- (\ \ --) &     --- &    --- & -- & ---  & ---  \\
  435  & Sgr OB7             &  -0.056 &  0.378 &  0.535 &  0.120 &   2 (\ \ --) &    --- &    --- & -- (\ \ --) &     --- &    --- & -- & ---  & ---  \\
  436  & Markarian 38        &   0.180 &  0.001 &  0.001 &  0.095 &   2 (\ \ --) &  0.445 &  0.025 &  2 (\ \ --) &     --- &    --- & -- & ---  & ---  \\
  444  & NGC 6618            &   0.132 &  0.110 &    --- &  0.110 &   1 (\ \ --) &  0.020 &    --- &  1 (\ \ --) &     --- &    --- & -- & ---  & ---  \\
  445  & Trumpler 33         &  -1.544 &  0.190 &    --- &  0.190 &   1 (\ \ --) &    --- &    --- & -- (\ \ --) &     --- &    --- & -- & ---  & ---  \\
  449  & IC 4725             &  -0.257 &  0.007 &  0.011 &  0.140 &   2 (\ \ --) &    --- &    --- & -- (\ \ --) &   0.170 &  0.060 & -- & HRS  & G00  \\
  452  & Ruprecht 145        &  -0.127 &  0.080 &    --- &  0.080 &   1 (\ \ --) & -0.390 &    --- &  1 (\ \ --) &     --- &    --- & -- & ---  & ---  \\
 1033  & ASCC 33             &  -0.166 &  0.080 &  0.138 &  0.090 &  --  (\ \ 3) & -0.144 &  0.134 & --  (\ \ 3) &     --- &    --- & -- & ---  & ---  \\
 1057  & ASCC 57             &  -0.296 &  0.076 &  0.132 &  0.095 &   3 (\ \ --) & -0.178 &  0.114 &  3 (\ \ --) &     --- &    --- & -- & ---  & ---  \\
 1078  & ASCC 78             &  -0.062 &  0.042 &  0.110 &  0.079 &   7 (\ \ --) & -0.042 &  0.038 &  7 (\ \ --) &     --- &    --- & -- & ---  & ---  \\
 1089  & Alessi 24           &  -0.133 &  0.080 &  0.113 &  0.089 &   2 (\ \ --) & -0.112 &  0.090 &  2 (\ \ --) &     --- &    --- & -- & ---  & ---  \\
 1091  & ASCC 91             &  -0.035 &  0.080 &    --- &  0.080 &   1 (\ \ --) & -0.250 &    --- &  1 (\ \ --) &     --- &    --- & -- & ---  & ---  \\
 1097  & Alessi 40           &   0.129 &  0.166 &  0.234 &  0.080 &   2 (\ \ --) & -0.160 &    --- &  1 (\ \ --) &     --- &    --- & -- & ---  & ---  \\

\end{longtable}
\end{center}
\end{landscape}
\clearpage

\end{document}